\documentclass[12pt,a4paper]{article}
\usepackage[latin1]{inputenc}
\usepackage[T1]{fontenc}
\usepackage{amsmath}
\usepackage{amsfonts}
\usepackage{amssymb}
\usepackage[english]{babel}
\usepackage[dvips]{graphicx,color}
\def\graph#1{
\begin{array}{c}\mbox{
\includegraphics*[height=1cm]{#1.eps}
}\end{array}
             }
\begin{document}
\newcommand {\ee}{\end{equation}}
\newcommand {\bea}{\begin{eqnarray}}
\newcommand {\eea}{\end{eqnarray}}
\newcommand {\nn}{\nonumber \\}
\newcommand {\Tr}{{\rm Tr\,}}
\newcommand {\tr}{{\rm tr\,}}
\newcommand {\e}{{\rm e}}
\newcommand {\etal}{{\it et al.}}
\newcommand {\m}{\mu}
\newcommand {\n}{\nu}
\newcommand {\pl}{\partial}
\newcommand {\p} {\phi}
\newcommand {\vp}{\varphi}
\newcommand {\vpc}{\varphi_c}
\newcommand {\al}{\alpha}
\newcommand {\be}{\beta}
\newcommand {\ga}{\gamma}
\newcommand {\Ga}{\Gamma}
\newcommand {\x}{\xi}
\newcommand {\ka}{\kappa}
\newcommand {\la}{\lambda}
\newcommand {\La}{\Lambda}
\newcommand {\si}{\sigma}
\newcommand {\sh}{\theta}
\newcommand {\Th}{\Theta}
\newcommand {\om}{\omega}
\newcommand {\Om}{\Omega}
\newcommand {\ep}{\epsilon}
\newcommand {\vep}{\varepsilon}
\newcommand {\na}{\nabla}
\newcommand {\del}  {\delta}
\newcommand {\Del}  {\Delta}
\newcommand {\mn}{{\mu\nu}}
\newcommand {\ls}   {{\lambda\sigma}}
\newcommand {\ab}   {{\alpha\beta}}
\newcommand {\half}{ {\frac{1}{2}} }
\newcommand {\third}{ {\frac{1}{3}} }
\newcommand {\fourth} {\frac{1}{4} }
\newcommand {\sixth} {\frac{1}{6} }
\newcommand {\sqg} {\sqrt{g}}
\newcommand {\sqtwo} {\sqrt{2}}
\newcommand {\fg}  {\sqrt[4]{g}}
\newcommand {\invfg}  {\frac{1}{\sqrt[4]{g}}}
\newcommand {\sqZ} {\sqrt{Z}}
\newcommand {\gbar}{\bar{g}}
\newcommand {\sqk} {\sqrt{\kappa}}
\newcommand {\sqt} {\sqrt{t}}
\newcommand {\reg} {\frac{1}{\epsilon}}
\newcommand {\fpisq} {(4\pi)^2}
\newcommand {\Lcal}{{\cal L}}
\newcommand {\Ocal}{{\cal O}}
\newcommand {\Dcal}{{\cal D}}
\newcommand {\Ncal}{{\cal N}}
\newcommand {\Mcal}{{\cal M}}
\newcommand {\scal}{{\cal s}}
\newcommand {\Dvec}{{\hat D}}   
\newcommand {\dvec}{{\vec d}}
\newcommand {\Evec}{{\vec E}}
\newcommand {\Hvec}{{\vec H}}
\newcommand {\Vvec}{{\vec V}}
\newcommand {\Btil}{{\tilde B}}
\newcommand {\ctil}{{\tilde c}}
\newcommand {\Ftil}{{\tilde F}}
\newcommand {\Stil}{{\tilde S}}
\newcommand {\Ztil}{{\tilde Z}}
\newcommand {\Ktil}  {{\tilde K}}
\newcommand {\Ltil}  {{\tilde L}}
\newcommand {\Qtil}  {{\tilde Q}}
\newcommand {\altil}{{\tilde \alpha}}
\newcommand {\betil}{{\tilde \beta}}
\newcommand {\latil}{{\tilde \lambda}}
\newcommand {\ptil}{{\tilde \phi}}
\newcommand {\Ptil}{{\tilde \Phi}}
\newcommand {\natil} {{\tilde \nabla}}
\newcommand {\ttil} {{\tilde t}}
\newcommand {\Rhat}{{\hat R}}
\newcommand {\Shat}{{\hat S}}
\newcommand {\shat}{{\hat s}}
\newcommand {\Dhat}{{\hat D}}   
\newcommand {\Vhat}{{\hat V}}   
\newcommand {\xhat}{{\hat x}}
\newcommand {\Zhat}{{\hat Z}}
\newcommand {\Gahat}{{\hat \Gamma}}
\newcommand {\nah} {{\hat \nabla}}
\newcommand {\gh}  {{\hat g}}
\newcommand {\labar}{{\bar \lambda}}
\newcommand {\cbar}{{\bar c}}
\newcommand {\bbar}{{\bar b}}
\newcommand {\Bbar}{{\bar B}}
\newcommand {\Dbar}{{\bar D}}
\newcommand {\Qbar}{{\bar Q}}
\newcommand {\Kbar}  {{\bar K}}
\newcommand {\Lbar}  {{\bar L}}
\newcommand {\Wbar}  {{\bar W}}
\newcommand {\albar}{{\bar \alpha}}
\newcommand {\psibar}{{\bar \psi}}
\newcommand {\Psibar}{{\bar \Psi}}
\newcommand {\sibar}{{\bar \si}}
\newcommand {\thbar}{{\bar \theta}}
\newcommand {\chibar}{{\bar \chi}}
\newcommand {\xibar}{{\bar \xi}}
\newcommand {\bbartil}{{\tilde {\bar b}}}
\newcommand {\aldot} {{\dot \al}}
\newcommand {\bedot} {{\dot \be}}
\newcommand {\deldot} {{\dot \delta}}
\newcommand {\gadot} {{\dot \ga}}
\newcommand  {\vz}{{v_0}}
\newcommand  {\ez}{{e_0}}
\newcommand {\intfx} {{\int d^4x}}
\newcommand {\inttx} {{\int d^2x}}
\newcommand {\change} {\leftrightarrow}
\newcommand {\ra} {\rightarrow}
\newcommand {\larrow} {\leftarrow}
\newcommand {\ul}   {\underline}
\newcommand {\pr}   {{\quad .}}
\newcommand {\com}  {{\quad ,}}
\newcommand {\q}    {\quad}
\newcommand {\qq}   {\quad\quad}
\newcommand {\qqq}   {\quad\quad\quad}
\newcommand {\qqqq}   {\quad\quad\quad\quad}
\newcommand {\qqqqq}   {\quad\quad\quad\quad\quad}
\newcommand {\qqqqqq}   {\quad\quad\quad\quad\quad\quad}
\newcommand {\qqqqqqq}   {\quad\quad\quad\quad\quad\quad\quad}
\newcommand {\lb}    {\linebreak}
\newcommand {\nl}    {\newline}

\newcommand {\vs}[1]  { \vspace*{#1 mm} }

\newcommand {\MPL}  {Mod.Phys.Lett.}
\newcommand {\IJMP}  {Int.Jour.Mod.Phys.}
\newcommand {\NP}   {Nucl.Phys.}
\newcommand {\PL}   {Phys.Lett.}
\newcommand {\PR}   {Phys.Rev.}
\newcommand {\PRL}   {Phys.Rev.Lett.}
\newcommand {\CMP}  {Commun.Math.Phys.}
\newcommand {\JMP}  {Jour.Math.Phys.}
\newcommand {\AP}   {Ann.of Phys.}
\newcommand {\PTP}  {Prog.Theor.Phys.}
\newcommand {\NC}   {Nuovo Cim.}
\newcommand {\CQG}  {Class.Quantum.Grav.}


\font\smallr=cmr5
\def\ocirc#1{#1^{^{{\hbox{\smallr\llap{o}}}}}}
\def\ogamma{\ocirc{\gamma}{}}
\def\oM{{\buildrel {\hbox{\smallr{o}}} \over M}}
\def\osigma{\ocirc{\sigma}{}}

\def\overleftrightarrow#1{\vbox{\ialign{##\crcr
 $\leftrightarrow$\crcr\noalign{\kern-1pt\nointerlineskip}
 $\hfil\displaystyle{#1}\hfil$\crcr}}}
\def\overnab{{\overleftrightarrow\nabslash}}

\def\va{{a}}
\def\vb{{b}}
\def\vc{{c}}
\def\tilpsi{{\tilde\psi}}
\def\tbpsi{{\tilde{\bar\psi}}}

\def\delL{{\delta_{LL}}}
\def\delG{{\delta_{G}}}
\def\delc{{\delta_{cov}}}

\newcommand {\sqxx}  {\sqrt {x^2+1}}   
\newcommand {\gago}  {\gamma^5}
\newcommand {\Rtil}  {{\tilde R}}
\newcommand {\Pp}  {P_+}
\newcommand {\Pm}  {P_-}
\newcommand {\GfMp}  {G^{5M}_+}
\newcommand {\GfMpm}  {G^{5M'}_-}
\newcommand {\GfMm}  {G^{5M}_-}
\newcommand {\Omp}  {\Omega_+}    
\newcommand {\Omm}  {\Omega_-}
\def\Aslash{{}\hbox{\hskip2pt\vtop
 {\baselineskip23pt\hbox{}\vskip-24pt\hbox{/}}
 \hskip-11.5pt $A$}}
\def\Rslash{{}\hbox{\hskip2pt\vtop
 {\baselineskip23pt\hbox{}\vskip-24pt\hbox{/}}
 \hskip-11.5pt $R$}}
\def\kslash{
{}\hbox       {\hskip2pt\vtop
                   {\baselineskip23pt\hbox{}\vskip-24pt\hbox{/}}
               \hskip-8.5pt $k$}
           }    
\def\qslash{
{}\hbox       {\hskip2pt\vtop
                   {\baselineskip23pt\hbox{}\vskip-24pt\hbox{/}}
               \hskip-8.5pt $q$}
           }    
\def\dslash{
{}\hbox       {\hskip2pt\vtop
                   {\baselineskip23pt\hbox{}\vskip-24pt\hbox{/}}
               \hskip-8.5pt $\partial$}
           }    
\def\dbslash{{}\hbox{\hskip2pt\vtop
 {\baselineskip23pt\hbox{}\vskip-24pt\hbox{$\backslash$}}
 \hskip-11.5pt $\partial$}}
\def\Kbslash{{}\hbox{\hskip2pt\vtop
 {\baselineskip23pt\hbox{}\vskip-24pt\hbox{$\backslash$}}
 \hskip-11.5pt $K$}}
\def\Ktilbslash{{}\hbox{\hskip2pt\vtop
 {\baselineskip23pt\hbox{}\vskip-24pt\hbox{$\backslash$}}
 \hskip-11.5pt ${\tilde K}$}}
\def\Ltilbslash{{}\hbox{\hskip2pt\vtop
 {\baselineskip23pt\hbox{}\vskip-24pt\hbox{$\backslash$}}
 \hskip-11.5pt ${\tilde L}$}}
\def\Qtilbslash{{}\hbox{\hskip2pt\vtop
 {\baselineskip23pt\hbox{}\vskip-24pt\hbox{$\backslash$}}
 \hskip-11.5pt ${\tilde Q}$}}
\def\Rtilbslash{{}\hbox{\hskip2pt\vtop
 {\baselineskip23pt\hbox{}\vskip-24pt\hbox{$\backslash$}}
 \hskip-11.5pt ${\tilde R}$}}
\def\Kbarbslash{{}\hbox{\hskip2pt\vtop
 {\baselineskip23pt\hbox{}\vskip-24pt\hbox{$\backslash$}}
 \hskip-11.5pt ${\bar K}$}}
\def\Lbarbslash{{}\hbox{\hskip2pt\vtop
 {\baselineskip23pt\hbox{}\vskip-24pt\hbox{$\backslash$}}
 \hskip-11.5pt ${\bar L}$}}
\def\Rbarbslash{{}\hbox{\hskip2pt\vtop
 {\baselineskip23pt\hbox{}\vskip-24pt\hbox{$\backslash$}}
 \hskip-11.5pt ${\bar R}$}}
\def\Qbarbslash{{}\hbox{\hskip2pt\vtop
 {\baselineskip23pt\hbox{}\vskip-24pt\hbox{$\backslash$}}
 \hskip-11.5pt ${\bar Q}$}}
\def\Acalbslash{{}\hbox{\hskip2pt\vtop
 {\baselineskip23pt\hbox{}\vskip-24pt\hbox{$\backslash$}}
 \hskip-11.5pt ${\cal A}$}}

\begin{flushright}
March 2006\\
UWThPh-2006-7\\
hep-th/0603214 \\
\end{flushright}

\vspace{0.5cm}

\begin{center}
{\Large\bf 
Graphical Representation of Supersymmetry 
}

\vspace{1.5cm}
{\large Shoichi ICHINOSE
         \footnote{
On leave of absence from Laboratory of Physics, 
School of Food and Nutritional Sciences, 
University of Shizuoka, Yada 52-1, Shizuoka 422-8526, Japan
(untill 31 March, 2006).\\          
E-mail address:\ ichinose@u-shizuoka-ken.ac.jp
                  }
}
\vspace{1cm}

{\large 
Institut f{\"u}r Theoretische Physik, Universit{\"a}t Wien\\
Boltzmanngasse 5, A-1090 Vienna, Austria  }


\end{center}

\vfill

{\large Abstract}\nl
A graphical representation of supersymmetry
is presented. It clearly expresses the chiral flow 
appearing in SUSY quantities, by representing
spinors by {\it directed lines} (arrows). The chiral suffixes
are expressed by the directions (up, down, left, right)
of the arrows. 
The SL(2,C) invariants are represented by {\it wedges}. 
Both the Weyl spinor and the Majorana spinor
are treated. 
We are free from the complicated symbols of spinor suffixes.
The method is applied to the 5D supersymmetry. 
Many applications are expected. 
The result is suitable for coding a computer program
and is highly expected to be applicable to 
various SUSY theories (including Supergravity) 
in various dimensions.

\vspace{0.5cm}
PACS NO:\ 
02.10.Ox,\ 
02.70.-c,\ 
02.90.+p,\ 
11.30.Pb,\ 
11.30.Rd,\ 
\nl
Key Words:\ Graphical representation, Supersymmetry, 
Spinor suffix, Chiral suffix, Graph index


\section{Introduction}
Since supersymmetry was born, more than quarter
century has passed. Although super particles
are not yet discovered in experiments, everybody now
admits its importance as one possible extension
beyond the standard model. Some important models,
such as 4 dimensional ${\cal N}=4$ SUSY YM 
, give us 
a deep insight in the non-perturbative aspects
of the quantum field theories. 
Because of the high 
symmetry, the dynamics is strongly constrained and
it makes possible to analyse the nonperturbative
aspects. The BPS state is such a
representative. 

The beauty of SUSY
theory comes from the harmony between bosons and fermions. 
At the cost of the high symmetry, the SUSY fields
generally carry many suffixes:\ 
chiral-suffixes ($\al$), anti-chiral suffixes ($\aldot$)
in addition to usual ones:\ 
gauge suffixes ($i,j,..$), Lorentz suffixes ($m,n,\cdots$).
The usual notation is, for example, $\psi^{i~\aldot}_{m\al}$.
Many suffixes are ``crowded'' within one character $\psi$.
Whether the meaning of a quantity is clearly read, 
sometimes crucially depends on
the way of description. In the case of quantities
with many suffixes,  
we are sometimes lost in the ``jungle'' of suffixes.
In this circumstance we propose a new representation
to express SUSY quantities.
It has the following properties.
\begin{enumerate}
\item
All suffix-information is expressed.
\item
Suffixes are suppressed as much as possible. 
Instead we use the "geometrical" notation:\ lines, arrows, $\cdots$.
\footnote
{In this sense, the use of ``differential forms'' instead of the
tensorial quantities is the similar line of
simplification.
}
Particularly, 
contracted suffixes (we call them ``dummy'' suffixes) are
expressed by vertices (for fermion suffixes) or lines (for vector suffixes and SU(2)$_R$-suffixes).
\item
The chiral flow is manifest.
\item
The {\it graphical indices} (defined in Sect.6) specify a spinorial quantity.
\end{enumerate}
The content of the present paper is an improved version of Ref.\cite{SI03}. 
Partial results are also reported in Ref.\cite{SI04}. 

Another quantity with many suffixes
is the Riemann tensor appearing in the general relativity.
It was already graphically represented \cite{SI95CQG}
and some applications have appeared\cite{II97JMP,
GI00CQG}. 

We follow the notation and the convention of the textbook by
Wess and Bagger\cite{WB92}. Many (graphical) relations
appearing in the present paper (except Sec.6 and Sec.8)
appear in the textbook.

The results of this paper have recently been applied to the
C-program of SUSY calculation\cite{SI06UW08}.

\section{Definition}

\subsection{Basic Ingredients}
Let us represent the Weyl fermion $\psi_\al,\psibar_\aldot$
(2 complex components, $\al,\aldot=1,2,$ ) 
and their 'suffixes-raised' partners as in Fig.1.
Raising and lowering the spinor suffixes is done by
the antisymmetric tensors $\ep^\ab, \ep_{\ab}$:\ 
\begin{eqnarray}
\psi^\al=\ep^\ab\psi_\be\com\q
\psibar^\aldot=\ep^{\aldot\bedot}\psibar_\bedot\com\q
\ep^{12}=\ep_{21}=1\com
\label{def1}
\end{eqnarray}
where $\ep^\ab$ and $\ep_\ab$ are in the inverse relation:\ 
$\ep^\ab\ep_{\be\ga}=\del^\al_\ga$. 
\begin{figure}
\begin{center}
\includegraphics*[height=5cm]{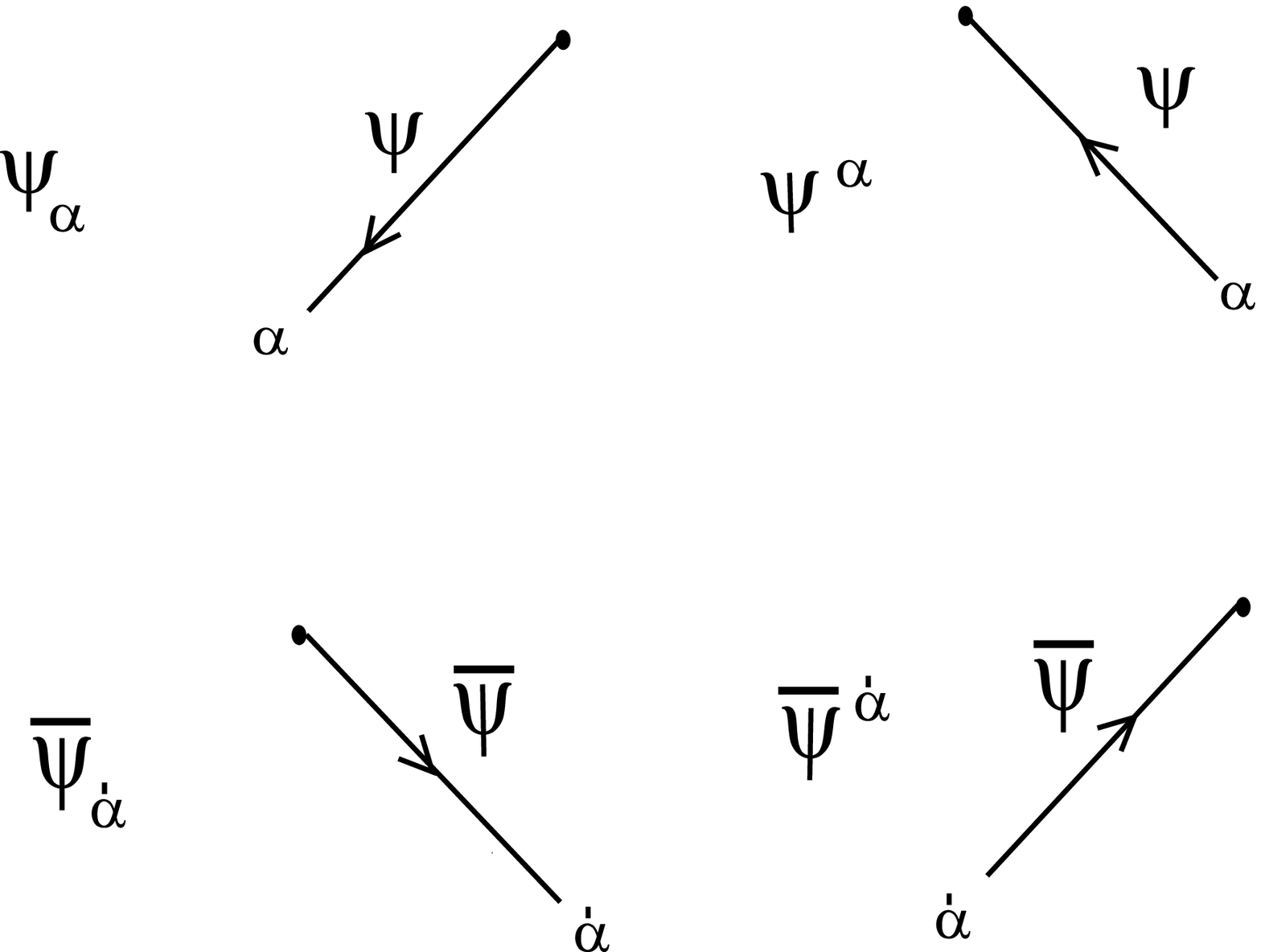}
\end{center}
\caption{
Weyl fermions. 
}
\end{figure}

\vs 2

{\bf Graphical Rule 1 ( Fig.1)} \nl
1. The arrow is pointed to the {\it left} for the chiral field $\psi$ and
to the {\it right} for the anti-chiral one $\psibar$.(Complex structure)\nl
2. The arrow is pointed to the {\it up} for the upper-suffix quantity and
to the {\it down} for lower-suffix quantity.(Symplectic structure)\nl
3. All spinor suffixes ($\al,\be,\cdots$;$\aldot,\bedot,\cdots$)
 are labeled at the lowest position of arrow lines.

\vs 2

The choice of 3 is fixed by the condition that, 
when we express 
the basic SL(2,C)(Lorentz) invariants 
$\psi^\al\chi_\al$(NW-SE convention)
=$\ep_\ab\psi^\al\psi^\be$
,
$\psibar_\aldot\chibar^\aldot$(SW-NE convention)
=$\ep^{\aldot\bedot}\psibar_\aldot\chibar_\bedot$
where suffixes are contracted by the anti-symmetric
tensor
\footnote{
NorthWest-SouthEast(NW-SE), SouthWest-NorthEast(SW-NE).
}
, 
the arrows {\it continuously} flow along the lines ( see Fig.4 
which will be explained later)
without changing the order of the spinor-field graphs. 

\vs 2

{\bf Graphical Rule 2} \nl
Every spinor graph is {\it anticommuting} 
in the horizontal direction. \nl

The derivatives of fermions are expressed as in Fig.2.
We give it as a rule.

\begin{figure}
\begin{center}
\includegraphics*[height=6cm]{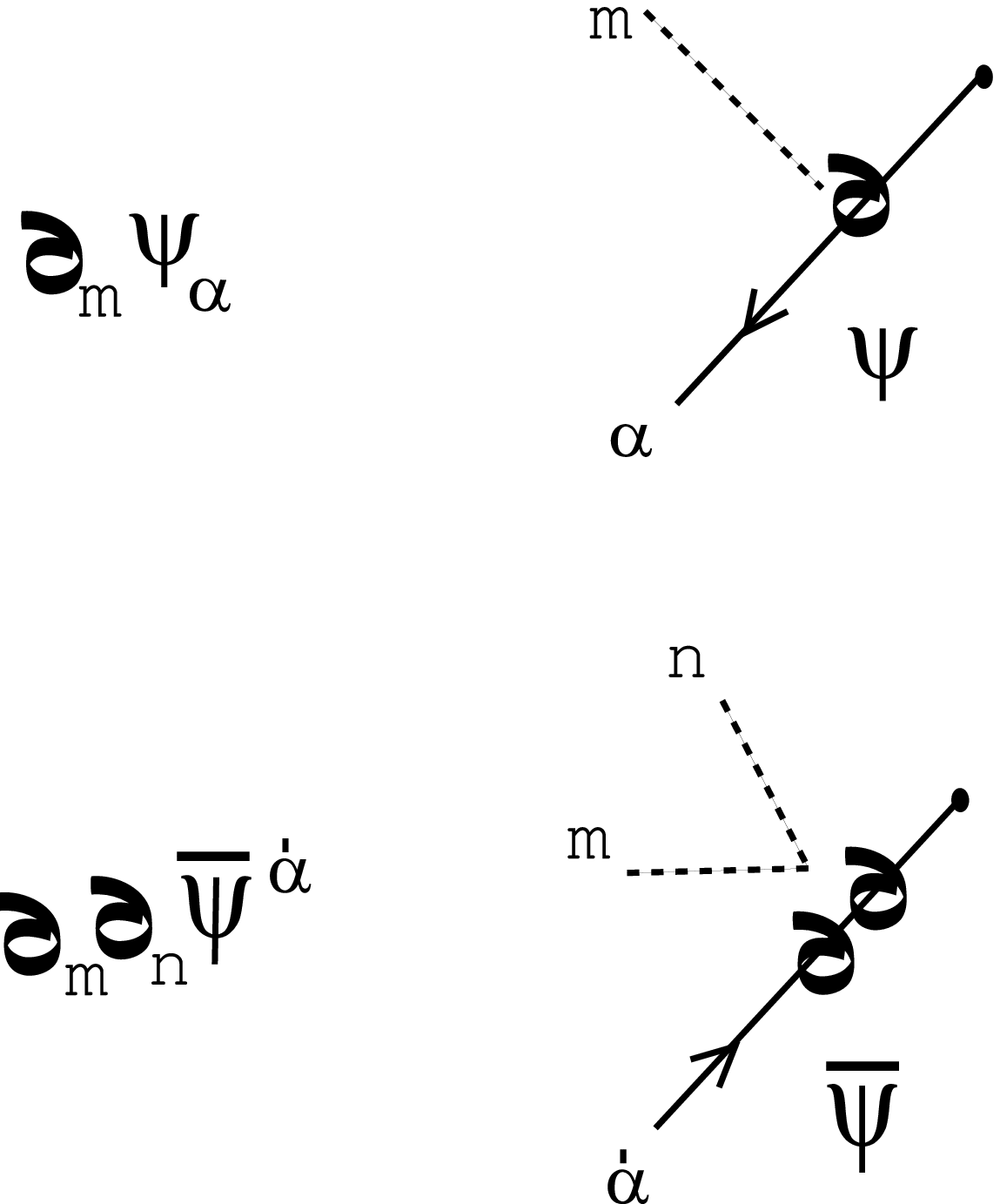}
\end{center}
\caption{
Derivatives of fermions. 
}
\end{figure}

\vs 2

{\bf Graphical Rule 3 ( Fig.2)} \nl
1. For each derivative, attach the derivative symbol "$\pl$"
to the corresponding spinor-arrow with a dotted line 
as in Fig.2. At the end of the line, the Lorentz suffix
of the derivative is described.\nl
2. The order of the derivative lines described above is irrelevant
because the derivative $\pl_\m$ is {\it commutative}.

\vs 2

Following the above rule, 
the elements of the SL(2,C) $\si$-matrix are expressed as in Fig.3. 
\begin{figure}
\begin{center}
\includegraphics*[height=8cm]{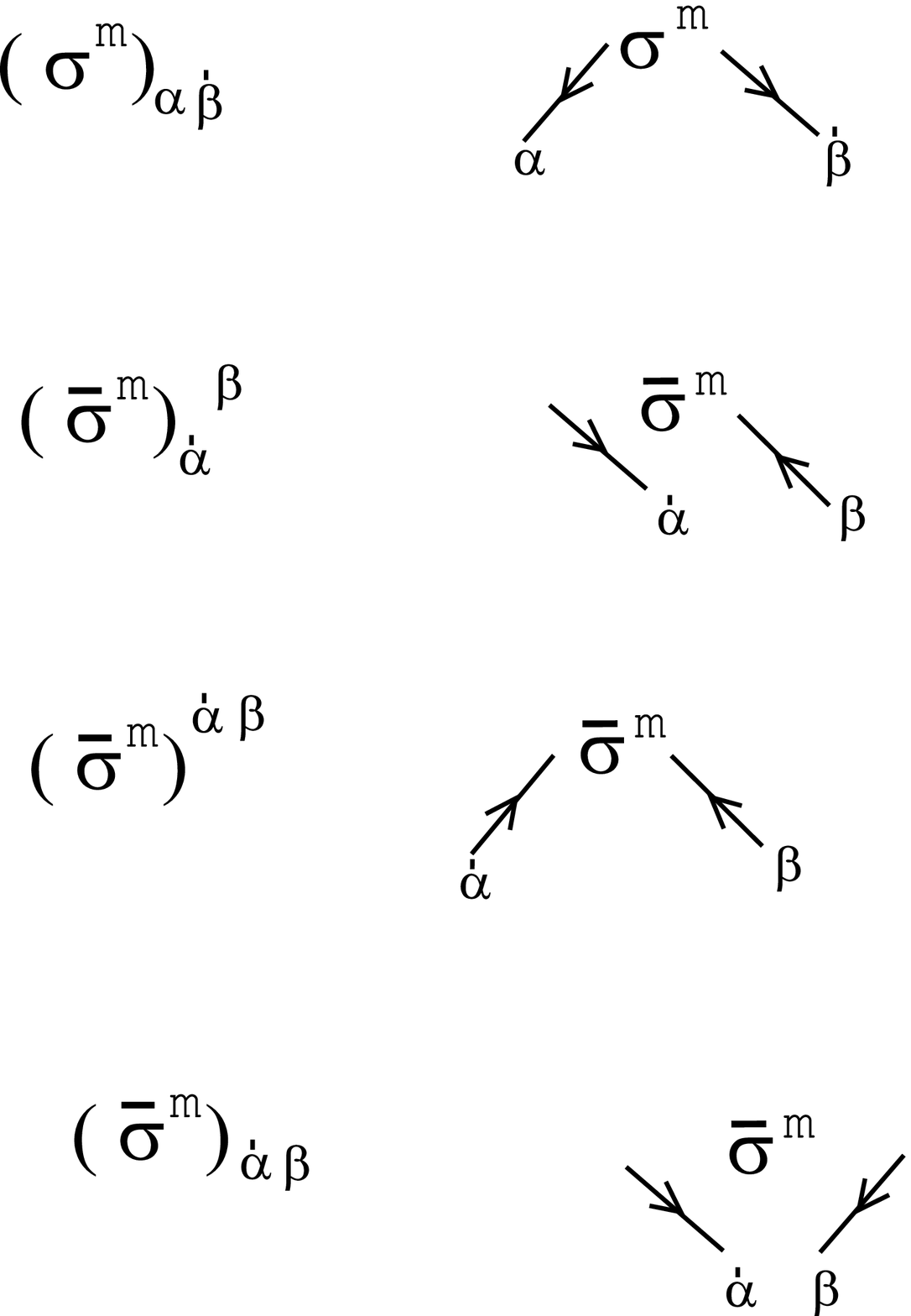}
\end{center}
   \caption{
Elements of SL(2,C) $\si$-matrices. 
$(\si^m)_{\al\bedot}$ and $(\sibar^m)^{\aldot\be}$ are
the standard form.
  }
\end{figure}
In Fig.3, the two arrows are directed 'horizontally outward' for $\si$, 
whereas 'horizontally inward' for $\sibar$. 
We consider $(\si^m)_{\al\bedot}$ and $(\sibar^m)^{\aldot\be}$ are
the standard form which is basically used in this text.

\subsection{Spinor suffix contraction}
Lorentz covariants and invariants are expressed by the {\it contraction} of 
the spinor suffixes. We take the convention of 
NW-SE contraction for the chiral suffix $\al$,
and SW-NE contraction for the anti-chiral one $\aldot$. 
They are expressed by connecting the corresponding suffix-ends
as in Fig.4. 
\begin{figure}
\begin{center}
\includegraphics*[height=4cm]{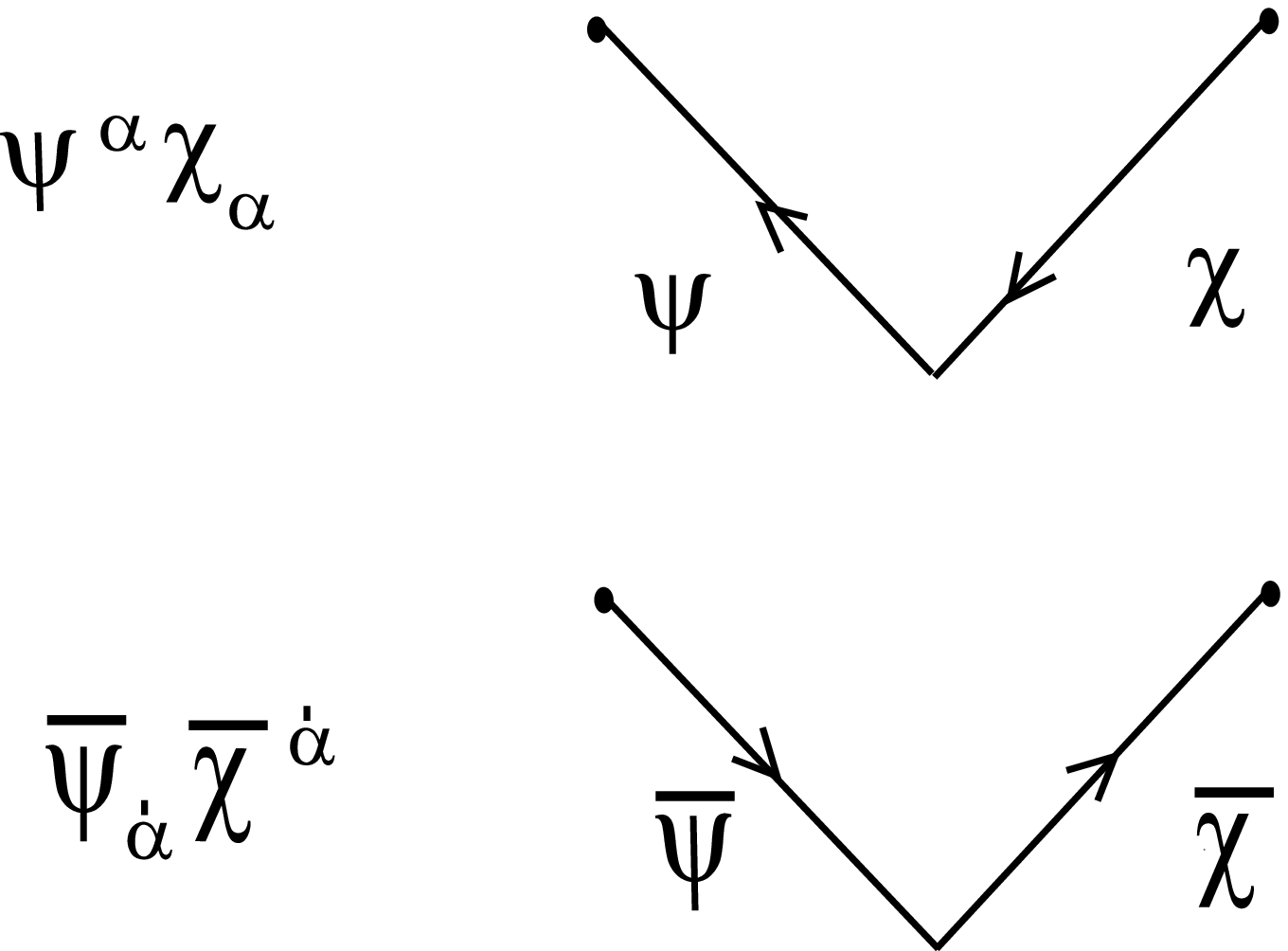}
\end{center}
   \caption{
Contraction of spinor suffixes. 
[above]\ N(orth)W(est)-S(outh)E(ast) contraction for the chiral suffix
$\al$
($
\psi^\al\chi_\al=\ep_\ab\psi^\al\psi^\be
$)
:\ 
[below]\ SW-NE contraction for the anti-chiral suffix $\aldot$
($
\psibar_\aldot\chibar^\aldot
=\ep^{\aldot\bedot}\psibar_\aldot\chibar_\bedot
$). 
   }
\end{figure}
The {\it wedge} structure, in Fig.4, characterizes all spinor
contractions in the following. 
For the {\it chiral}-suffixes contraction,
the wedge 'runs' to the {\it left}, whereas the {\it anti-chiral} ones 
to the {\it right}.

\vs 2

{\bf Graphical Rule 4:\ Spinor Suffix Contraction ( Fig.4)} \nl
The contraction is expressed by connecting the corresponding
suffix-ends. 

\vs 2

Two Lorentz vectors,
$\chi^\al (\si^m)_{\al\bedot}\psibar^\bedot$ and
$\psibar_\aldot(\sibar^m)^{\aldot\be}\chi_\be$
are expressed as in Fig.5. 
\begin{figure}
\begin{center}
\includegraphics*[height=4cm]{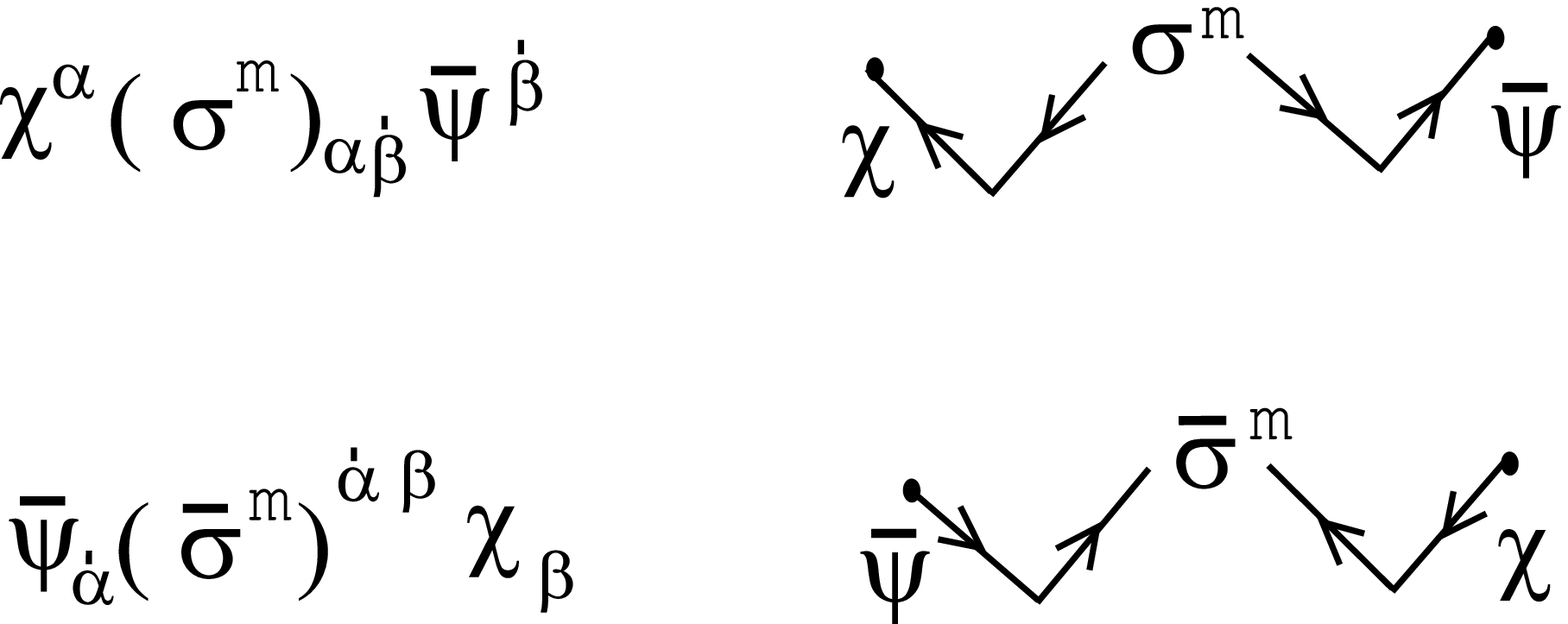}
\end{center}
   \caption{
Two Lorentz vectors 
$\chi^\al (\si^m)_{\al\bedot}\psibar^\bedot$ and
$\psibar_\aldot(\sibar^m)^{\aldot\be}\chi_\be$. 
The double wedge structure appears. 
   }
\end{figure}
The {\it double-wedge} structure, in Fig.5, characterizes
the {\it vector} quantities which involve one $\si^m$ or one $\sibar^m$. 

Next we take examples with two $\si$'s. 
$(\si^n)_{\al\aldot}(\sibar^m)^{\aldot\be}$ and 
$(\sibar^n)^{\aldot\al}(\si^m)_{\al\bedot}$
are expressed as in Fig.6. 
\begin{figure}
\begin{center}
\includegraphics*[height=6cm]{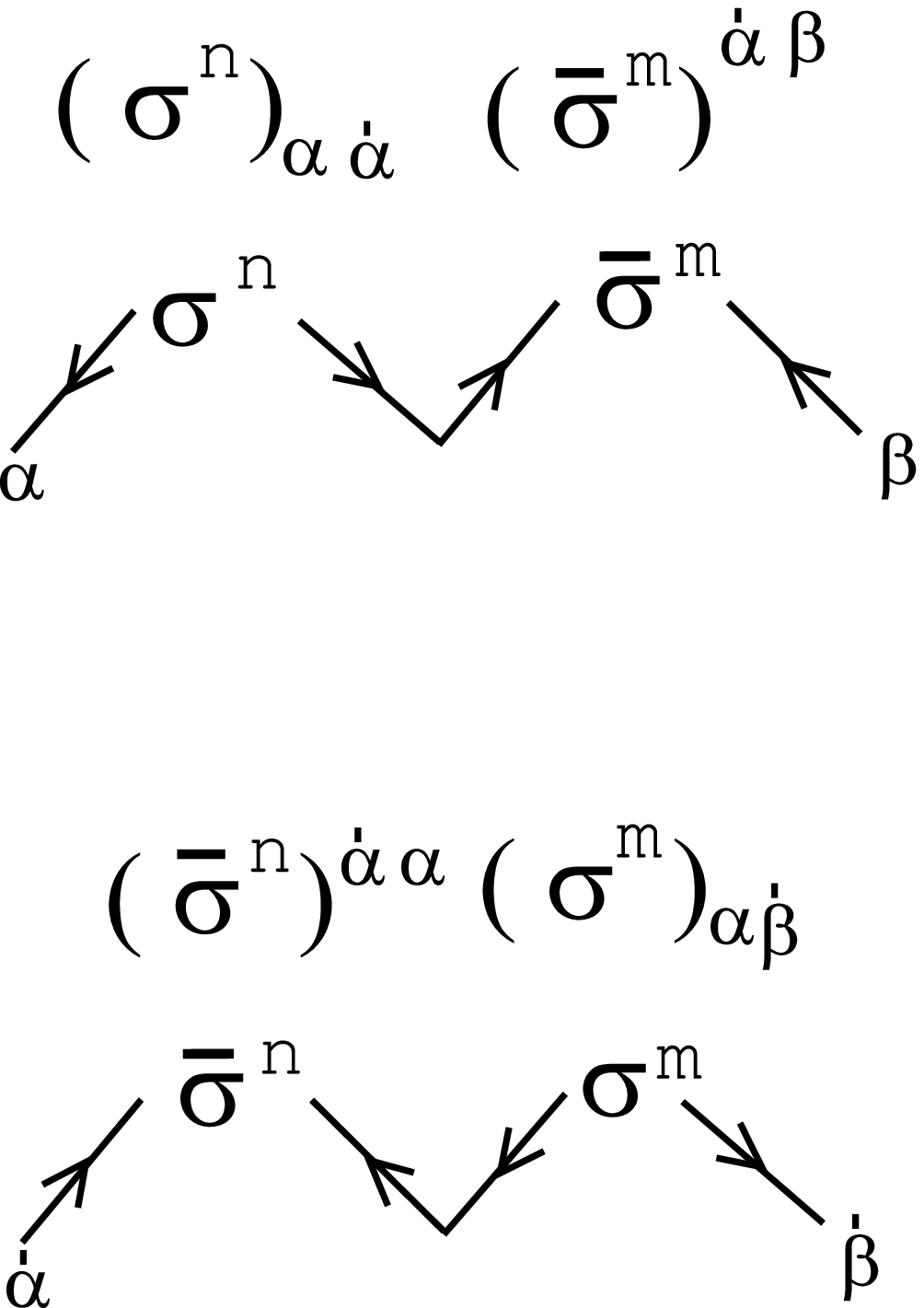}
\end{center}
   \caption{
$(\si^n)_{\al\aldot}(\sibar^m)^{\aldot\be}$ and 
$(\sibar^n)^{\aldot\al}(\si^m)_{\al\bedot}$. 
The "spinor contraction" between $\si$ and $\sibar$ is
expressed as a wedge.
   }
\end{figure}
The "spinor contraction" between $\si$ and $\sibar$ is also
expressed as a wedge. 

\vs 2

{\bf Graphical Rule 5:\ Space-Time Suffixes Contraction}\nl
The contraction of space-time suffix $m$ is expressed
by a dotted line.

\vs 2
 
An example $\xi^\al (\si^m)_{\al\bedot}\pl_m\psibar^\bedot$
is expressed as in Fig.7. 
\begin{figure}
\begin{center}
\includegraphics*[height=2cm]{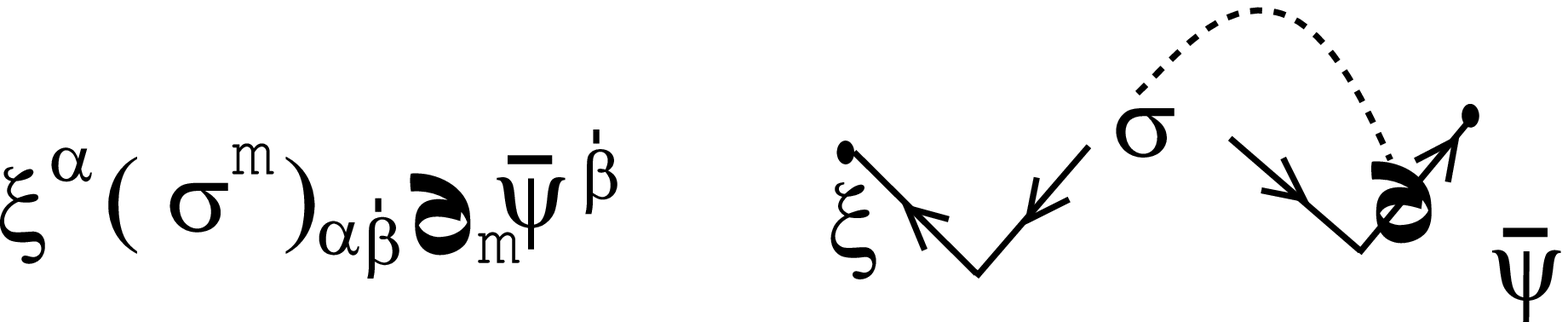}
\end{center}
   \caption{
An Lorentz invariant:\ 
$\xi^\al (\si^m)_{\al\bedot}\pl_m\psibar^\bedot$.
   }
\end{figure}
Note that \nl
\vs 2
all {\it dummy} suffixes do {\it not} appear
in the final invariant quantities. \nl
\vs 2
The {\it contraction}
is expressed by the {\it directed wedges}
and the {\it dotted curved lines}. This makes the expression
very transparent.

As the partially contracted examples, we take
$(\sibar^m)^{\aldot\al}\xi_\al$ and
$(\sibar^m)_\bedot^{~\al}\xi_\al$. See Fig.8.
\begin{figure}
\begin{center}
\includegraphics*[height=4cm]{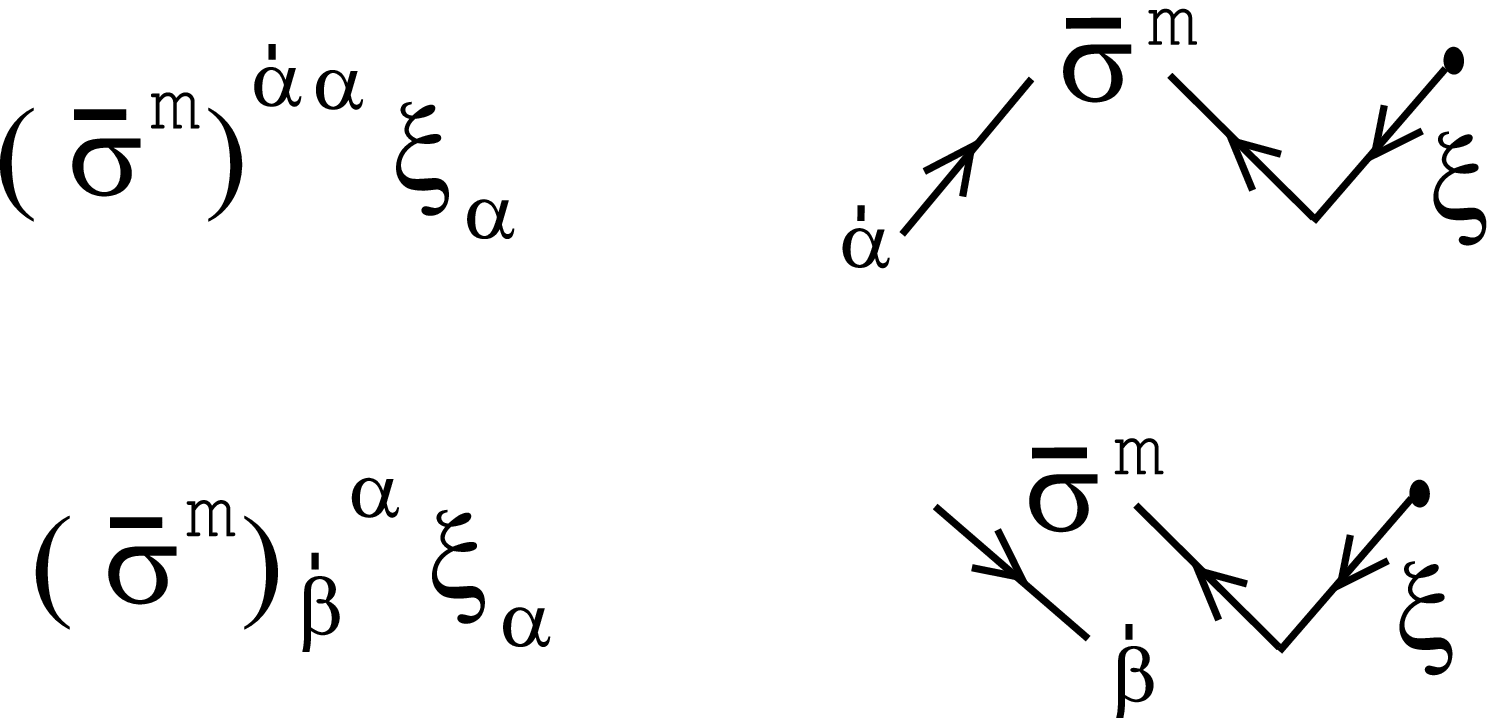}
\end{center}
   \caption{
Partially contracted cases. 
$(\sibar^m)^{\aldot\al}\xi_\al$ and
$(\sibar^m)_\bedot^{~\al}\xi_\al$. 
   }
\end{figure}
The spinor suffixes $\aldot, \bedot$ and the space-time
suffix $m$ remain and wait for further contraction.

\subsection{Graphical Formula}
An important advantage of the graphical representation
is the usage of graphical formulae (relations). 
It helps so much in practical calculation of SUSY quantities.
Some demonstrations will be given later. 
We express the formula:\ 
$\psi^\al\chi_\al
=-\psi_\al\chi^\al
=\chi^\al\psi_\al$, 
$\psibar_\aldot\chibar^\aldot
=-\psibar^\aldot\chibar_\aldot
=\chibar_\aldot\psibar^\aldot$
in Fig.9. The last equalities in the both lines
of the figure comes from the anti-commutatibity of 
the spinor graphs(GR2). 
\begin{figure}
\begin{center}
\includegraphics*[height=4cm]{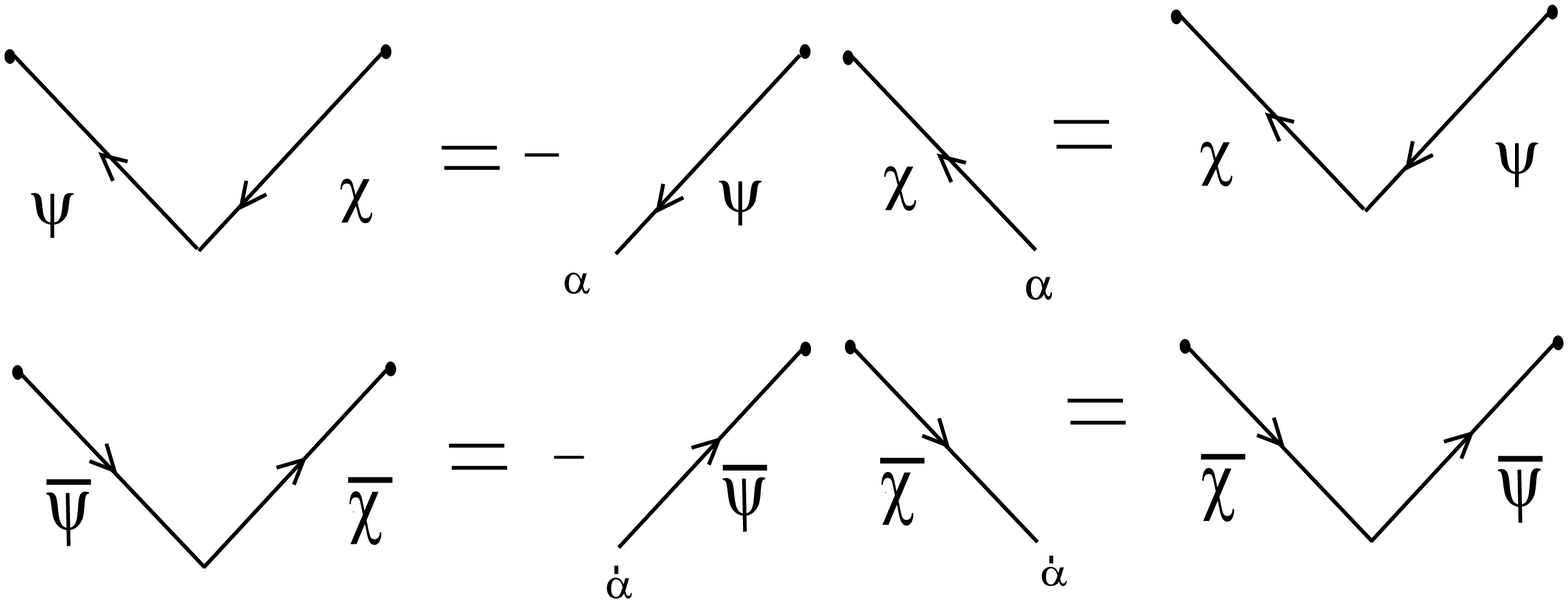}
\end{center}
   \caption{
A graphical formula. 
$\psi^\al\chi_\al=-\psi_\al\chi^\al=\chi^\al\psi_\al$, and 
$\psibar_\aldot\chibar^\aldot=-\psibar^\aldot\chibar_\aldot
=\chibar_\aldot\psibar^\aldot$
. 
   }
\end{figure}
We can express all formulae graphically.
In this subsection, we list only basic ones.
In Fig.10, the symmetric combination of $\sibar^m\si^n$
are shown as the basic spinor algebra.
\begin{figure}
\begin{center}
\includegraphics*[height=4cm]{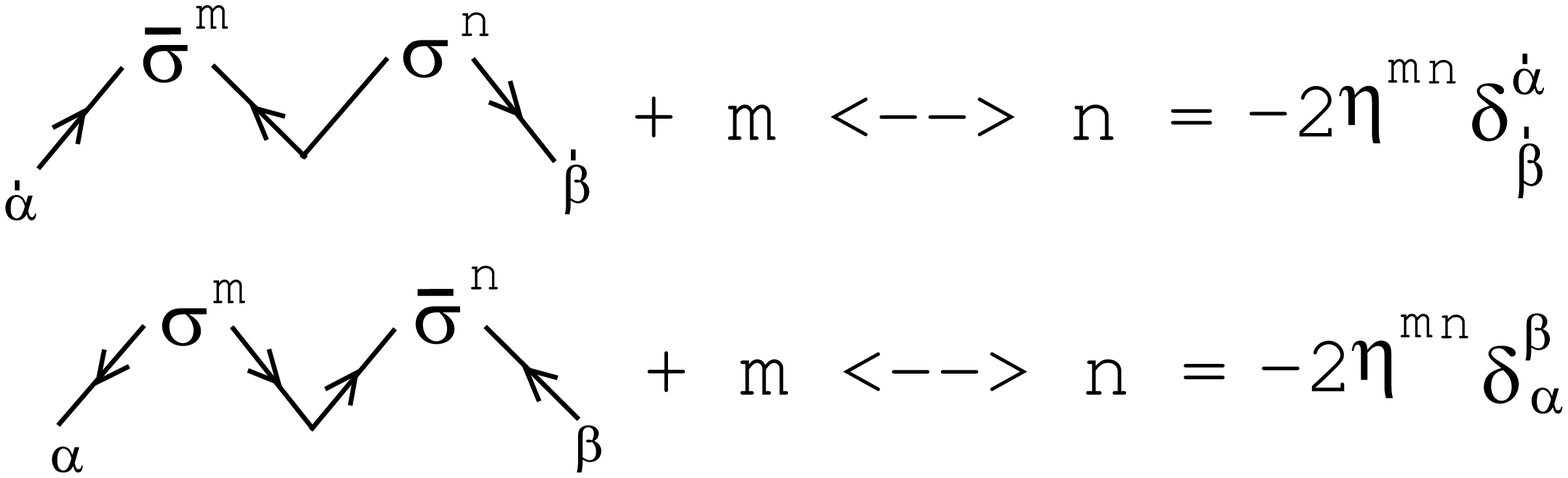}
\end{center}
   \caption{
A graphical formula of the basic spinor algebra. 
$(\sibar^m)^{\aldot\be}(\si^n)_{\be\bedot}+m\change n
=-2\eta^{mn}\del^\aldot_\bedot$, and 
$(\si^m)_{\al\aldot}(\sibar^n)^{\aldot\be}+m\change n
=-2\eta^{mn}\del^\be_\al$. 
   }
\end{figure}
The antisymmetric combination gives the generators
of the Lorentz group, $\si^{nm},\sibar^{nm}$.
\begin{eqnarray}
(\si^{nm})_\al^{~\be}=\fourth\left\{
\graph{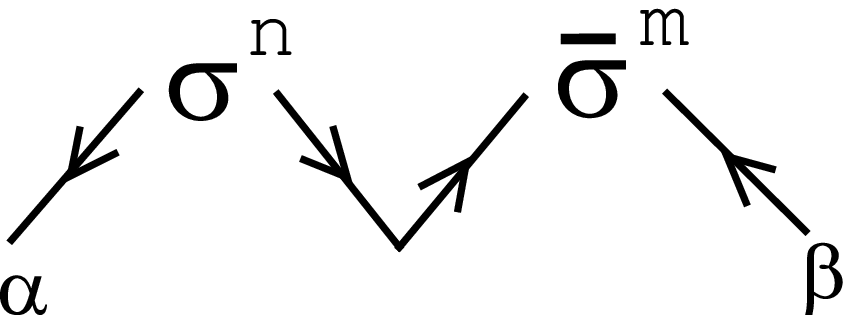}-m\change n\right\}\com\nn
(\sibar^{nm})^\aldot_{~\bedot}=
\fourth\left\{
\graph{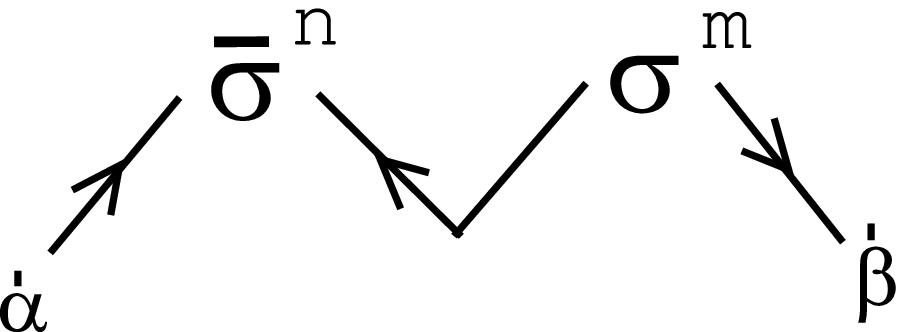}-m\change n\right\}\com
\label{def2}
\end{eqnarray}

Although we have already used $\sibar^m$, its definition 
in terms of $\si^m$ and
the "inverse" relation are displayed in Fig.11.
\begin{figure}
\begin{center}
\includegraphics*[height=4cm]{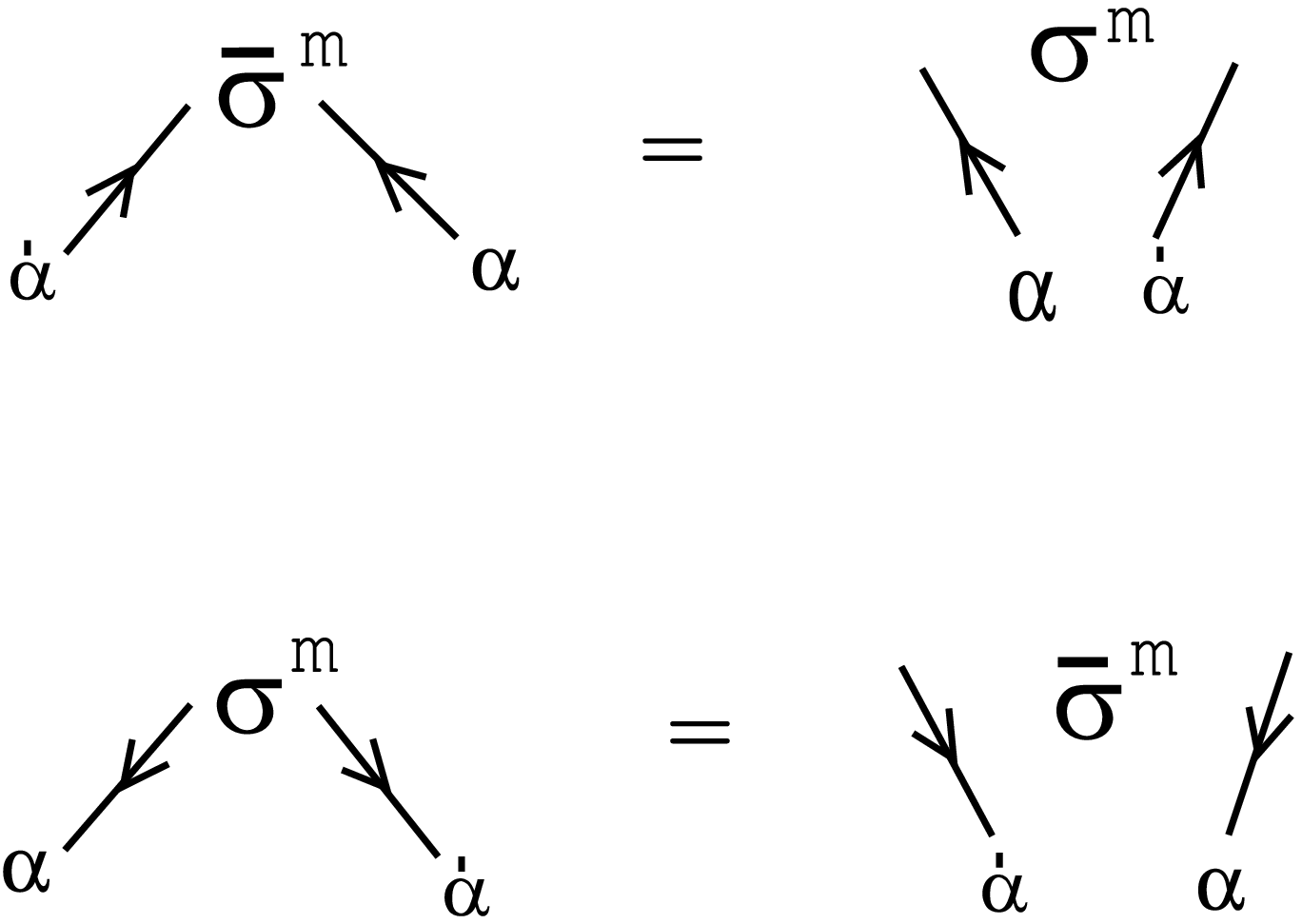}
\end{center}
   \caption{
Definition of $\sibar^m$,
$(\sibar^m)^{\aldot\al}\equiv \ep^{\ab}\ep^{\aldot\bedot}
(\si^m)_{\be\bedot}=(\si^m)^{\al\aldot}$, and its "inverse" relation,
$(\si^m)_{\al\aldot}=\ep_{\aldot\bedot}\ep_\ab 
(\sibar^m)^{\bedot\be}=(\sibar^m)_{\aldot\al}     $.
   }
\end{figure}
Here we do the upward and downward changes, within the $\si$ and $\sibar$, 
by $\ep_\ab$ and $\ep^\ab$ as explained for the spinor 
in (\ref{def1}).
Using the relation Fig.11, we can obtain the following useful relation.
\begin{eqnarray}
\mbox{Graphical Formula:\ Figure\ 11B}\q\q\q\q\q\q\q\q\nn
\graph{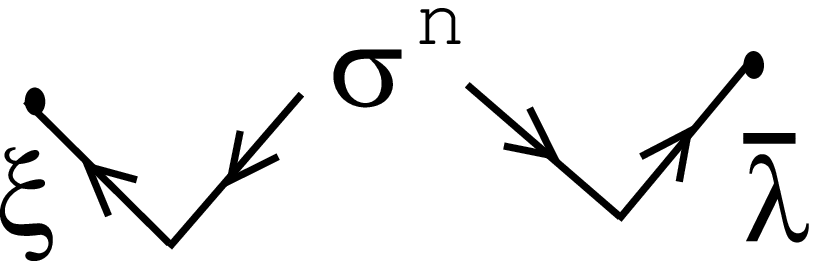}=\graph{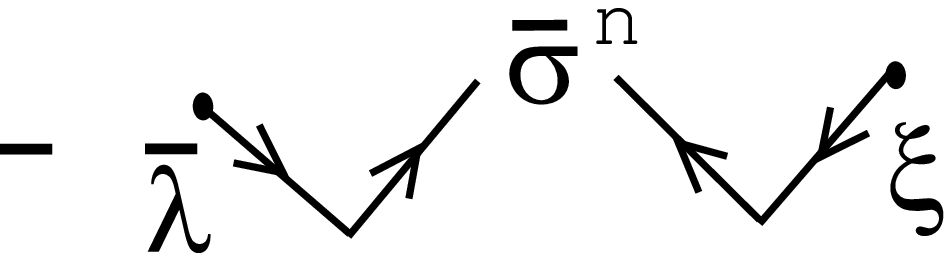}
\label{fig11b}
\end{eqnarray}
The "reduction" formulae (from the cubic $\si'$s to 
the linear one) are expressed as in Fig.12. 
\begin{figure}
\centerline{ 
\includegraphics*[height=8cm]{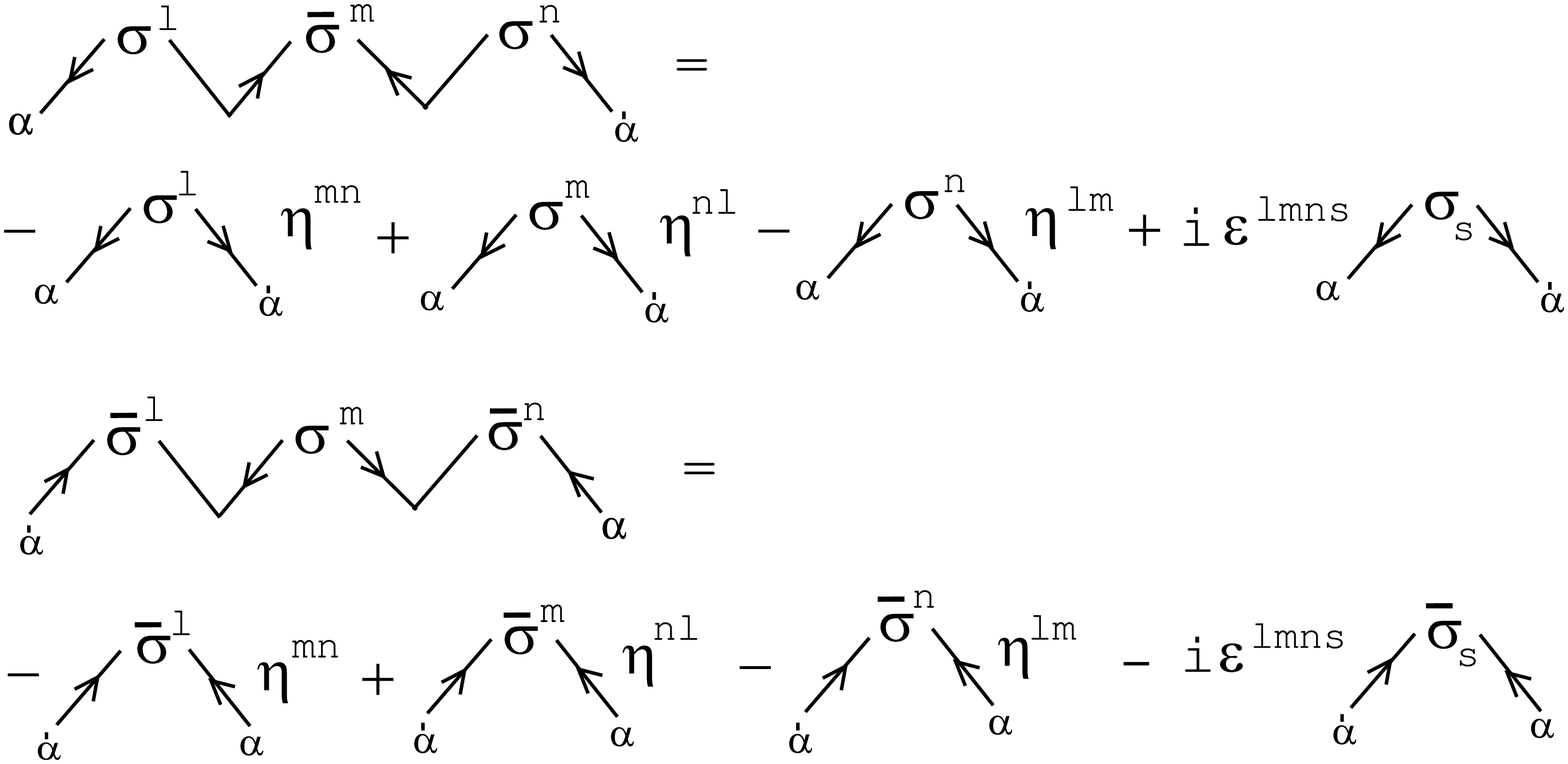}
            }
   \caption{
Two relations:\ 
1) $\si^l\sibar^m\si^n=-\si^l\eta^{mn}+\si^m\eta^{nl}
-\si^n\eta^{lm}+i\ep^{lmns}\si_s$,\ 
2) $\sibar^l\si^m\sibar^n=-\sibar^l\eta^{mn}+\sibar^m\eta^{nl}
-\sibar^n\eta^{lm}-i\ep^{lmns}\sibar_s$.
   }
\end{figure}
From Fig.12, we notice any chain of $\si'$s
can always be expressed by less than three $\si'$s. 
The appearance of the 4th rank anti-symmetric tensor $\ep^{abcd}$
is quite illuminating. 
The {\it completeness} relations are expressed as in Fig.13.
\begin{figure}
\begin{center}
\includegraphics*[height=4cm]{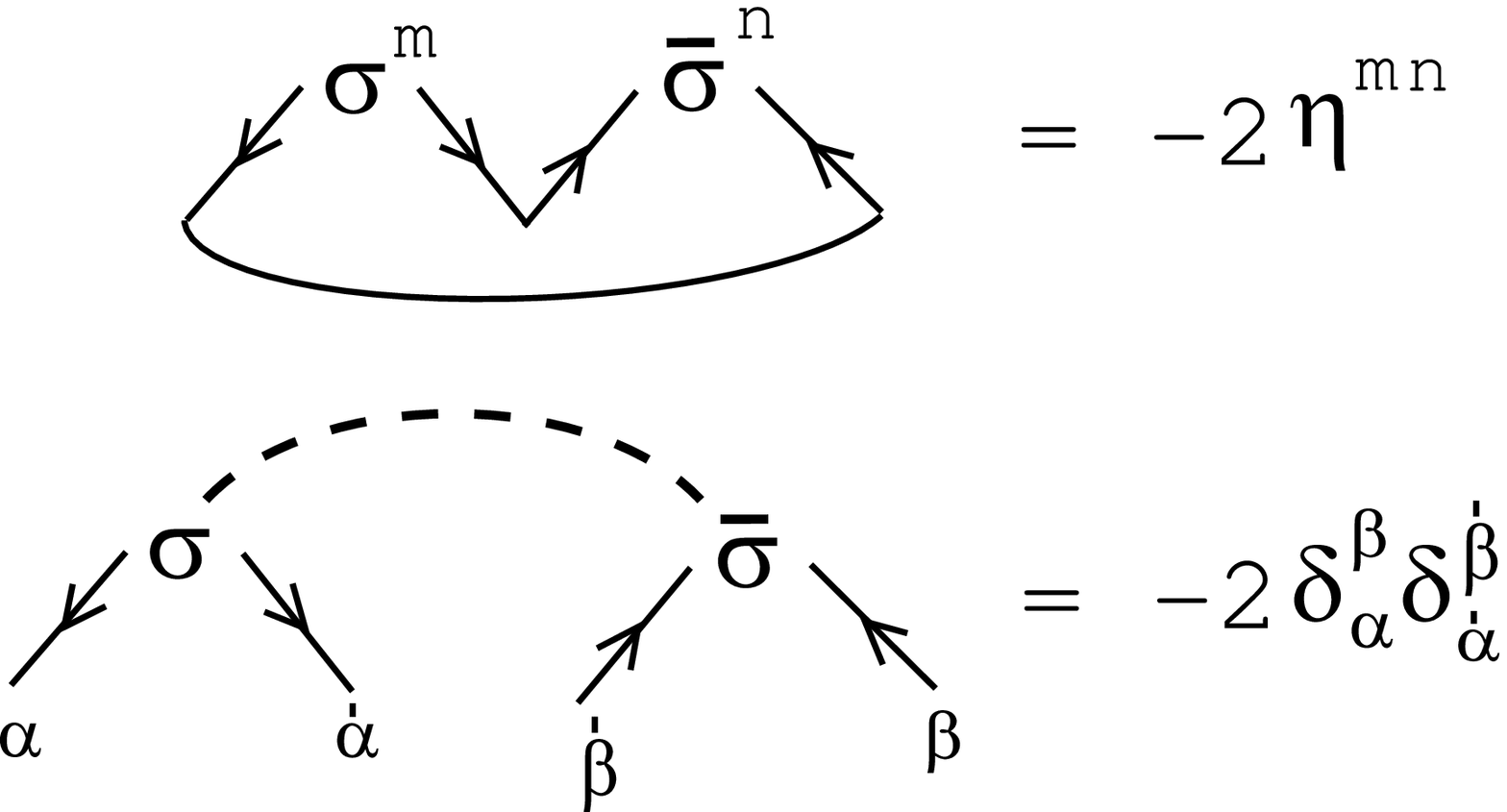}
\end{center}
   \caption{
Completeness relations:\ 
1) $\del^\al_\be(\si^m)_{\al\aldot}(\sibar^n)^{\aldot\be}
=-2\eta^{mn}$,\ 
2) $(\si^m)_{\al\aldot}(\sibar_m)^{\bedot\be}
=-2\del^\be_\al\del^\bedot_\aldot$.\ 
The contraction shown by a curve is the matrix trace.
The relation 1) is also obtained from Fig.10.
   }
\end{figure}
The contraction, expressed by the directed curve in Fig.13, 
is the matrix trace.
The Fierz identity is shown as.
\begin{eqnarray}
\graph{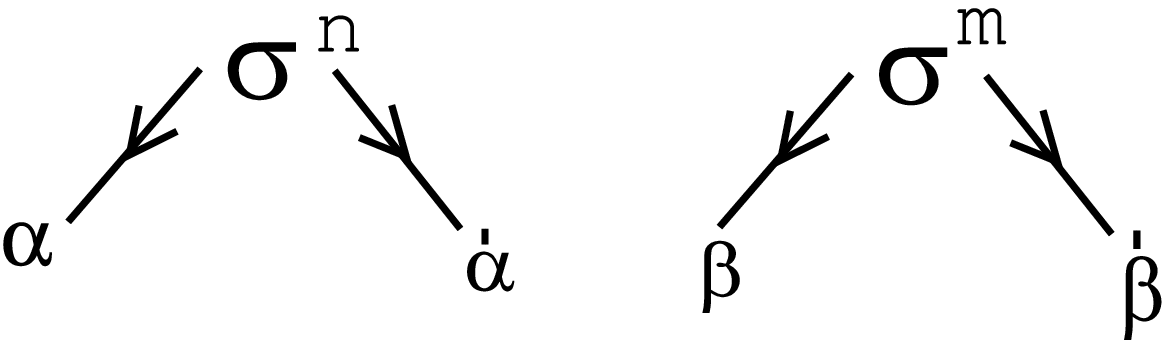}=\nn
\fourth\left\{
-\graph{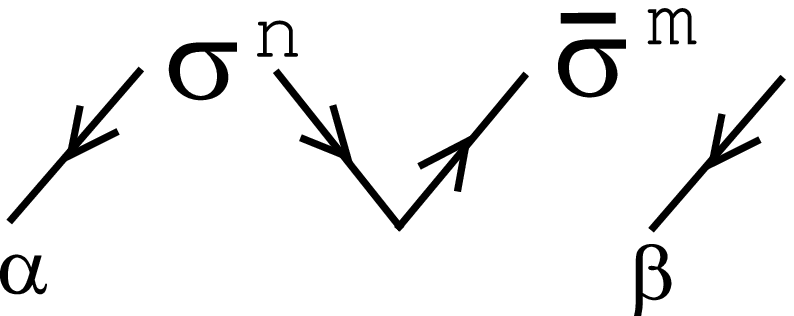}\ep_{\aldot\bedot}+\graph{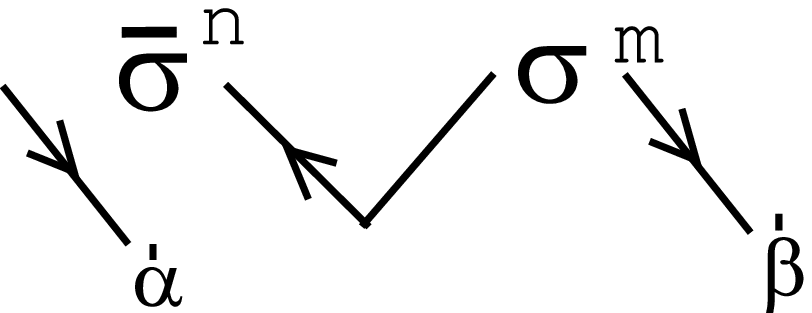}\ep_\ab
-m\change n\right\}\nn
-\half\eta^{nm}\ep_\ab \ep_{\aldot\bedot}
-\frac{1}{8}\left\{\graph{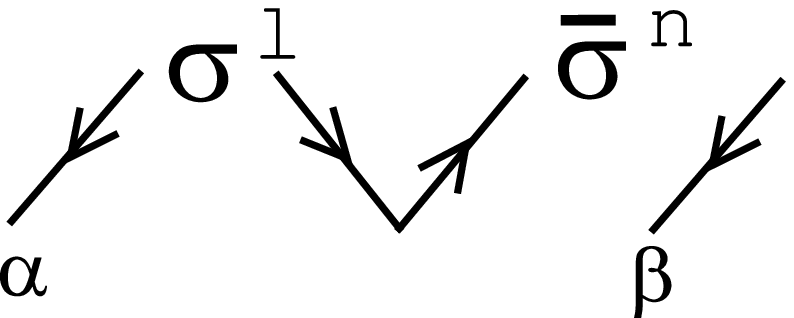}-l\change n\right\}
\left\{\graph{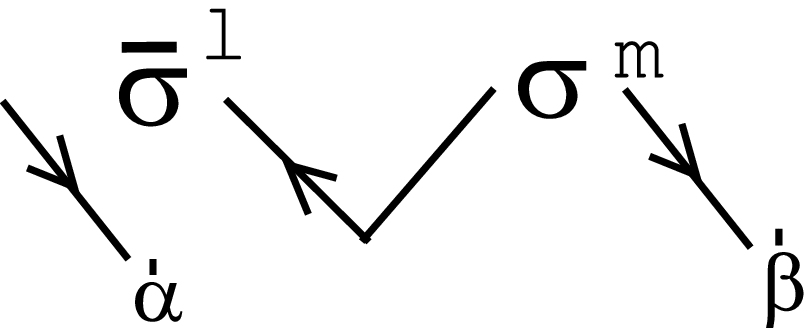}-l\change m\right\},
\label{def3}
\end{eqnarray}
where the relation:\  
$(\si^n)_{\al\aldot}(\si^m)_{\be\bedot}=
-\ep_{\be\ga}(\si^{nm})_\al^{~\ga}\ep_{\aldot\bedot}
+\ep_{\aldot\gadot}(\sibar^{nm})^\gadot_{~\bedot}\ep_{\ab}
-\half\eta^{nm}\ep_\ab\ep_{\aldot\bedot}
-2\ep_{\be\ga}(\si^{ln})_\al^{~\ga}\ep_{\aldot\gadot}
(\sibar_l^{~m})^\gadot_{~\bedot}$, is expressed.

Finally we display 4$\si$'s cotraction formula.
\begin{eqnarray}
\graph{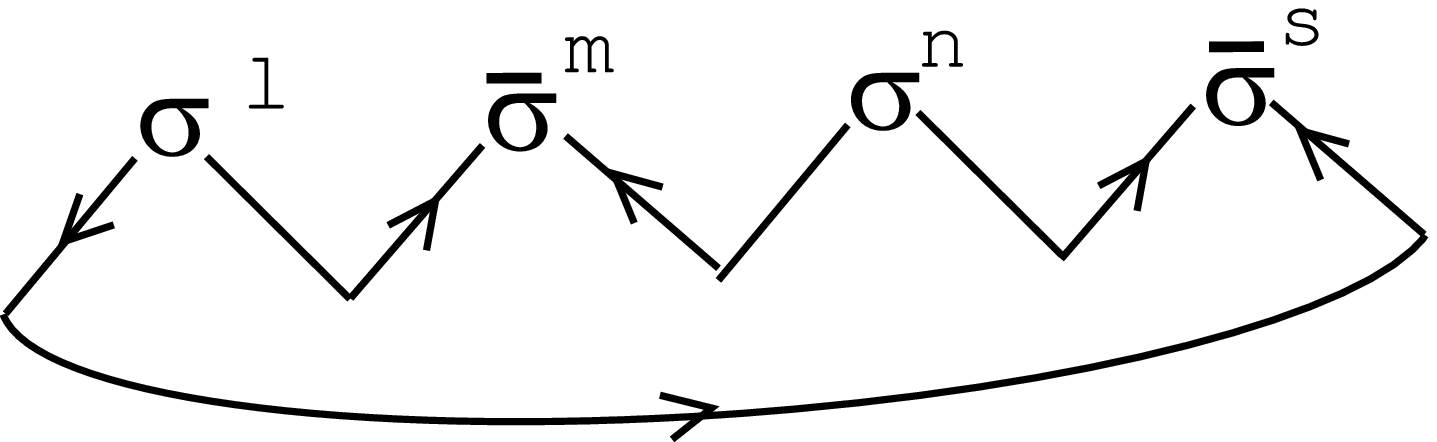}=
2(\eta^{lm}\eta^{ns}-\eta^{ln}\eta^{ms}+\eta^{ls}\eta^{mn}-i\ep^{lmns})
\label{ssbssbBR}
\end{eqnarray}
Its complex conjugate one is given by
\begin{eqnarray}
\graph{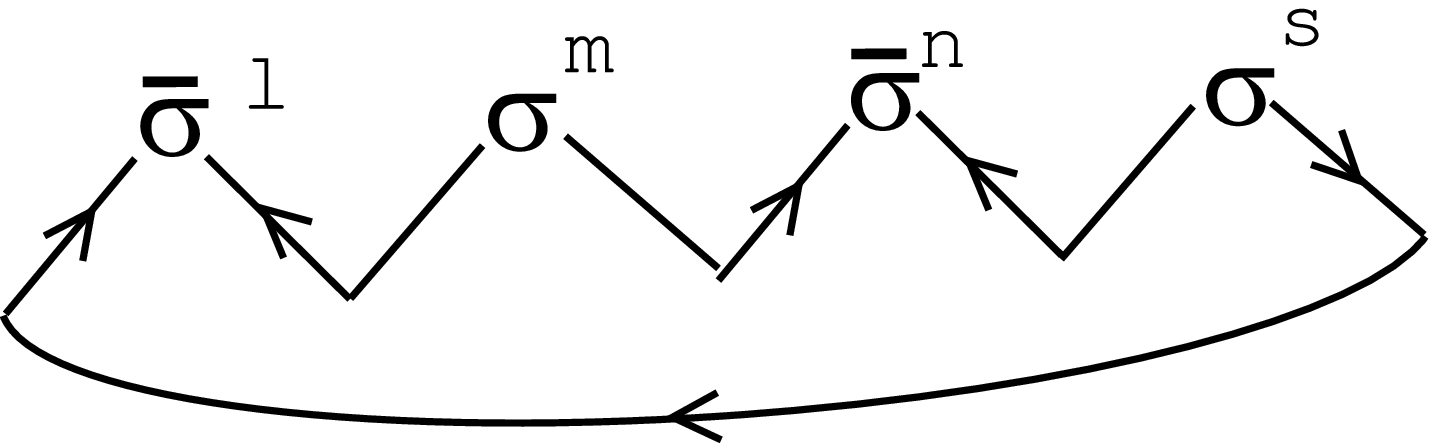}=
2(\eta^{lm}\eta^{ns}-\eta^{ln}\eta^{ms}+\eta^{ls}\eta^{mn}+i\ep^{lmns})
\label{sbssbsR}
\end{eqnarray}
The above 2 terms appear in the calculation of $W_\al W^\al$ and $\Wbar_\aldot \Wbar^\aldot$
respectively. ($W_\al$ and $\Wbar_\aldot$ are superfields of field strength. )

\section{4D Chiral Multiplet}
As the first example, we take the 4D chiral multiplet.
It is made of a complex scalar field $A$, a Weyl spinor $\psi_\al$
and an auxiliary field (complex scalar) $F$. Their transformations 
are expressed as follows.
\begin{eqnarray}
\del_\xi A=\sqtwo\xi^\al\psi_\al=\sqtwo\graph{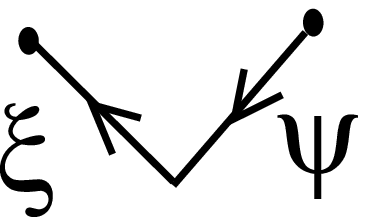}\com\nn
\del_\xi \psi_\al=i\sqtwo {\si^m}_{\al\aldot}\xibar^\aldot
\pl_mA+\sqtwo\xi_\al F
=i\sqtwo\graph{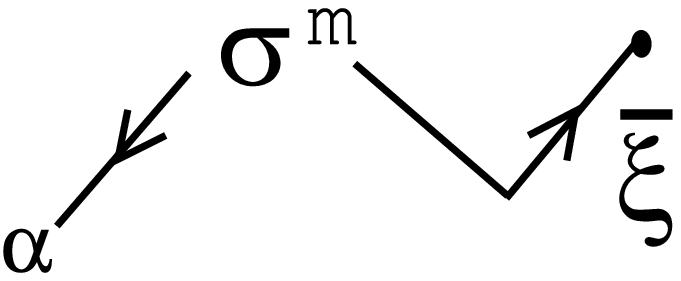}\pl_mA+\sqtwo\graph{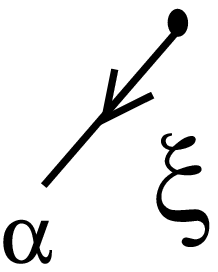}F
\com\nn
\del_\xi F=i\sqtwo\xibar_\aldot (\sibar^m)^{\aldot\be}\pl_m\psi_\be
=i\sqtwo\graph{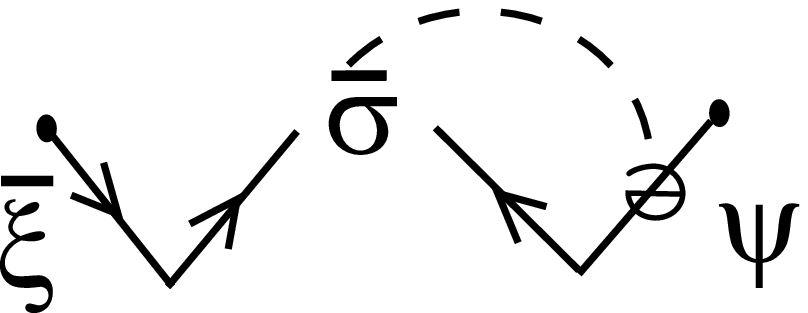}
\pr
\label{chi1a}
\end{eqnarray}
The complex conjugate ones are given as
\begin{eqnarray}
\del_\xi A^*=\sqtwo\xibar_\aldot\psibar^\aldot
=\sqtwo\graph{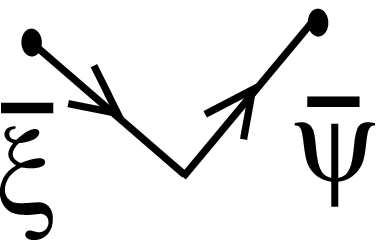}\com\nn
\del_\xi \psibar_\aldot=i\sqtwo ({\sibar^m})^{~\al}_\aldot \xi_\al
\pl_mA^*+\sqtwo\xibar_\aldot F^*
=i\sqtwo\graph{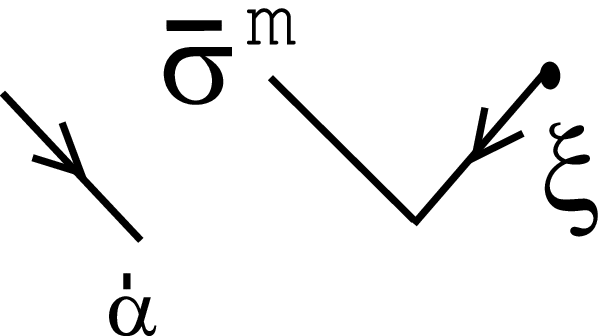}\pl_mA^*+\sqtwo\graph{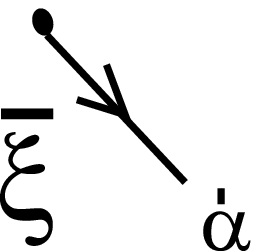}F^*
\com\nn
\del_\xi F^*=i\sqtwo\xi^\al (\si^m)_{\al\bedot}\pl_m\psibar^\bedot
=i\sqtwo\graph{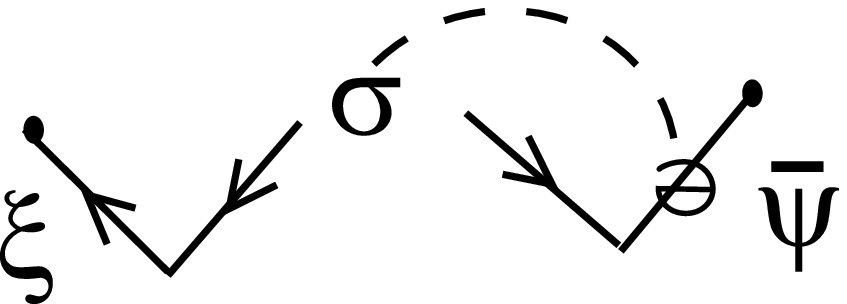}
\pr
\label{chi1b}
\end{eqnarray}
We can read the graphical rule of the complex conjugate
operation by comparing (\ref{chi1a}) and (\ref{chi1b}).\nl

{\bf Graphical Rule 6:\ Complex Conjugation Operation}
\begin{eqnarray}
\graph{F1chi1a}\q\ra\q\graph{F1chi1b}\com\nn
\graph{F4chi1a}\q\ra\q -\graph{F4chi1b}
\pr
\label{chi1c}
\end{eqnarray}
In order to 
show the graphical representation, presented in Sec.2, 
satisfies the SUSY representation and the usage of the graphical rules
and formulae,
 we graphically show the
SUSY symmetry of the Lagrangian. 
\begin{eqnarray}
\Lcal=i\pl_n\psibar_\aldot(\sibar^n)^{\aldot\be}\psi_\be
+A^*\Box A+F^*F\com\nn
=\graph{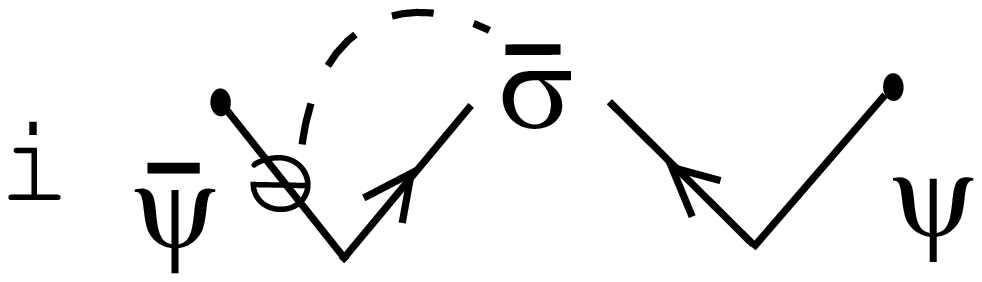}+A^*\Box A+F^*F\com
\label{chi2}
\end{eqnarray}
(Hermiticity of the first term, up to a total derivative, can be confirmed by
the use of GR6 anf Fig.11B.)
The three terms in the Lagrangian transform as
\begin{eqnarray}
\del\left(i\graph{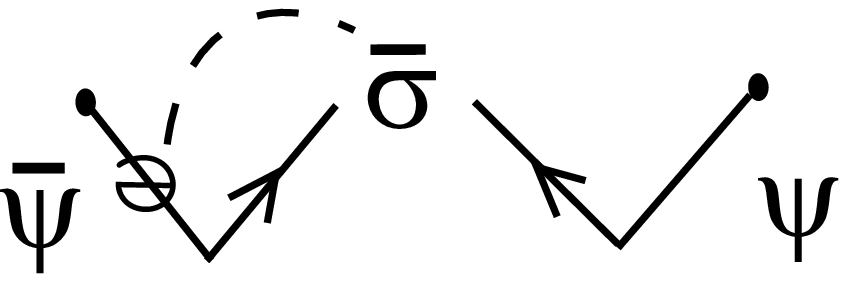}\right)=\nn
i\graph{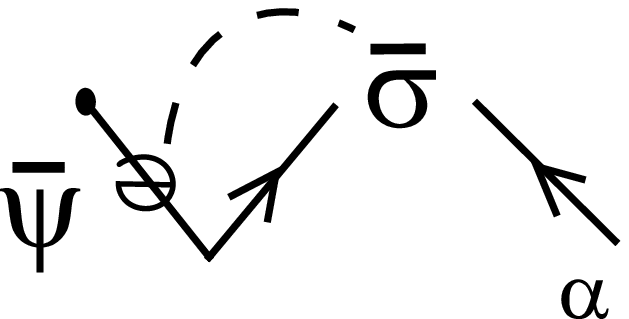}\left\{
\begin{array}{c} i\sqtwo\graph{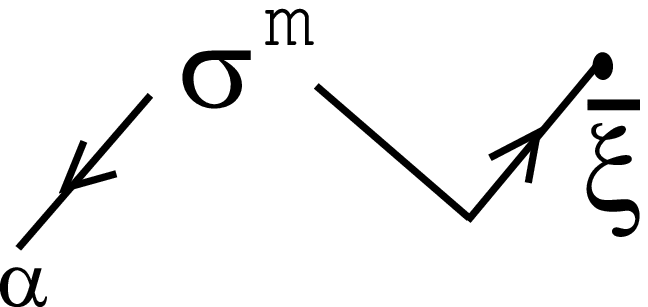}\pl_mA\\<2> \end{array}
+
\begin{array}{c} \sqtwo\graph{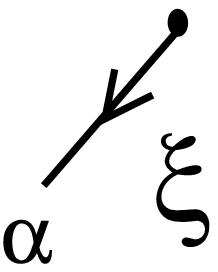}F\\<1> \end{array} \right\}
\nn
+i\pl_n\left\{
\begin{array}{c} i\sqtwo\graph{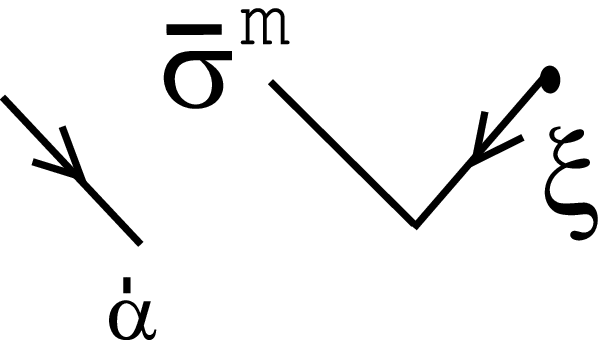}\pl_mA^*\\<4> \end{array}
+
\begin{array}{c}\sqtwo\graph{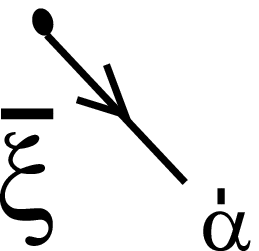}F^*\\<3>\end{array}\right\}
\graph{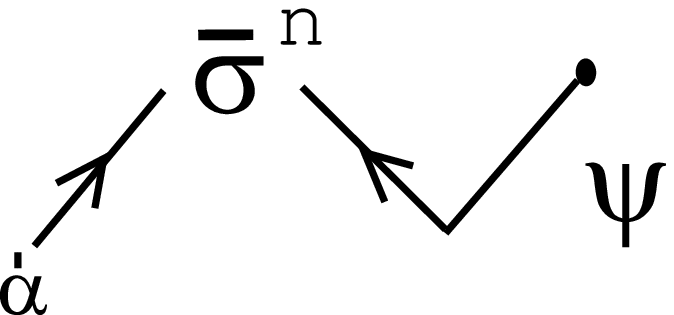}\com\nn
\del(A^*\Box A)=
\begin{array}{c}\sqtwo\graph{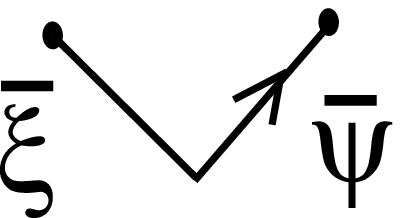}\Box A\\<2'>\end{array}
+
\begin{array}{c}\sqtwo A^*\Box\left(\graph{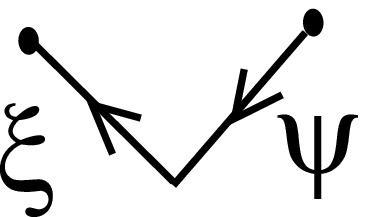}\right)\\<4'>\end{array}
\com\nn
\del(F^*F)=
\begin{array}{c}i\sqtwo\graph{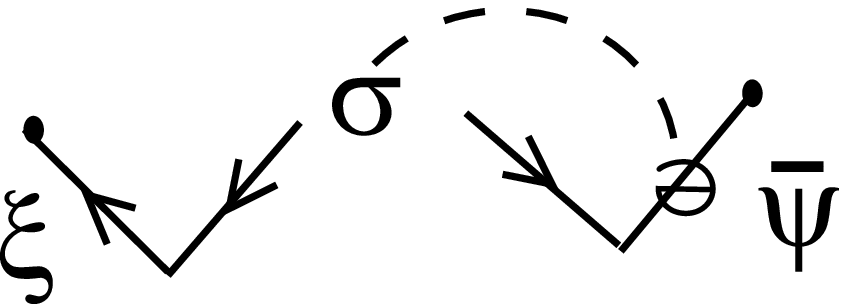}F\\<1'>\end{array}
+
\begin{array}{c}i\sqtwo F^*\graph{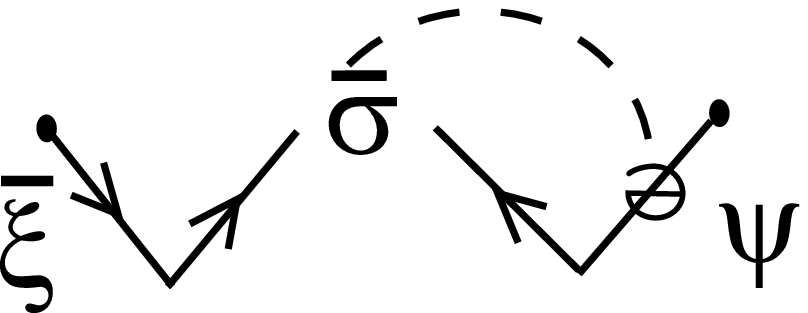}\\<3'>\end{array}\pr
\label{chi3a}
\end{eqnarray}
$<1>+<1'>=0$ can be shown as
\begin{eqnarray}
<1>=i\sqtwo\pl_n\left(\graph{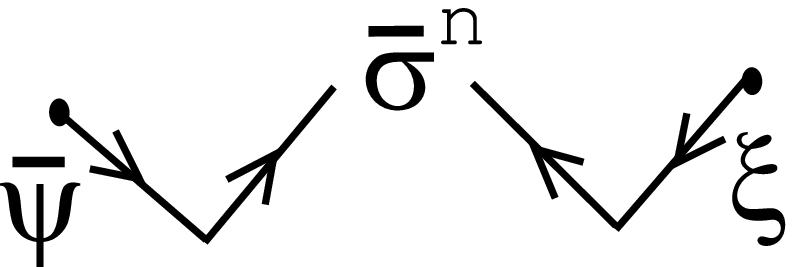}\right)F  \nn
=i\sqtwo\pl_n\left(-\graph{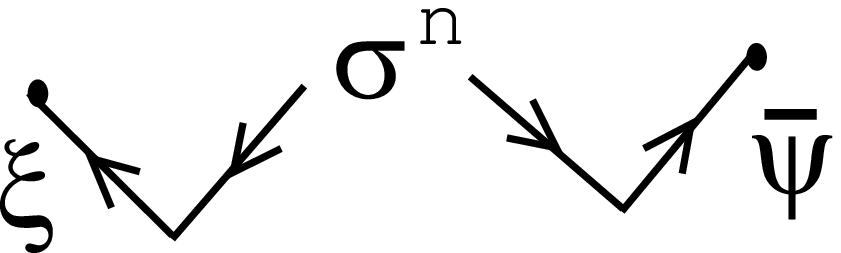}\right)F
=-<1'>
\com\label{chi3b}
\end{eqnarray}
where graph formula Fig.11B is used. 
$<2>+<2'>=$ a total derivative is shown as
\begin{eqnarray}
<2>=-\sqtwo\graph{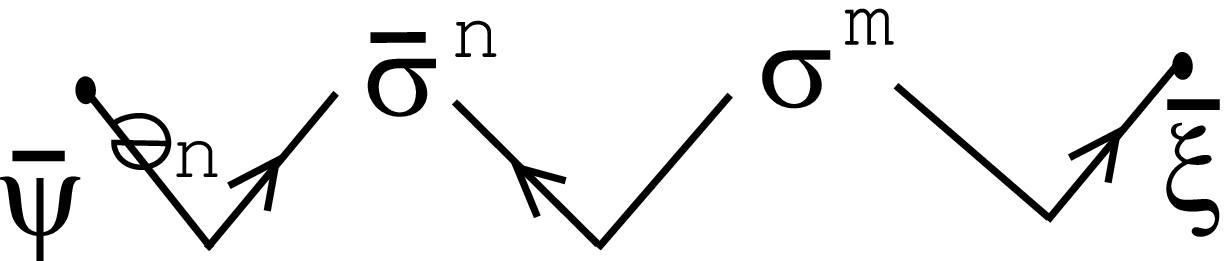}\pl_mA=\nn
-\sqtwo\pl_n\left(\graph{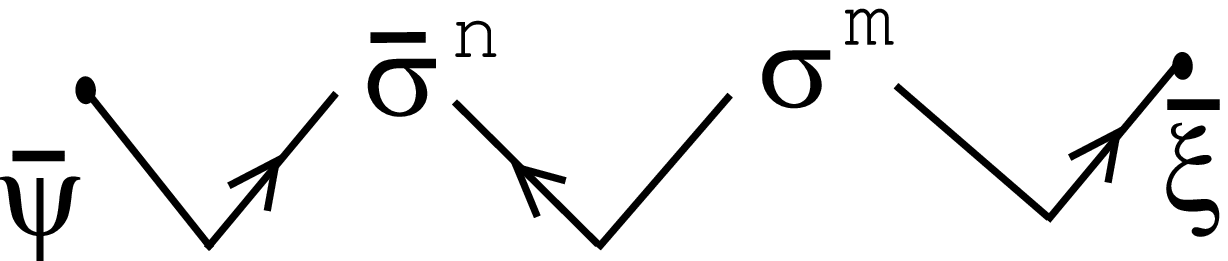}\pl_mA\right)\nn
+\sqtwo\graph{F15chi3}\pl_n\pl_mA=\nn
\mbox{ }\prime\prime \mbox{ }\q\q-\sqtwo\graph{F8chi3}\Box A
=\mbox{ }\prime\prime \mbox{ }\q\q-<2'>
\com\label{chi3c}
\end{eqnarray}
where "$\mbox{ }\prime\prime \mbox{ }$" means the corresponding previous
graph. In the above the relations Fig.10 and Fig.9 are used.
$<4>+<4'>=$ a total derivative can be shown as
\begin{eqnarray}
<4>=-\sqtwo\graph{F5chi3}\pl_n\pl_mA \graph{F7chi3}\nn
=\sqtwo\graph{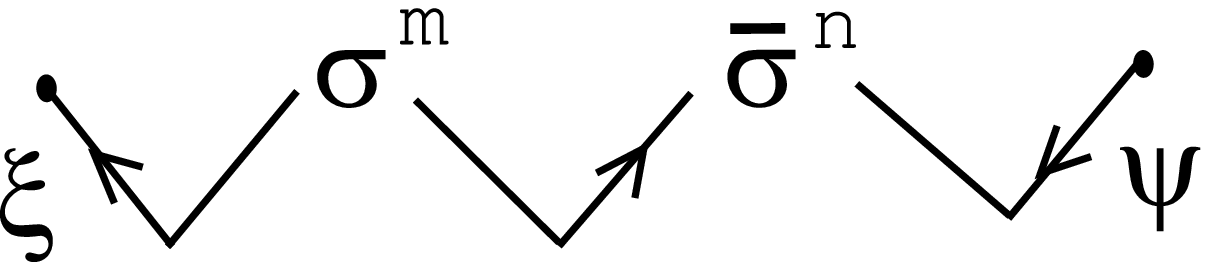}\pl_n\pl_mA^*\nn
=-\sqtwo\graph{F9chi3}\Box A^*\com\nn
<4'>=\pl_n\left\{
\sqtwo A^*\pl^n(\graph{F9chi3})-\sqtwo\pl^nA^*\graph{F9chi3}
           \right\}                   \nn
+\sqtwo\graph{F9chi3}\Box A^*
\com\label{chi3d}
\end{eqnarray}
where some modification of Fig.11B is used in the first line,
and Fig.10 is used in the second line. 
$<3>+<3'>=$ a total derivative can be obtained as
\begin{eqnarray}
<3>=i\sqtwo\pl_n\left(F^*\graph{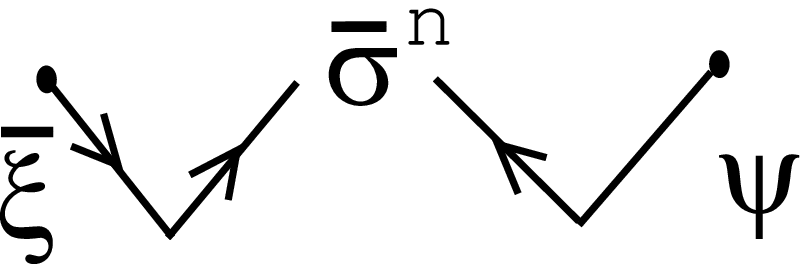}\right)
-i\sqtwo F^*\graph{F11chi3}\nn
=\mbox{ }\prime\prime \mbox{ }\q-<3'>
\ .\label{chi3e}
\end{eqnarray}
Summing the above results, we finally obtain
the result without ignoring total drivatives.
\begin{eqnarray}
\del_\xi\Lcal=
\pl_n\left\{
-\sqtwo\graph{F15chi3}\pl_mA
+\sqtwo A^*\pl^n(\graph{F9chi3})\right.\nn
\left.-\sqtwo\pl^n A^*\graph{F9chi3}
+i\sqtwo F^*\graph{F18chi3}
\right\}
\pr\label{chi3f}
\end{eqnarray}
Hence the Lagrangian indeed invariant {\it up to a
total derivative}.

\section{4D Vector Multiplet}
The super elctromagnetic theory, in the WZ gauge, is given by
\begin{eqnarray}
\Lcal_{EM}=\half D^2-\fourth v^{mn}v_{mn}
\graph{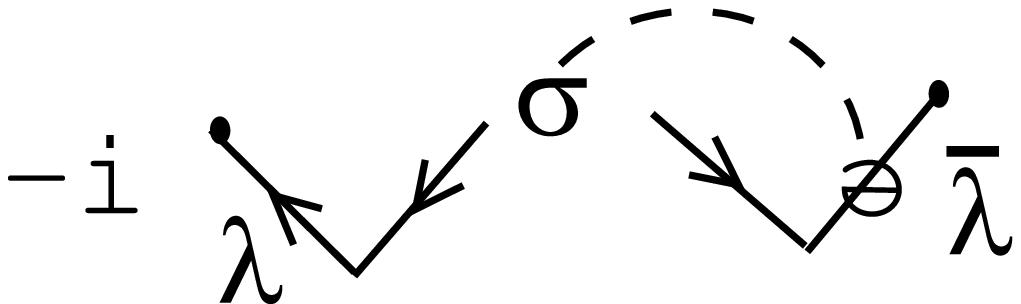}\com
\label{vec1}
\end{eqnarray}
where $v_{mn}=\pl_mv_n-\pl_nv_m$.
$v_m$ is the vector field, $\la$ is the Weyl fermion, $D$
is a scalar auxiliary field. (\ref{vec1}) is invariant
under the SUSY transformation.
\begin{eqnarray}
\del_\xi D=
\graph{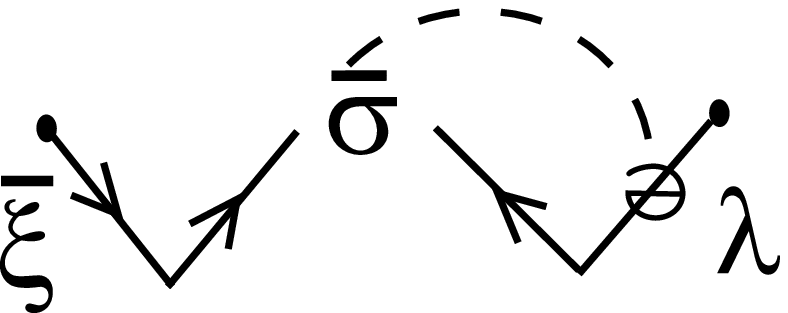} \graph{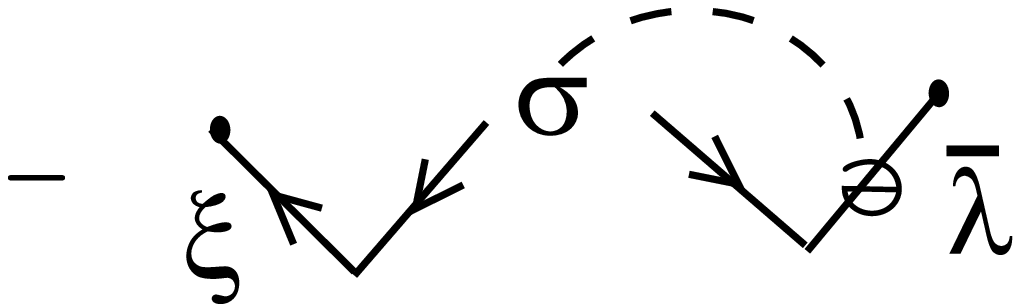}\com\nn
\del_\xi
\graph{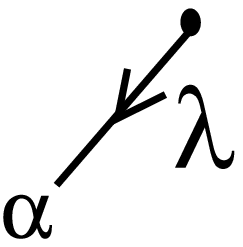}=
\graph{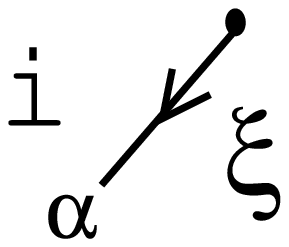}D+\fourth\left\{
\graph{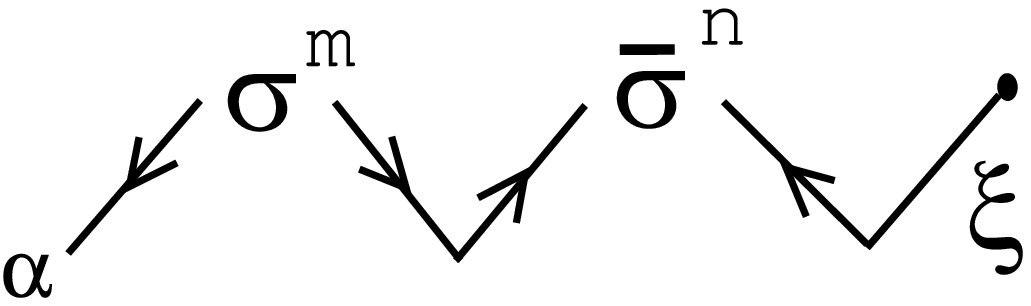}-m\change n
                     \right\}v_{mn}\com\nn
\del_\xi v_{mn}=\graph{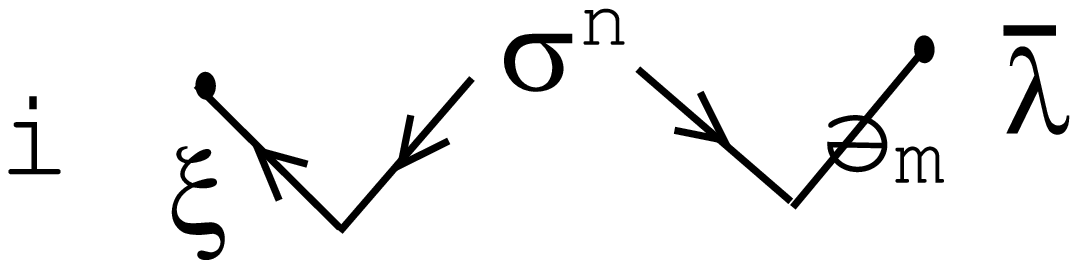} \graph{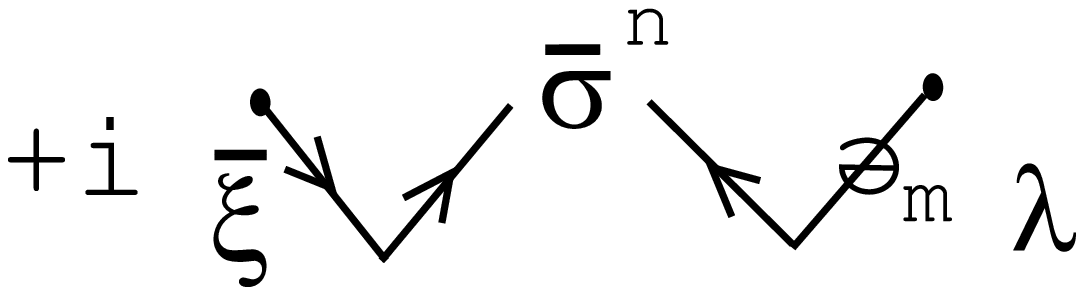}-m\change n\com\nn
\del_\xi v_m=\graph{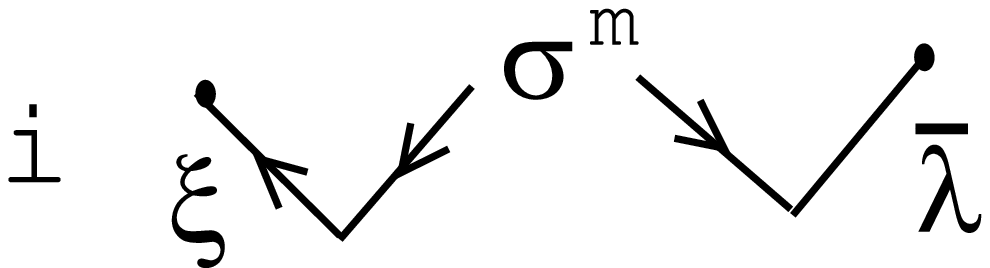} \graph{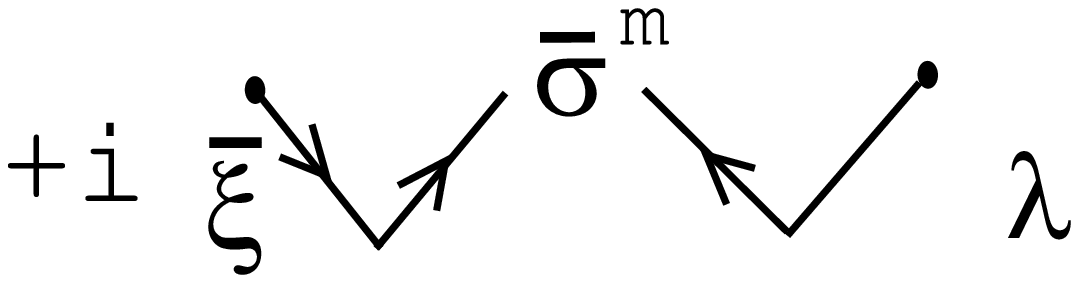}
\com
\label{vec2}
\end{eqnarray}

The SUSY invariance of (\ref{vec1}) can be graphically shown
by using the relations of Fig.9,Fig.11, Fig.12 and
Fig.11B.
The result is
\begin{eqnarray}
\del\Lcal_{EM}=\pl_l\left[
\graph{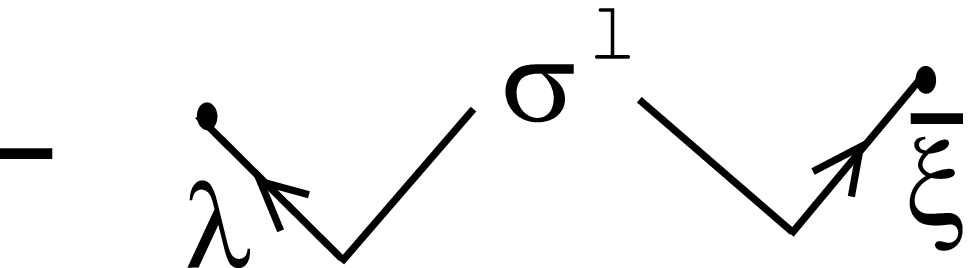}D \graph{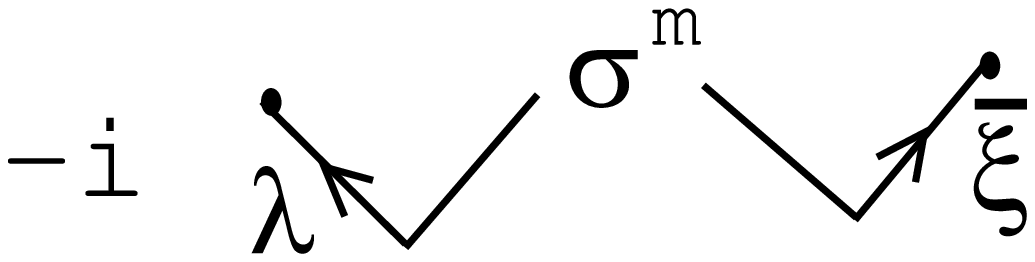}v_{ml}\right.\nn
\left.\graph{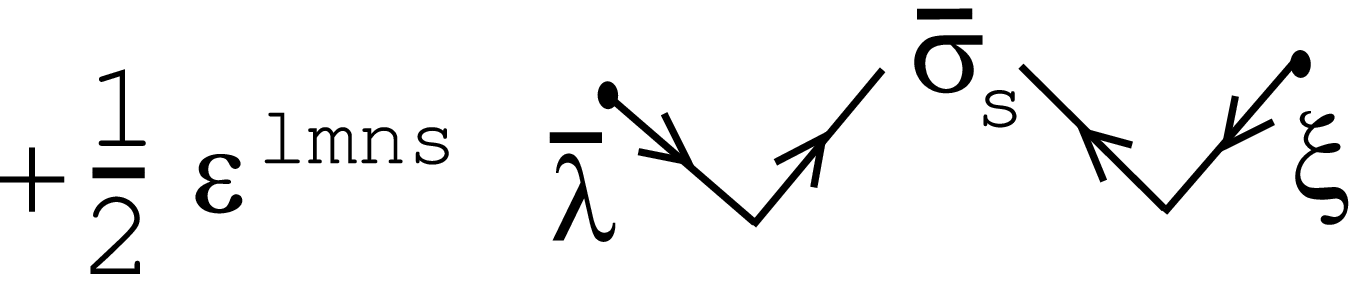}v_{mn}
\right]
\com
\label{vec4}
\end{eqnarray}
which expresses a {\it total derivative}. 
The appearance of the totally anti-symmetric tensor $\ep^{lmns}$
shows that the invariance crucially depends on the space-time dimensionality 4
in the case of vector multiplet.
This fact makes the dimensional regularization difficult
in the SUSY quantum calculation\cite{JJ97}. 

\section{Majorana spinor}
Another useful way to represent the supersymmetry
is the use of the Majorana spinor which is based on 
SO(1,3) (not SL(2,C)) structure. 
We define the graphical representation for 
the Majorana spinor $\Psi$ and its conjugate $\Psibar=\Psi^\dag\ga_0$
as in Fig.14.  
\begin{figure}
\begin{center}
\includegraphics*[height=2cm]{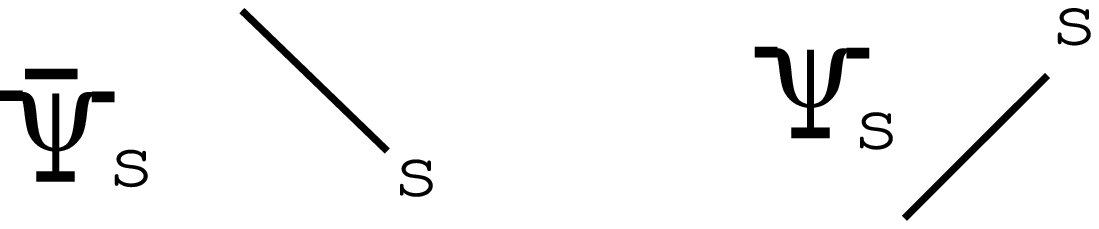}
\end{center}
   \caption{
The graphical representation for the
Majorana spinor, $\Psi$, and its conjugate $\Psibar=\Psi^\dag\ga_0$.
$s=1,2,3,4.$
   }
\end{figure}
The SO(1,3) invariants $\Psibar\Psi$ and $\Psibar\gago\Psi$ are
represented as in Fig.15.
\begin{figure}
\begin{center}
\includegraphics*[height=4cm]{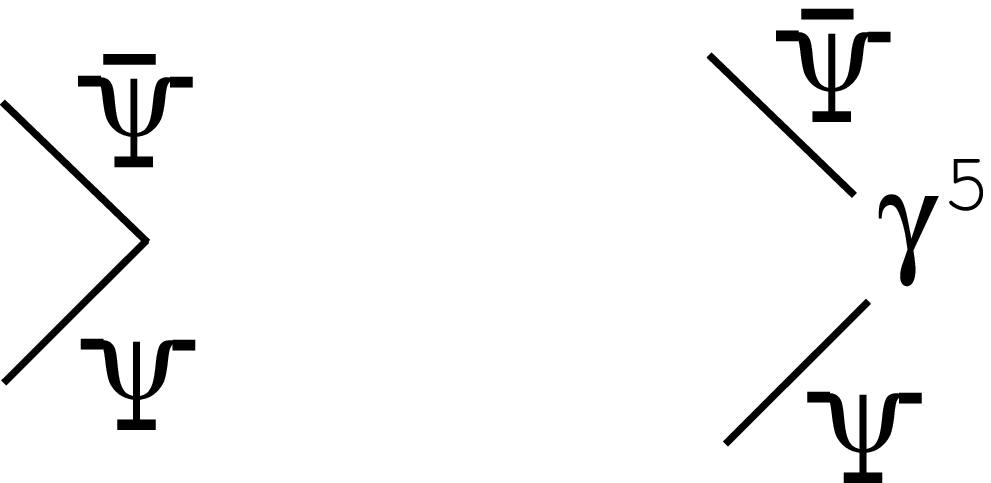}
\end{center}
   \caption{
The graphical representation for 
the SO(1,3) invariants made of the
Majorana spinors:\ $\Psibar\Psi$ and $\Psibar\gago\Psi$.
   }
\end{figure}
They are graphically much simpler than the Weyl case
( no arrows, the single (vertical) wedge structure with spinor matrices
placed at the vertex )
because only the adjoint structure is necessary to be build in
the graph. Remaining information, such as hermiticity and chiral properties, is in the 4$\times$4
matrix elements (made of $\ga^m$ matrices).

The relation between the Weyl and Majorana spinors is 
described in textbooks\cite{Wein02, Freund86, West90
}.
To show the precise relation, at the graphical level,
and to show some usage of the graph method,
we derive the relation using the previously defined contents.
The chiral multiplet of Sec.3 is taken for the explanation.

First we introduce 4 real fields $P,Q,G,H$, instead of
($A,A^*,F,F^*$).
\begin{eqnarray}
P=\frac{1}{\sqtwo}(A+A^*)\com\q Q=\frac{1}{\sqtwo i}(A-A^*)\com\nn
G=\frac{1}{\sqtwo}(F+F^*)\com\q H=\frac{1}{\sqtwo i}(F-F^*)\pr
\label{majo1}
\end{eqnarray}
As for spinor quantities, we introduce 
4 components spinor quantities $\al, \albar, \Psi, \Psibar$
instead of the 2 components ones ($\xi,\xibar,\psi,\psibar$). 
\begin{eqnarray}
\al=
\left(\begin{array}{c}  \xi_\al \\ \xibar^\aldot  \end{array}
\right)\ 
\com\q
\Psi=\left(\begin{array}{c}  \psi_\al \\ \psibar^\aldot  \end{array}
\right)\ 
\com\nn
\albar=
\left(\begin{array}{cc}  \xi^\al & \xibar_\aldot  \end{array}
\right)
\com\q
\Psibar=
\left(\begin{array}{cc}  \psi^\al & \psibar_\aldot  \end{array}
\right)
\pr
\label{majo2}
\end{eqnarray}
Using these quantities the SUSY transformation (of the chiral multiplet)
is obtained as

\begin{eqnarray}
\del_\xi P=\xi^\al\psi_\al+\xibar_\aldot\psibar^\aldot
=\albar\Psi\nn
=\graph{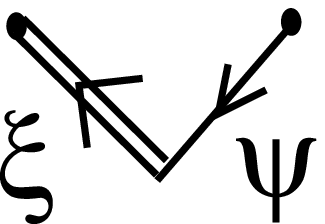}+\graph{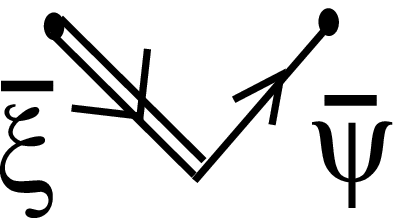}=\graph{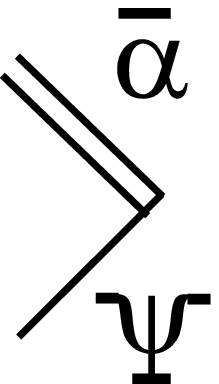}\com\nn
\del_\xi Q=\frac{1}{i}(\xi^\al\psi_\al-\xibar_\aldot\psi^\aldot)
=\albar\ga^5\Psi\nn
=-i\graph{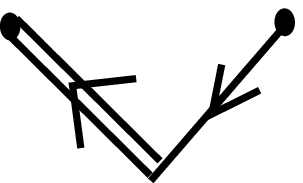}+i\graph{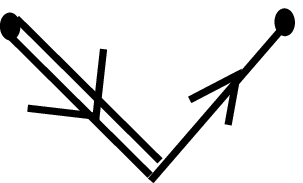}=\graph{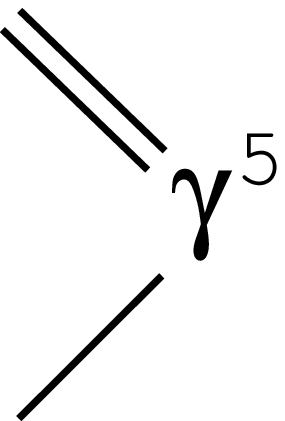}\com\nn
\del_\xi G=i\xibar_\aldot (\sibar^m)^{\aldot\be}\pl_m\psi_\be
+i\xi^\al (\si^m)_{\al\bedot}\pl_m\psibar^\bedot
=i\albar\ga^m\pl_m\Psi\nn
=i\graph{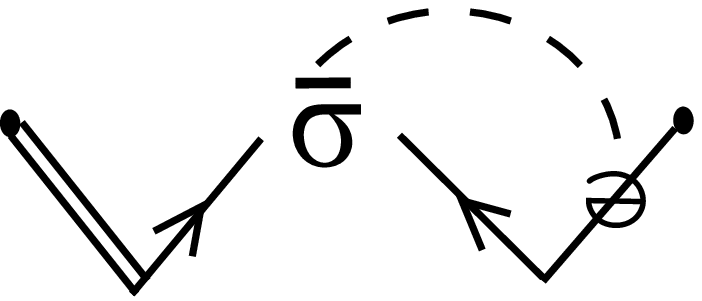}+i\graph{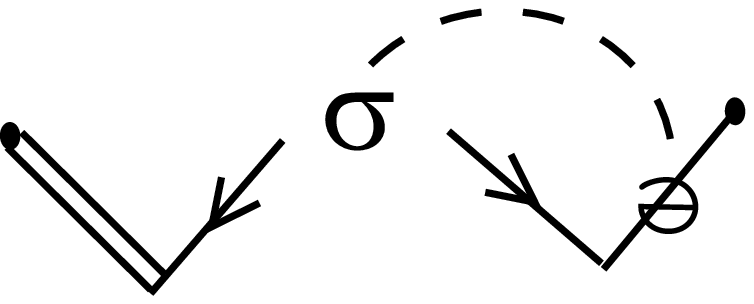}=i\graph{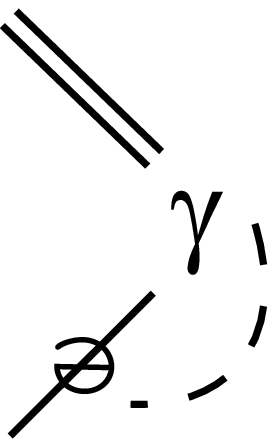}\com\nn
\del_\xi H=\xibar_\aldot (\sibar^m)^{\aldot\be}\pl_m\psi_\be
-\xi^\al (\si^m)_{\al\bedot}\pl_m\psibar^\bedot
=-i\albar\ga^5\ga^m\pl_m\Psi\nn
=\graph{F7majo3}-\graph{F8majo3}=-i\graph{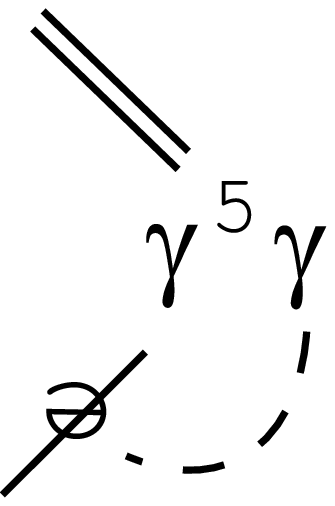}\com\nn
\del_\xi\Psi=
\left(\begin{array}{c}  \del_\xi\psi_\al \\ \del_\xi\psibar^\aldot  \end{array}
\right)\ 
=\left(\begin{array}{c}  i\sqtwo\graph{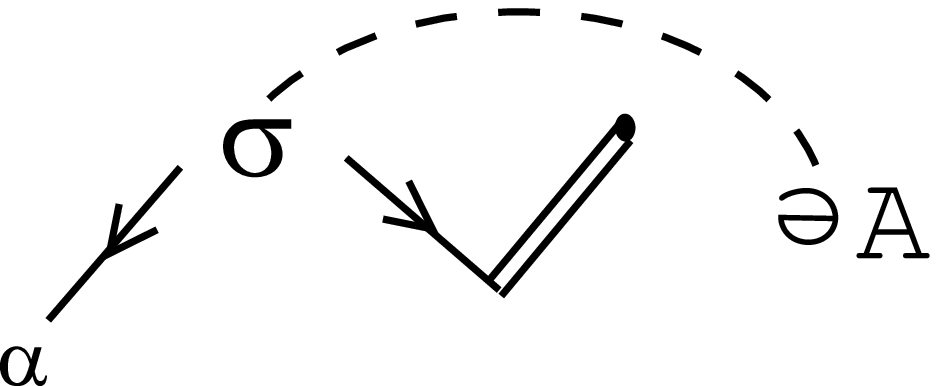}+\sqtwo\graph{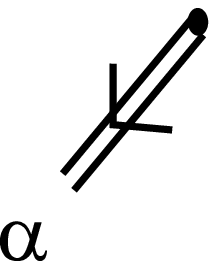}F \\ 
i\sqtwo\graph{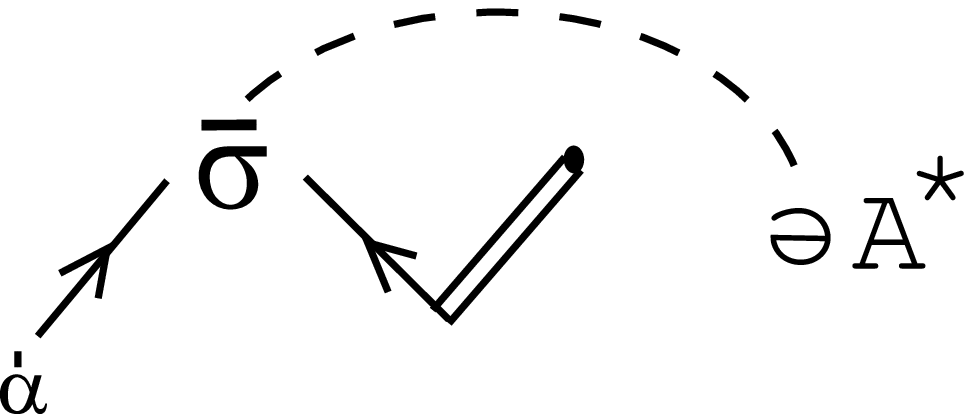}+\sqtwo\graph{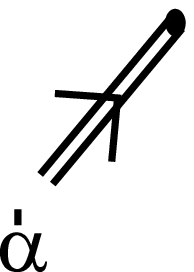}F^*  \end{array}
\right)\nn
=i\pl_m (P-\ga^5 Q)\ga^m\al+(G-\ga^5 H)\al\nn
=i\graph{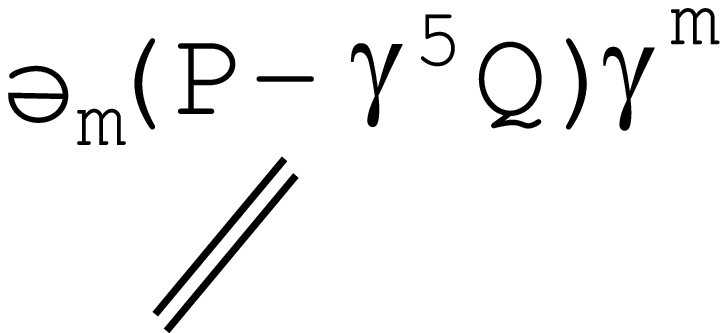}+\graph{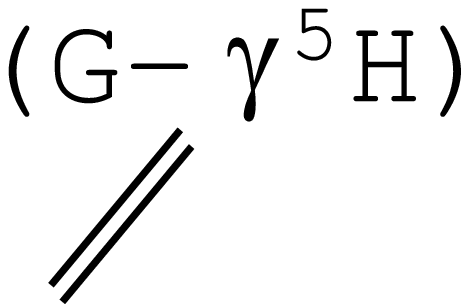}
\com
\label{majo3}
\end{eqnarray}
where the double lines are used to express the SUSY
parameters and the following gamma matrices are taken\cite{WB92}: 
\begin{eqnarray}
\ga^m=
\left(\begin{array}{cc}
0 & (\si^m)_{\al\bedot} \\
(\sibar^m)^{\aldot\be} & 0 \end{array}
\right)
\com\q
\ga^5=\ga^0\ga^1\ga^2\ga^3=
\left(\begin{array}{cc}
-i & 0 \\
0 & i \end{array}
\right)
\pr\label{majo4}
\end{eqnarray}
We show, in (\ref{majo3}), the graphical expressions for 
the SO(1,3) invariants made of the
Majorana spinors:\ 
$\albar\Psi,\albar\gago\Psi, i\albar\ga^m\pl_m\Psi,
-i\albar\gago\ga^m\pl_m\Psi,
i\pl_m (P-\ga^5 Q)\ga^m\al, (G-\ga^5 H)\al$.
The above graphical relations manifestly show the $\gago$ matrix
controls the chirality in the Majorana spinor,
whereas it is shown by the left-right direction
(dot-undotted suffixes)in the Weyl case.
The above relations can be used in the transformation
between both expressions even at the graphical level.

The fermion kinetic term of the chiral Lagrangian
(\ref{chi2}) is graphically
transformed into the Majorana expression as follows.
\begin{eqnarray}
i\graph{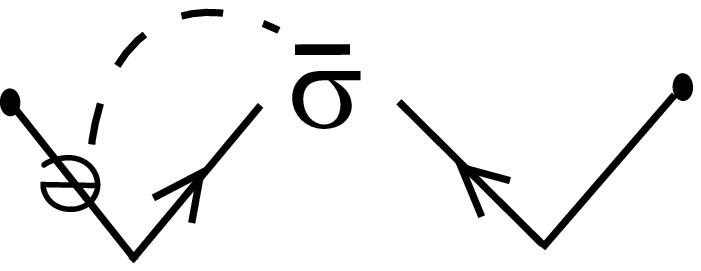}=\frac{i}{2}\graph{F1majo5}-\frac{i}{2}\graph{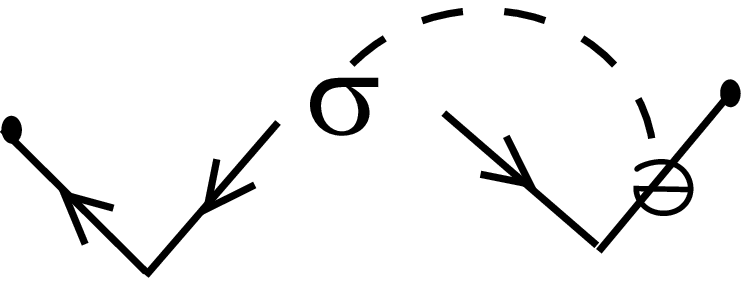}\nn
=\frac{i}{2}\pl_m\left(\graph{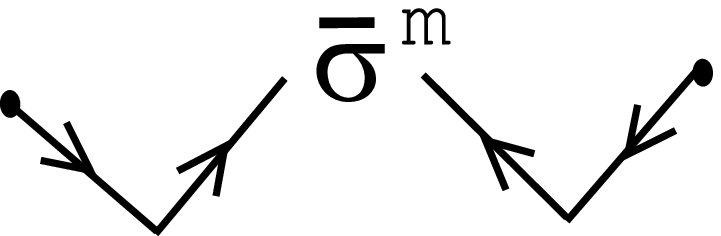}\right)
-\frac{i}{2}\graph{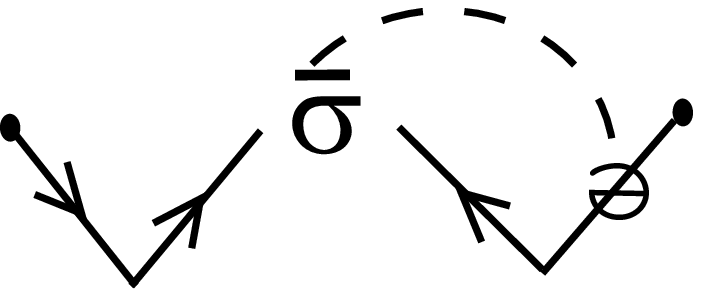}-\frac{i}{2} \graph{F2majo5}\nn
=\frac{i}{2}\pl_m(\mbox{ }\prime\prime \mbox{ })
-\frac{i}{2}
\left(\begin{array}{c}  \psi_\al \\ \psibar^\aldot  \end{array}\right)\ 
\left(\begin{array}{cc}
0 & (\si^m)_{\al\bedot} \\
(\sibar^m)^{\aldot\be} & 0 \end{array}
\right)
\pl_m
\left(\begin{array}{cc}  \psi^\al & \psibar_\aldot  \end{array}\right)
                                                    \nn
=\frac{i}{2}\pl_m(\mbox{ }\prime\prime \mbox{ })
-\frac{i}{2}\Psibar\ga^m\pl_m\Psi
=\frac{i}{2}\pl_m(\mbox{ }\prime\prime \mbox{ })
-\frac{i}{2}\graph{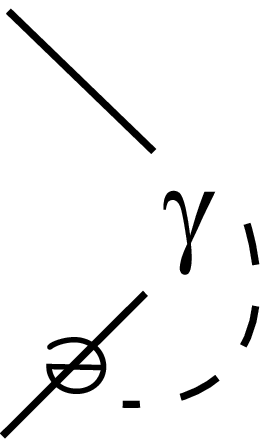}
\com
\label{majo5}
\end{eqnarray}
where the relation of Fig.11B is used in the first line.


\section{Indices of Graph}
We introduce some {\it indices} of a graph. They classify graphs.
Its use is another advantage of the graphical representation.

(i) {\it Left Chiral Number} and {\it Right Chiral Number}\nl
We assign $\half$ for each one step leftward arrow 
and define its total sum within a graph
as {\it Left Chiral Number}(LCN). In the same way,
we assign $\half$ for each one step rightward arrow 
and define its total sum within a graph as {\it Right Chiral Number}(RCN).

(ii) Up-Down Counting\nl
We assign $+\half$ for one step of the upward arrow
and $-\half$ for the one step of the downward arrow. Then we define
{\it Left Up-Down Number}(LUDN) as the total sum for all leftward arrows
within a graph, and {\it Right Up-Down Number}(RUDN) as the total sum for 
all rightward arrows within a graph. For SL(2,C) invariants, these indices
vanish. 

In order to count the number of the suffix contraction
we introduce the following ones.

(iii) {\it Left Wedge Number} and {\it Right Wedge Number}\nl
We assign $1$ for each piece of $\graph{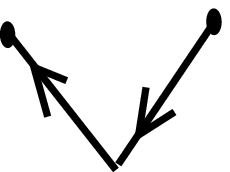}$ 
and define the total sum within a graph as {\it Left Wedge Number}(LWN).
In the same way,  
we assign $1$ for each piece of
$\graph{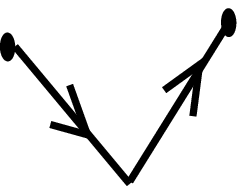}$ and define the total sum within a graph 
as {\it Right Wedge Number}(RWN).

(iv) {\it Dotted Line Number}\nl
We assign $1$ for one dotted line which shows a 
space-time suffix contraction. We define the total sum within a graph
as {\it Dotted Line Number}(DLN).

In addition to the graph-related indices, we introduce\nl
a) Physical Dimension (DIM);\ 
b) Number of the differentials (DIF);\ 
c) Number of $\si$ or $\sibar$ (SIG)

We list the above indices for basic spinor quantities in Table 1
and for the operators appearing the chiral multiplet Lagrangian (Sec.3)
in Table 2.

\vspace{2 cm}

\begin{tabular}{|c|c|c|c|c|}
\hline
             & $\graph{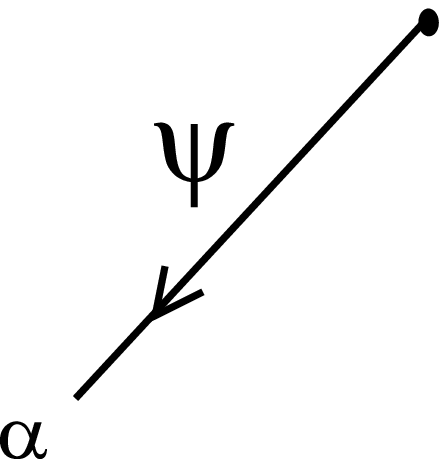}$  & $\graph{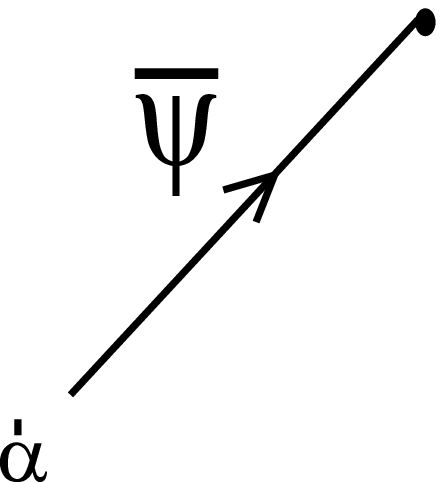}$ & $\graph{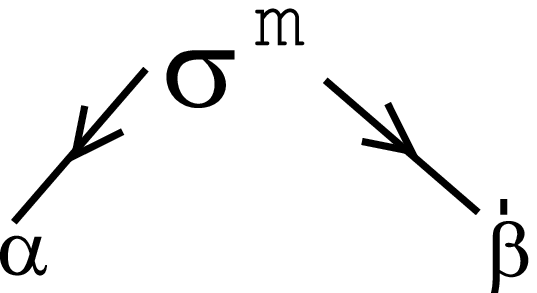}$  & 
                                                   $\graph{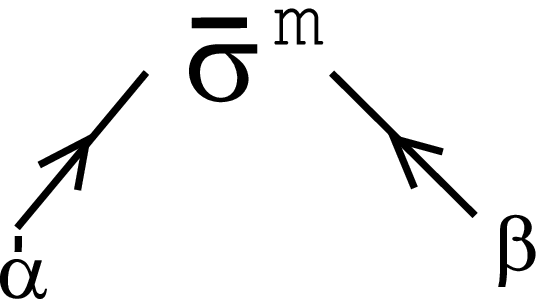}$            \\
\hline
(LCN, RCN)  & $(\half,0)$ & $(0,\half)$ & $(\half,\half)$& $(\half,\half)$\\
(LUDN, RUDN)  & $(-\half,0)$ & $(0,+\half)$ & $(-\half,-\half)$& 
                                                     $(\half,\half)$\\
(LWN, RWN)  &   0         &     0     &    0    &  0      \\
DLN         &  0         &     0     &    0    &  0      \\
DIM         &   $\frac{3}{2}$  &$\frac{3}{2}$  & 0  &  0  \\
DIF     &   0         &     0     &    0    &  0      \\
SIG     &   0     &   0     &   1  &   1  \\
\hline
\multicolumn{5}{c}{\q}                                                 \\
\multicolumn{5}{c}{Table 1\ \ Indices for basic spinor
quantities:\ $\psi_\al, \psibar^\aldot, (\si^m)_{\al\bedot}$
and $(\sibar^m)^{\aldot\be}$.  
}\\
\end{tabular}

\vspace{1 cm}
\begin{tabular}{|c|c|c|c|}
\hline
             & $\graph{Fchi2}$  & $A^*\Box A$ & $F^*F$            \\
\hline
(LCN, RCN)  & $(1,1)$ & / & /\\
(LUDN, RUDN)  & $(0,0)$ & / & /\\
(LWN, RWD)  &   (1,1)         &    /     &    /     \\
DLN         &  1         &     /    &    /       \\
DIM         &   $4$  &$4$  & $4$   \\
DIF     &   $1$         &   $2$     &    $0$        \\
SIG     &   $1$     &   $0$     &   $0$  \\
\hline
\multicolumn{4}{c}{\q}                                                 \\
\multicolumn{4}{c}{Table 2\ Indices for operators appearing in
the chiral multiplet Lagrangian.  
}\\
\end{tabular}
\vspace{1 cm}

We can identify every term appearing in the theory
in terms of some of these indices. We list some indices for
all spinorial operators in Super QED. We see all terms
are classified by the indices and the field contents.

\vs 2

\begin{tabular}{|c|c|c|c|c|c|}
\hline
       &      &(LCN,RCN)&   &      &                \\
       &      &=(LWN,RWN)&(LUDN,RUDN) & DIF  &  Fields        \\
\hline
1  & $-i\graph{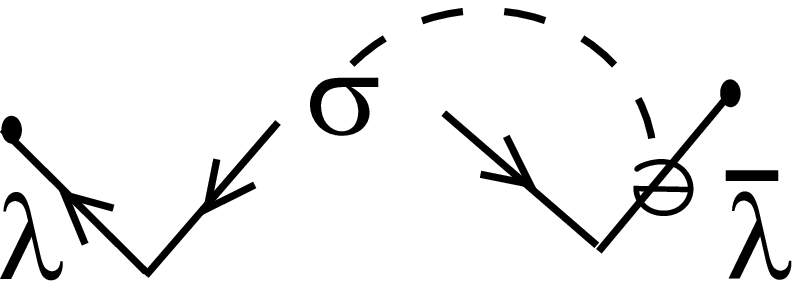}$ & $(1,1)$ & $(0,0)$ & $1$ & $\la,\labar$\\
2  & $i\graph{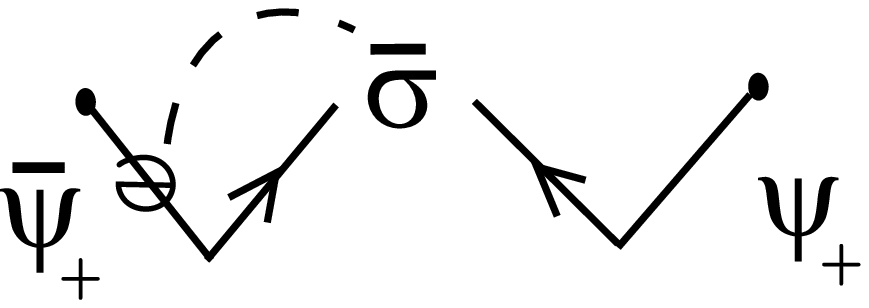}$ & $(1,1)$ & $(0,0)$ & $1$ & $\psi_+,\psibar_+$\\
3  & $i\graph{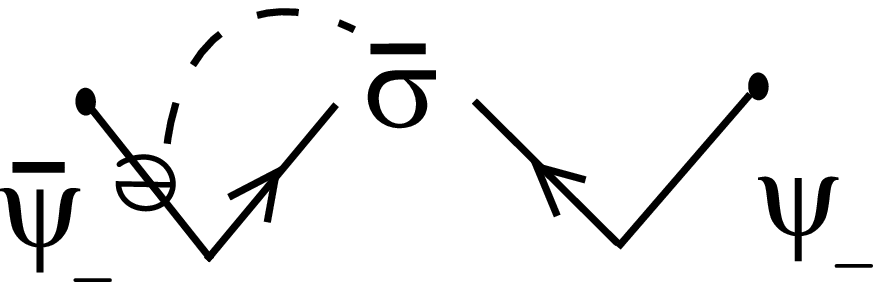}$ & $(1,1)$ & $(0,0)$ & $1$ & $\psi_-,\psibar_-$\\
\hline
4  & $\frac{e}{2}\graph{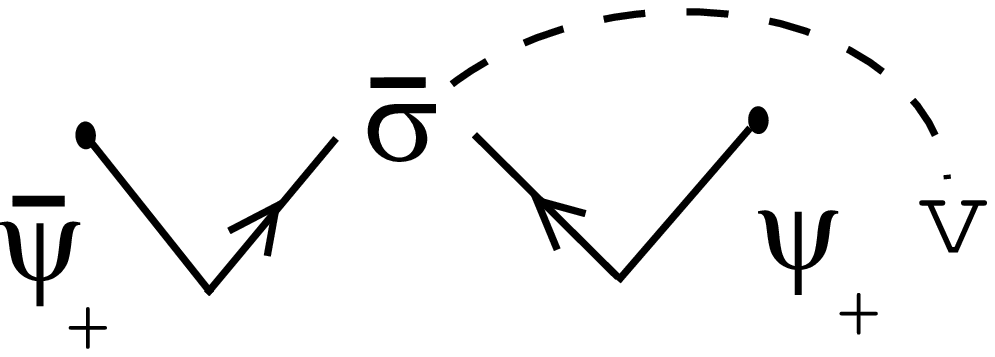}$ & $(1,1)$ & $(0,0)$ & $0$ & $\psi_+,\psibar_+,v^m$\\
5  & $-\frac{e}{2}\graph{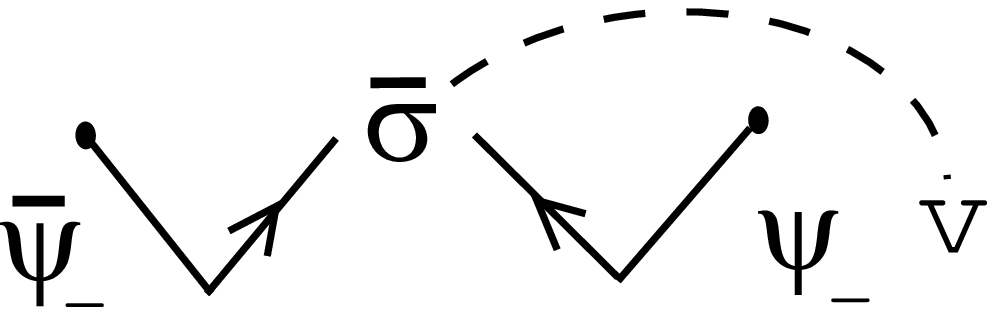}$ & $(1,1)$ & $(0,0)$ & $0$ & $\psi_-,\psibar_-,v^m$\\
\hline
6  & $-\frac{ie}{\sqtwo}A_+
\graph{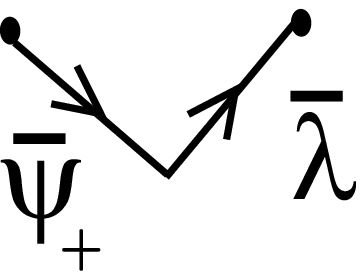}$ & $(0,1)$ & $(0,0)$ & $0$ & $A_+,\psibar_+,\labar$\\
7  & $+\frac{ie}{\sqtwo}A_-
\graph{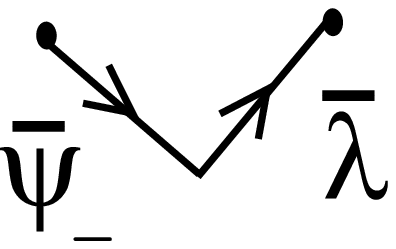}$ & $(0,1)$ & $(0,0)$ & $0$ & $A_-,\psibar_-,\labar$\\
\hline
8  & $+\frac{ie}{\sqtwo}A_+^*
\graph{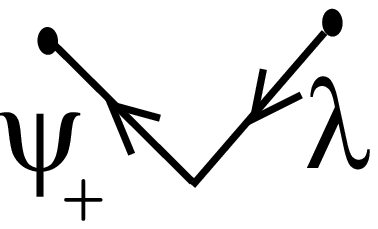}$ & $(1,0)$ & $(0,0)$ & $0$ & $A^*_+,\psi_+,\la$\\
9  & $-\frac{ie}{\sqtwo}A_-^*
\graph{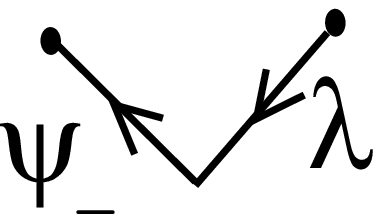}$ & $(1,0)$ & $(0,0)$ & $0$ & $A^*_-,\psi_-,\la$\\
\hline
10  & $-m\graph{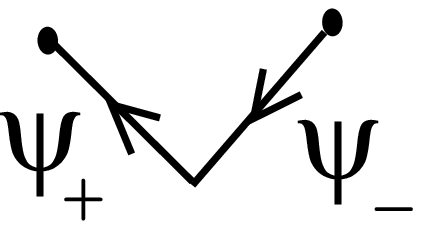}$ & $(1,0)$ & $(0,0)$ & $0$ & $\psi_+,\psi_-$\\
11  & $-m\graph{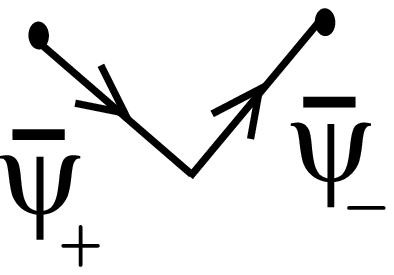}$ & $(0,1)$ & $(0,0)$ & $0$ & $\psibar_+,\psibar_-$\\
\hline
\multicolumn{6}{c}{\q}                                                 \\
\multicolumn{6}{c}{Table 3\ \ List of indices for all spinor operators in
the super QED Lagrangian.}\\
\multicolumn{6}{c}{$(\la,\labar)$:\ photino;\ $v^m$:\ photon;\ 
$(\psi_+,\psibar_+)$:\ $+e$ chiral fermion; $(\psi_-,\psibar_-)$:\ 
$-e$ chiral fermion. 
}\\
\end{tabular}

\vspace{5mm}

The product of $\si$-matrices appear in 
the intermediate stage of SUSY calculation. 
Its
classification is an important subject because
that fixes the right (graphical) formula to be
used for reduction of SUSY quantities. 
We list 
the result, by using the graph indices defined in this section, 
in TABLE 4, 5, 6 for the case of
2$\si$'s, 3$\si$'s and 4$\si$'s respectively. 
This result is exploited in the C-program calculation
of SUSY\cite{SI06UW08}.

\begin{table}
\begin{center}
\begin{tabular}{c|c|c|c}
\hline
DLN & LWN & RWN & figure\\
\hline
      &   0     &    0     &  $\graph{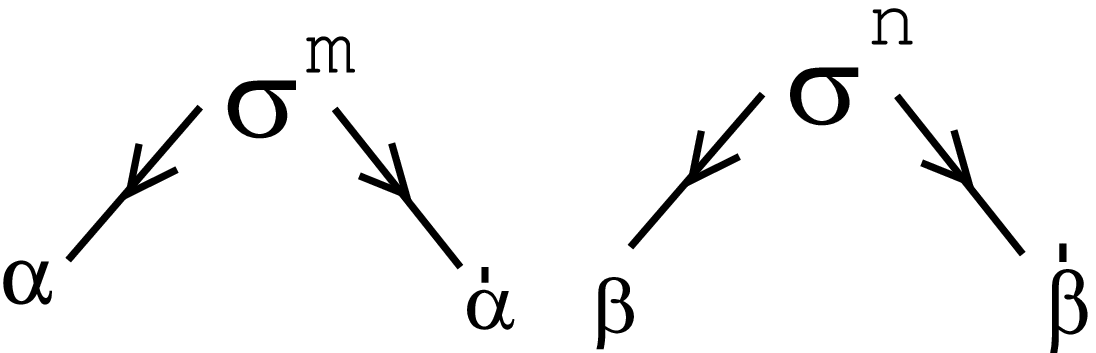}$  \\
\cline{2-4}
  0   &   0      &   1     &  $\graph{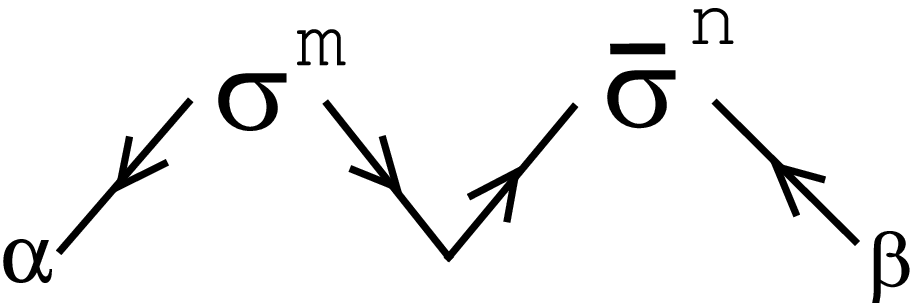}$  \\
\cline{2-4}
       &   1     &   0      &  $\graph{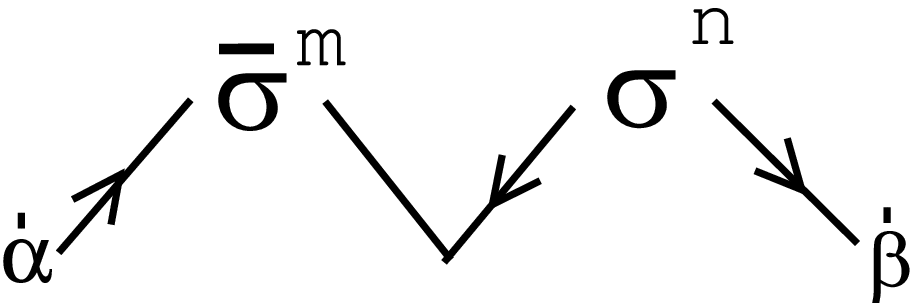}$  \\
\cline{2-4}
       &   1      &   1     &  $\graph{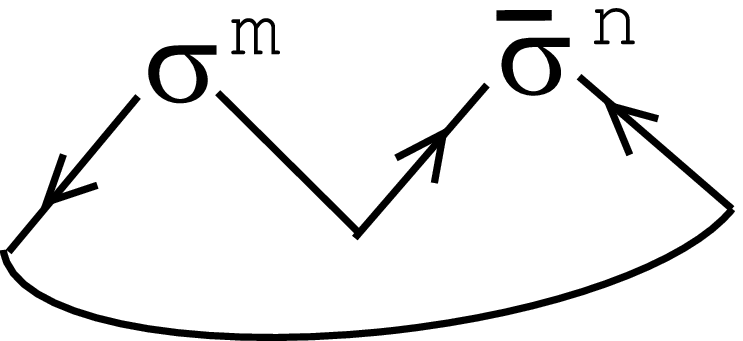}\ =-2\eta^{mn}$  \\
\hline
       &   0       &  0      &  $\graph{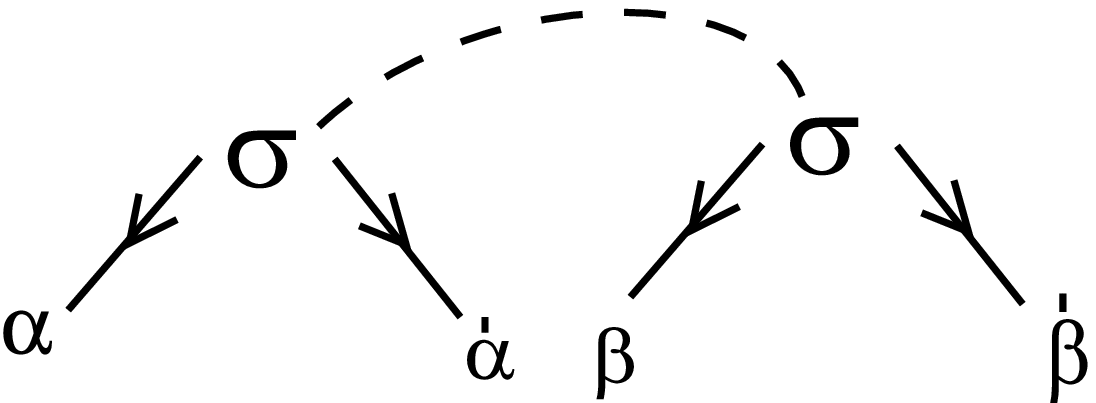}\ =-2\ep_{\al\be}\ep_{\aldot\bedot}$  \\
\cline{2-4}
   1   &   0      &    1     &  $\graph{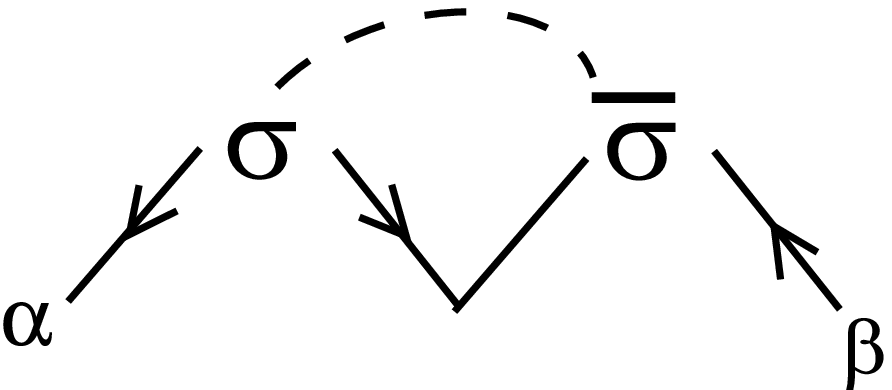}\ =-4\del^\be_\al$  \\
\cline{2-4}
       &    1      &    0     &  $\graph{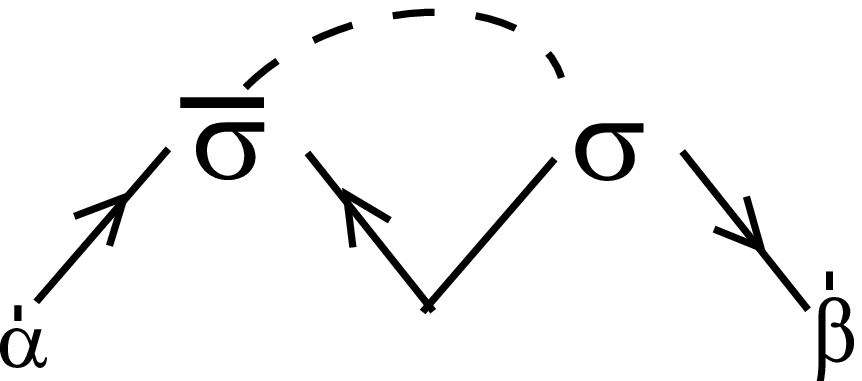}\ =-4\del_\bedot^\aldot$  \\
\cline{2-4}
        &    1     &    1     &   $\graph{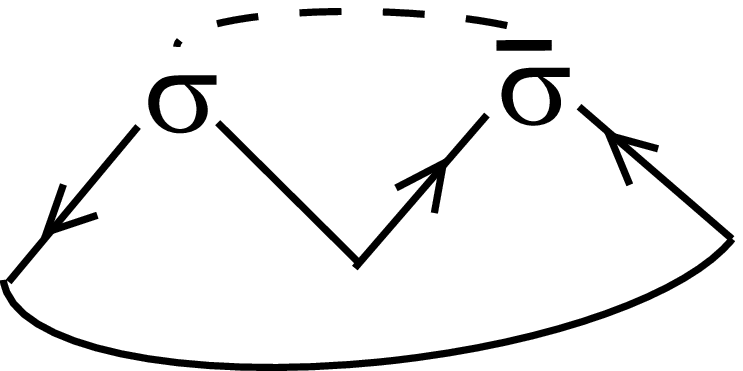}\ =-8$  \\
\hline
\multicolumn{4}{c}{}\\
\multicolumn{4}{c}{TABLE 4\ Classification of the product of 2 sigma matrices (nsi=2).}
\end{tabular}
\end{center}
\end{table}

\begin{table}
\begin{center}
\begin{tabular}{c|c|c|c}
\hline
DLN & LWN & RWN & figure\\
\hline
      &   0     &    0     &  $\graph{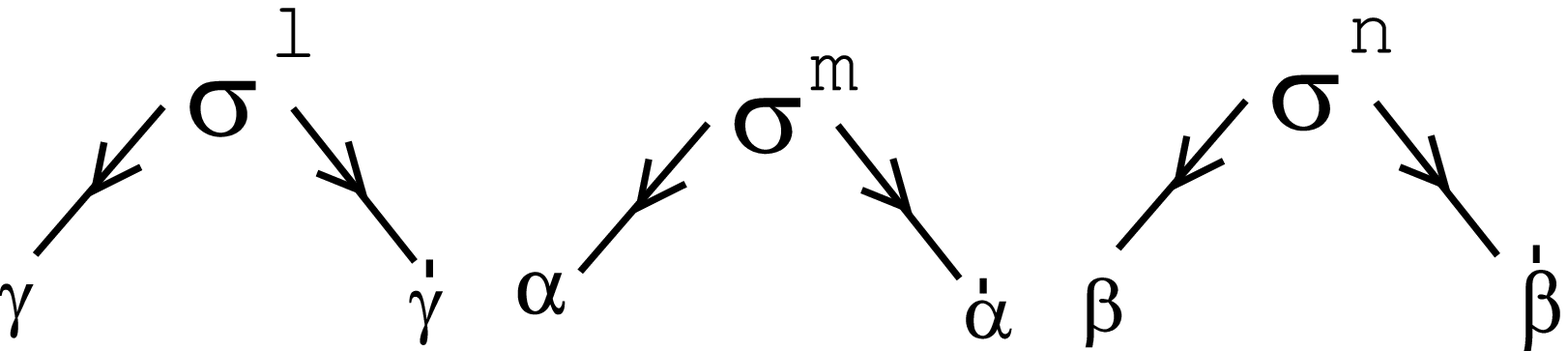}$  \\
\cline{2-4}
  0   &   0      &   1     &  $\graph{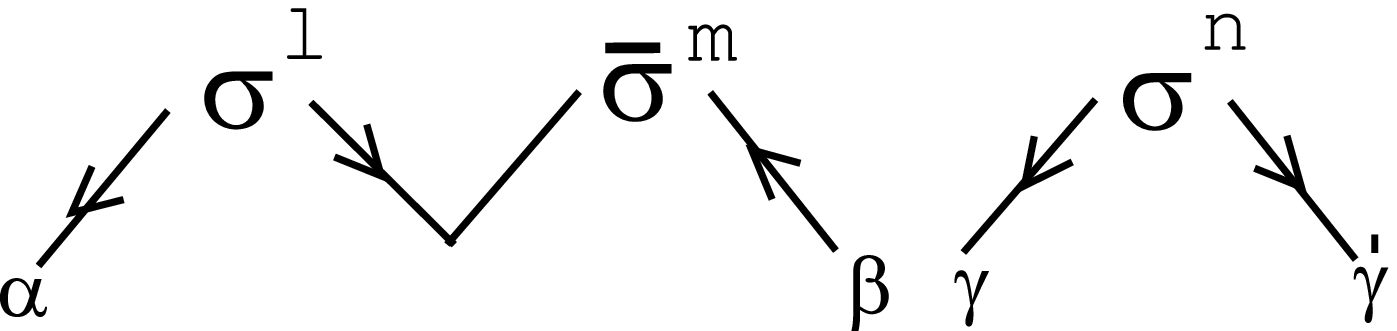}$  \\
\cline{2-4}
       &   1     &   0      &  $\graph{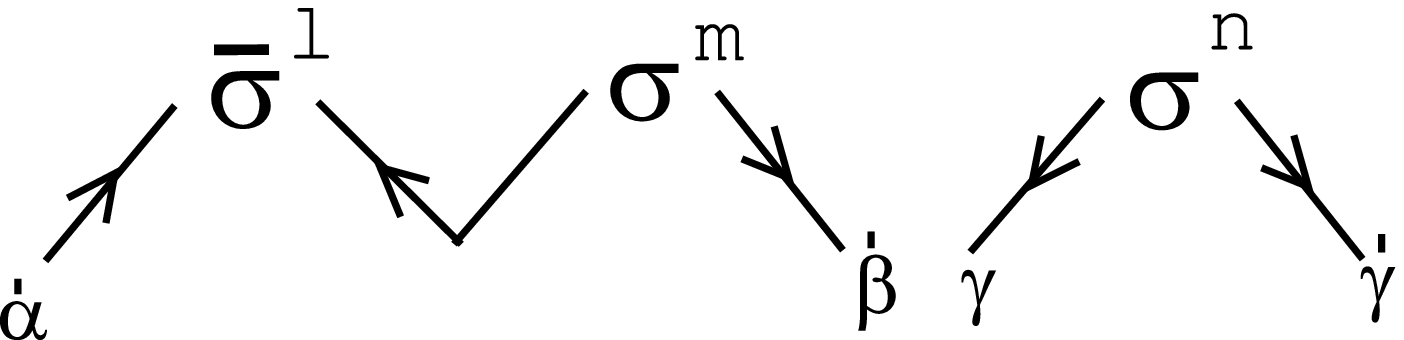}$  \\
\cline{2-4}
       &   1      &   1     &  
\shortstack[l]{
closed-chiral-loop No =1\\
$\graph{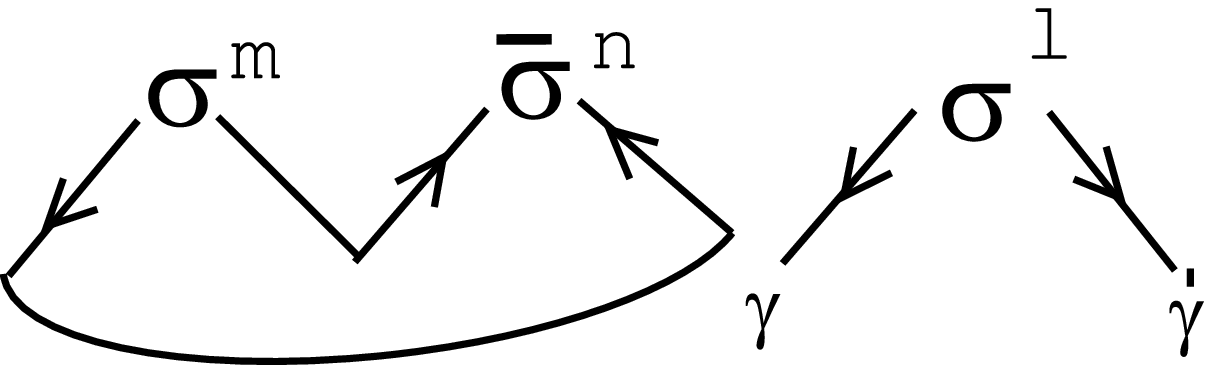}$  }
                                  \\
\cline{4-4}
       &          &         &   
\shortstack[l]{
closed-chiral-loop No =0\\
$\graph{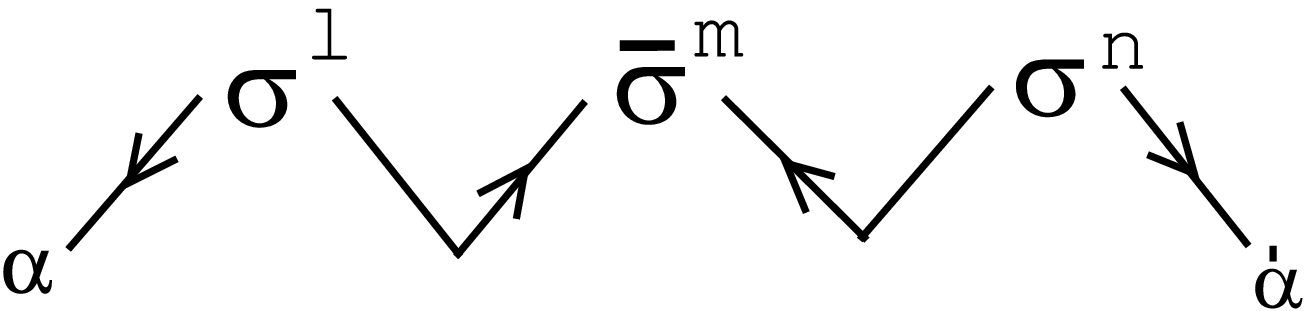}$  }
                                  \\
\hline
       &   0       &  0      &  $-2\ep_{\al\be}\ep_{\aldot\bedot}\ \graph{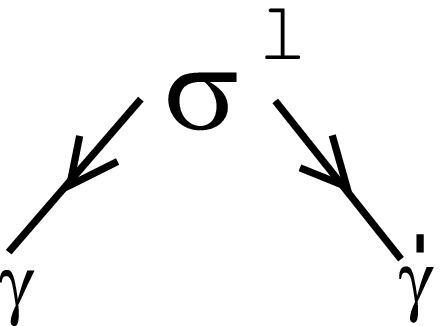}$  \\
\cline{2-4}
   1   &   0      &    1     &  $-4\del^\be_\al\ \graph{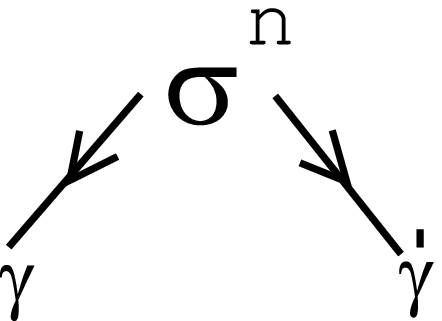}$  \\
\cline{2-4}
       &    1      &    0     &  $-4\del^\aldot_\bedot\ \graph{sig3Dr}$  \\
\cline{2-4}
        &    1     &    1     &   $-8\ \graph{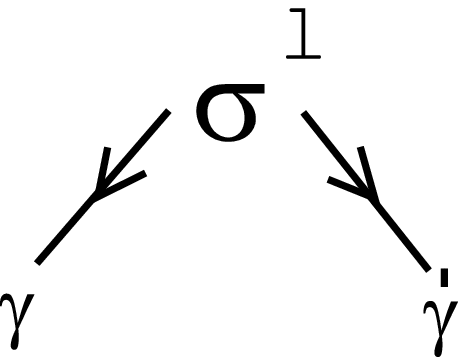}$  \\
\hline
\multicolumn{4}{c}{}\\
\multicolumn{4}{c}{TABLE 5\ Classification of the product of 3 sigma matrices (nsi=3).}
\end{tabular}
\end{center}
\end{table}

\begin{table}
\begin{center}
\begin{tabular}{c|c|c|c}
\hline
DLN & LWN & RWN & figure\\
\hline
      &   0     &    0     &  $\graph{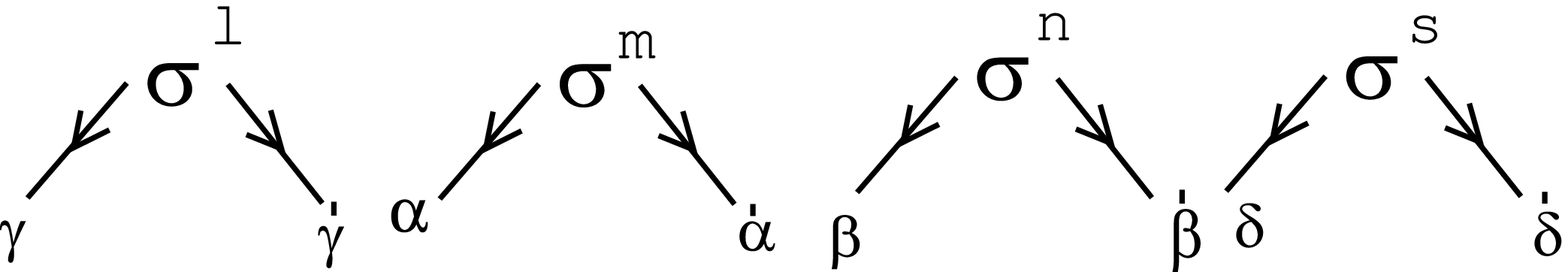}$  \\
\cline{2-4}
      &   0      &   1     &  $\graph{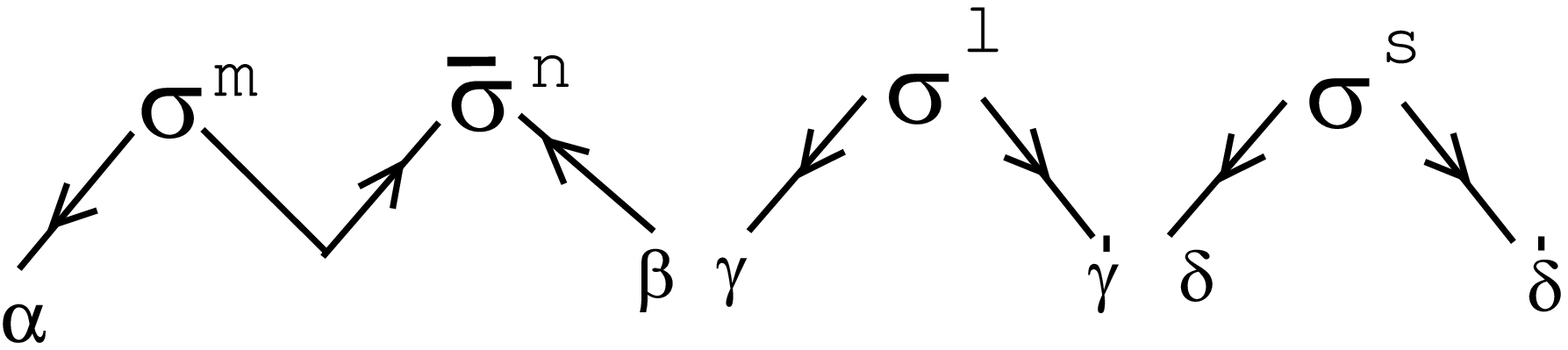}$  \\
\cline{2-4}
      &   1      &   0     &  $\graph{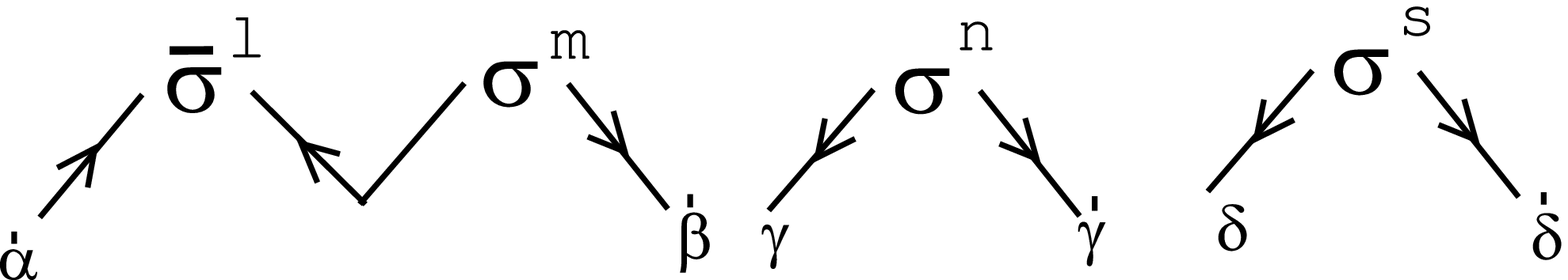}$  \\
\cline{2-4}
      &          &         &  
\shortstack[l]{
GrNum=2, Division=(2,2)\\
$\graph{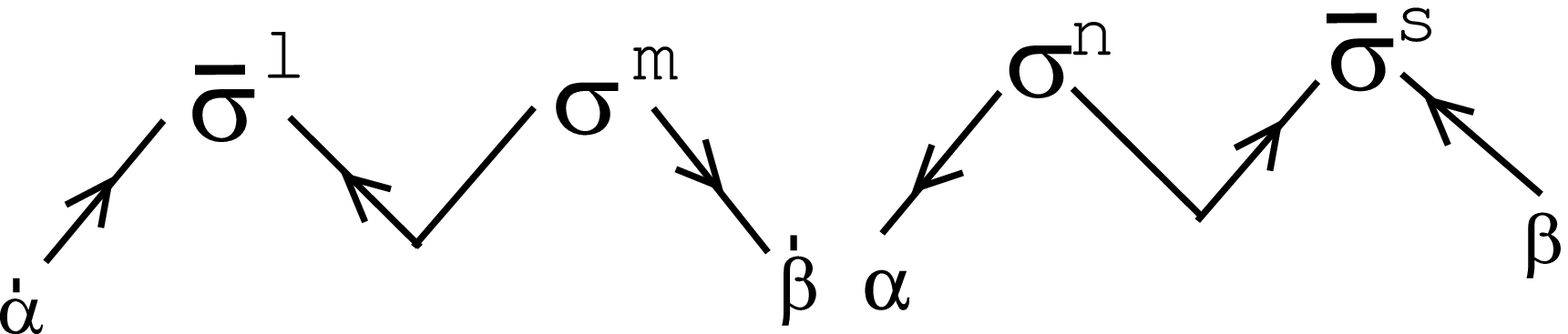}$ } 
                                                    \\
\cline{4-4}
      &   1      &   1     &  
\shortstack[l]{
GrNum=2, Division=(3,1)\\
$\graph{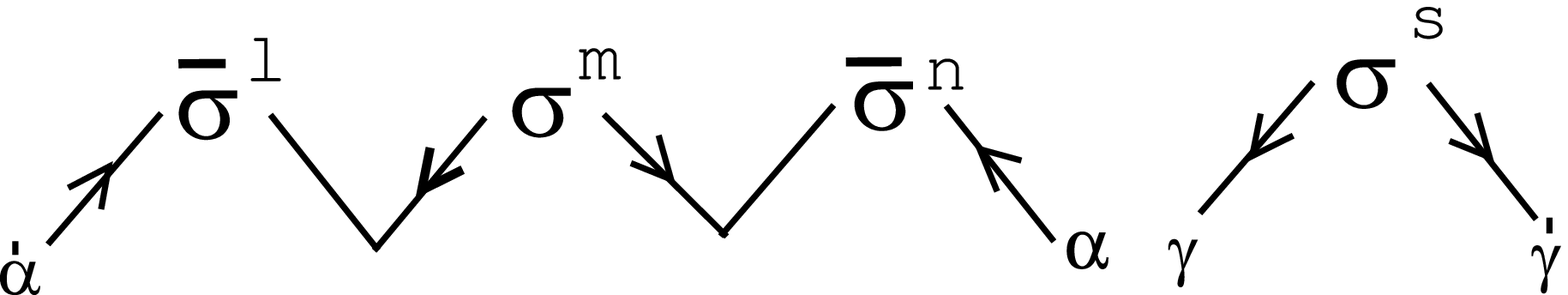}$}                                   \\
\cline{4-4}
      &        &       &  
\shortstack[l]{
GrNum=3, Division=(2,1,1)\\
$\graph{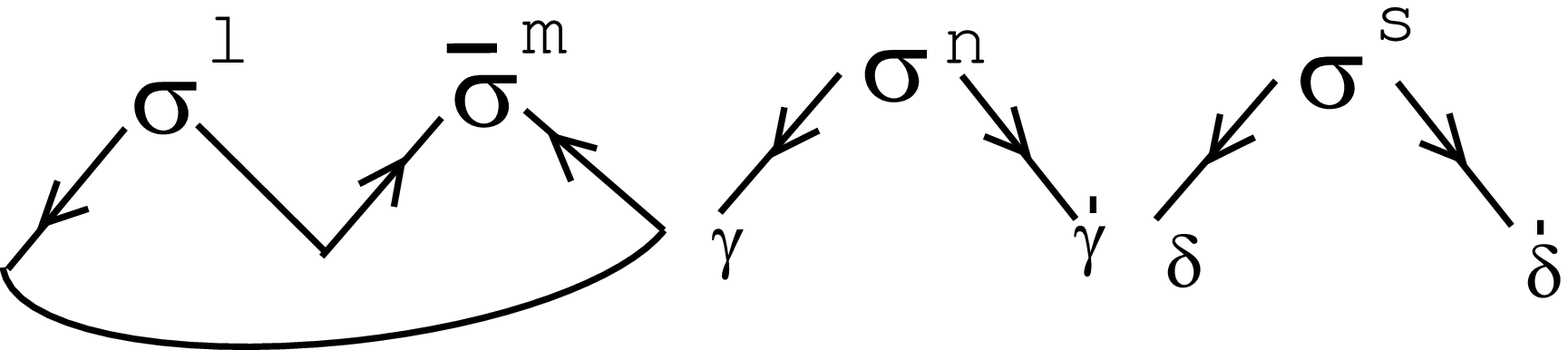}$ }                               \\
\cline{2-4}
  0   &   2      &   0     &  $\graph{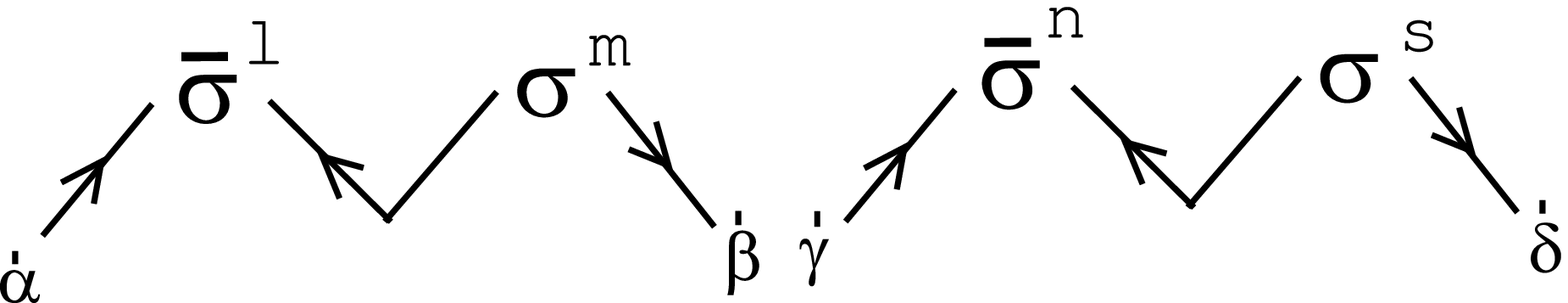}$  \\
\cline{2-4}
       &   0     &   2      &  $\graph{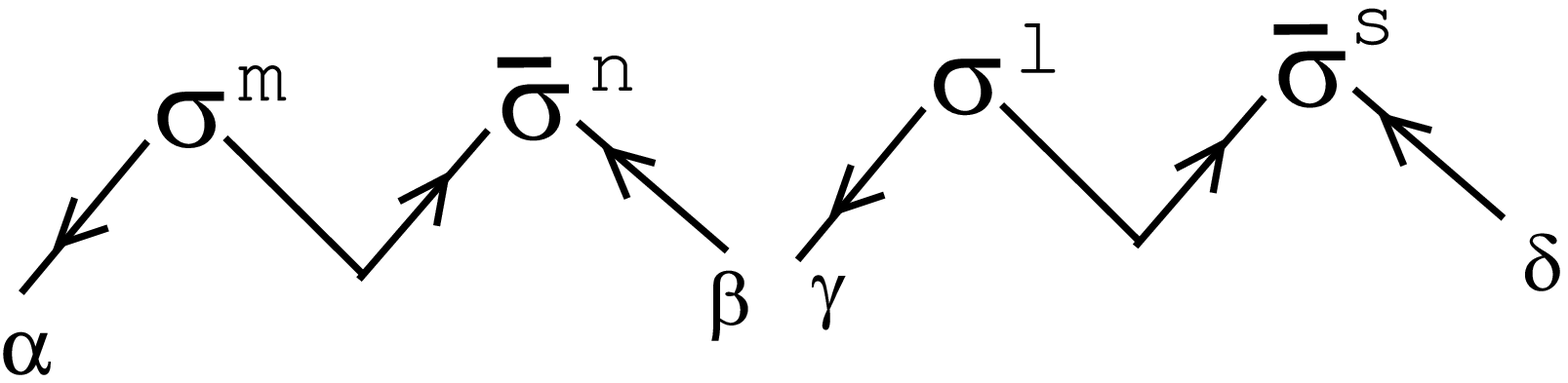}$  \\
\cline{2-4}
       &   1      &   2     &   GrNum=1, $\graph{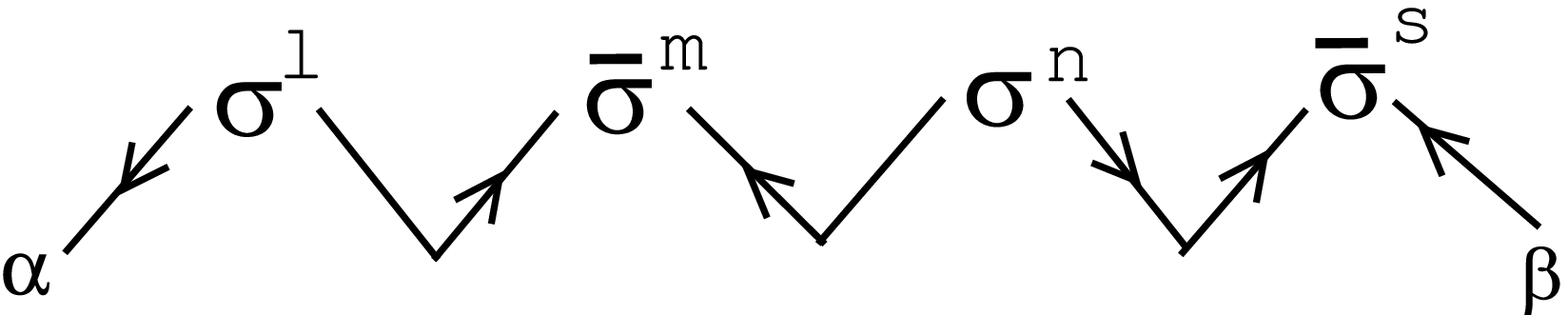}$  \\
\cline{4-4}
       &         &        &  GrNum=2, $\graph{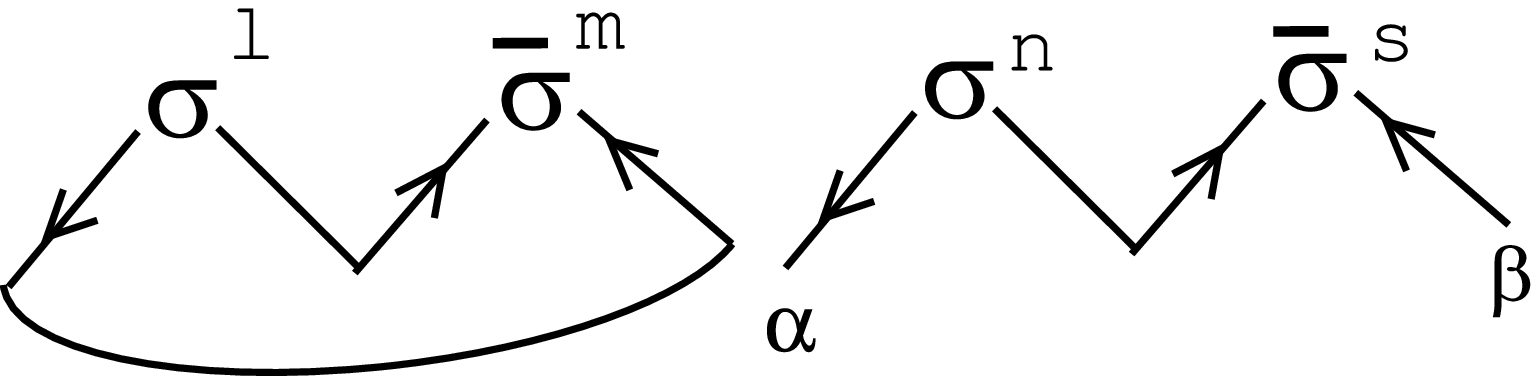}$     \\
\cline{2-4}
       &    2    &     1   &  GrNum=1, $\graph{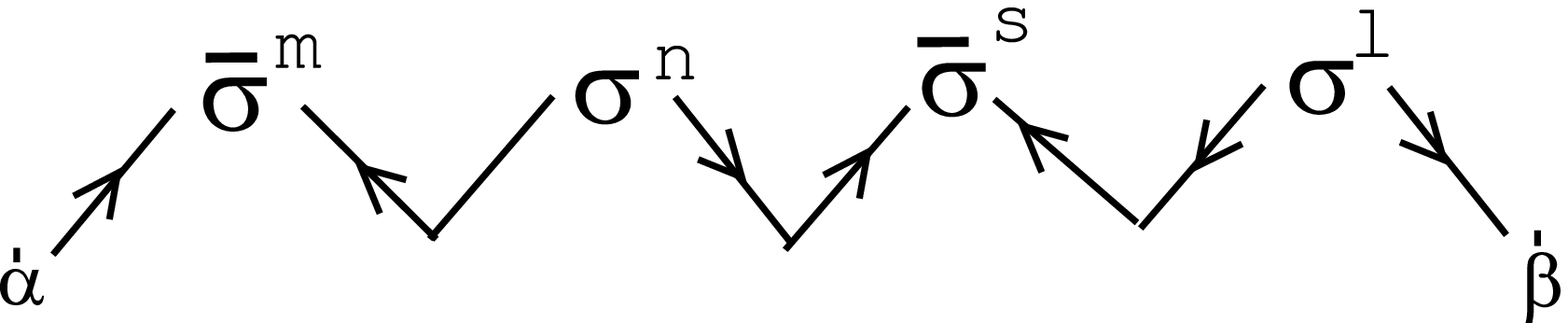}$  
                                                   \\
\cline{4-4}
       &          &         &   GrNum=2, $\graph{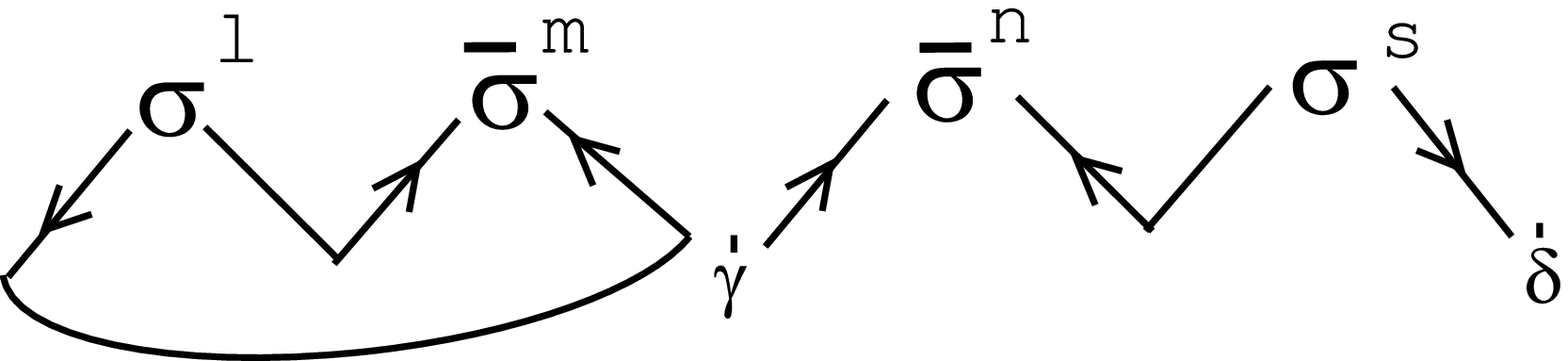}$  
                                                     \\
\cline{2-4}
       &    2     &   2     &   GrNum=1, $\graph{ssbssbBR}$  
                                                   \\
\cline{4-4}
       &          &         &   GrNum=2,\ $\graph{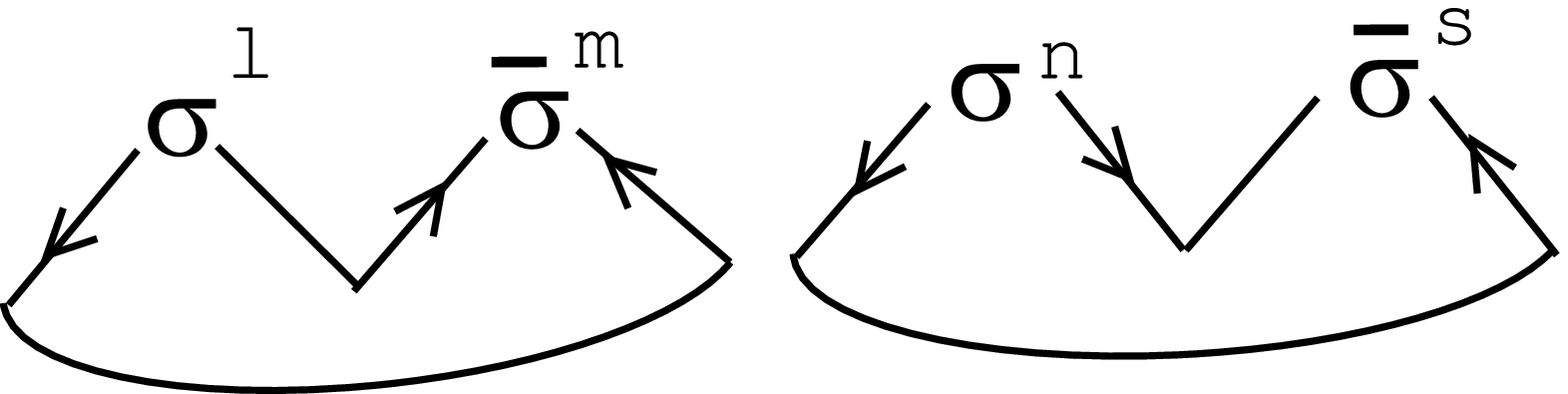}$  
                                                     \\
\hline
\multicolumn{4}{c}{}\\
\multicolumn{4}{c}{TABLE 6\ Classification of the product of 4 sigma matrices with}\\
\multicolumn{4}{c}{\q\q no vector-suffix contraction (nsi=4, DLN=0). }
\end{tabular}
\end{center}
\end{table}

\section{Superspace Quantities}
We know the SUSY symmetry is most naturally viewed in the
superspace ($x^m,\sh,\thbar$). Here
we introduce anti-commuting parameters $\sh_\al=\ep_\ab\sh^\be,
\sh^\al,\thbar_\aldot,\thbar^\aldot=\ep^{\aldot\bedot}\thbar_\bedot$
as the basic spinor coordinate. We show them graphically
in Fig.16.
\begin{figure}
\begin{center}
\includegraphics*[height=6cm]{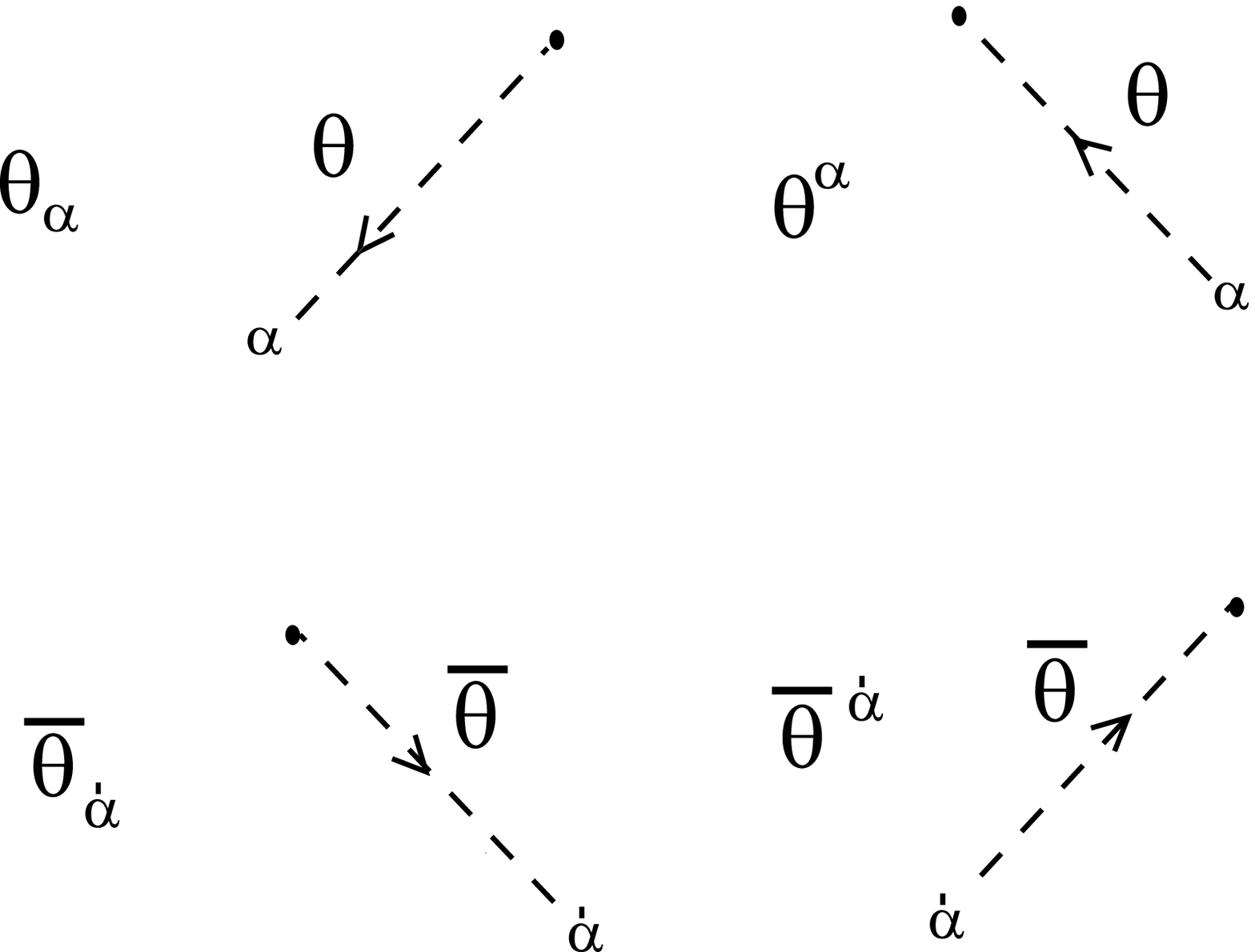}
\end{center}

   \caption{
The graphical representation for the
spinor coordinates in the superspace:\ $\sh_\al, \sh^\al, \thbar_\aldot$ and 
$\thbar^\aldot$.
   }
\end{figure}
They satisfy the graphical relations of Fig.17.
\begin{figure}
\begin{center}
\includegraphics*[height=8cm]{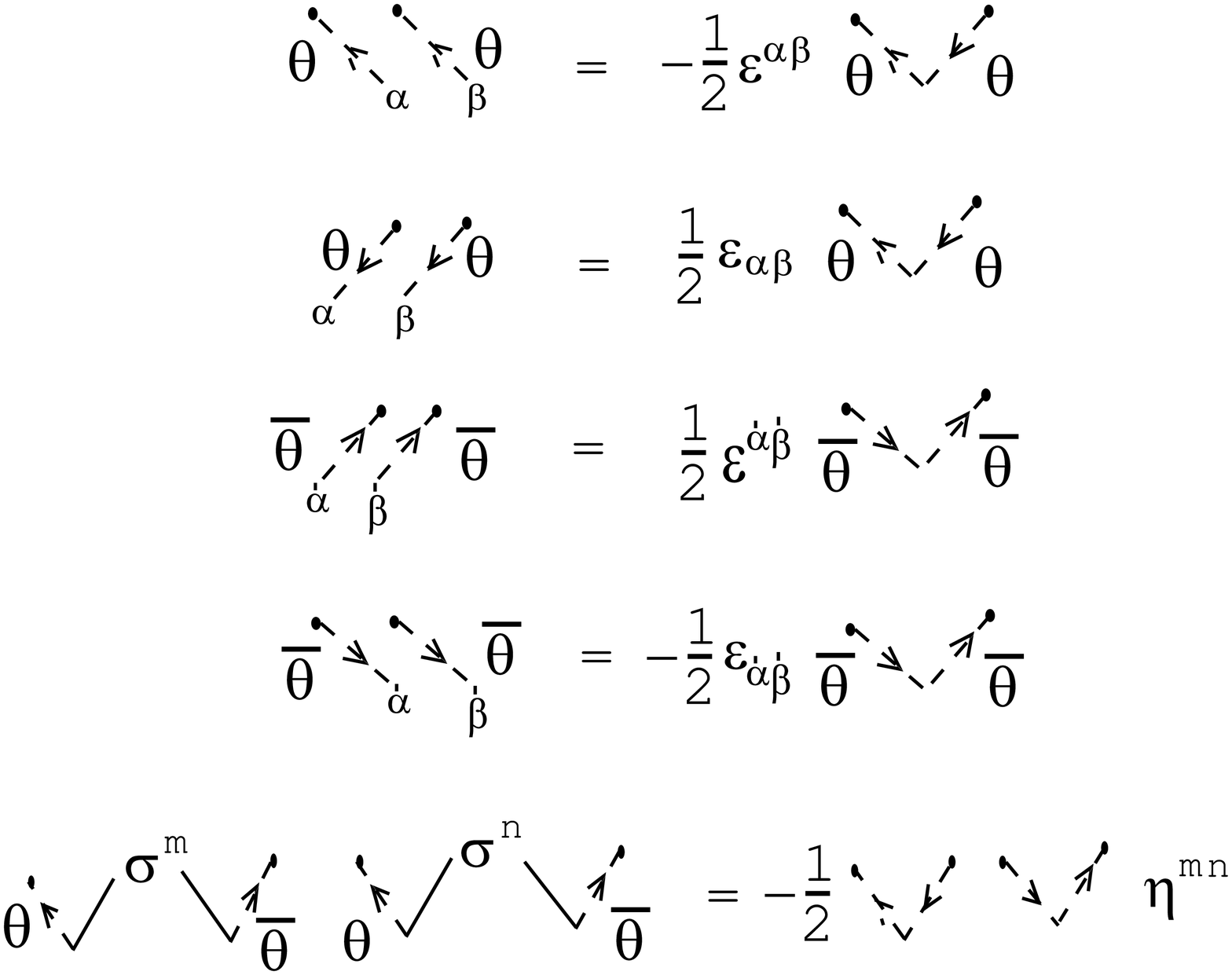}
\end{center}
   \caption{
The graphical rules for the spinor coordinates:\ 
$\sh^\al\sh^\be=-\half\ep^\ab\sh\sh,~\sh_\al\sh_\be=\half\ep_\ab\sh\sh,~
\thbar^\aldot\thbar^\bedot=\half\ep^{\aldot\bedot}\thbar\thbar,~
\thbar_\aldot\thbar_\bedot=-\half\ep_{\aldot\bedot}\thbar\thbar,~
\sh\si^m\thbar\sh\si^n\thbar=-\half\sh\sh\thbar\thbar\eta^{mn}$.
   }
\end{figure}

The general superfield $F(x,\sh,\thbar)$, in terms of
components fields\nl
($f(x), \phi(x), \chibar(x), m(x), n(x), v_m(x), \labar(x), \psi(x), d(x)
$)
, is shown as
\begin{eqnarray}
F(x,\sh,\thbar)=\nn
f(x)+
\graph{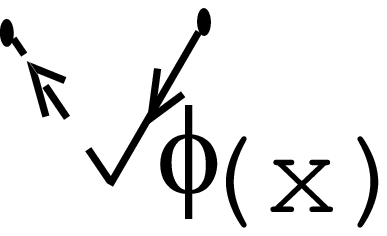}+\graph{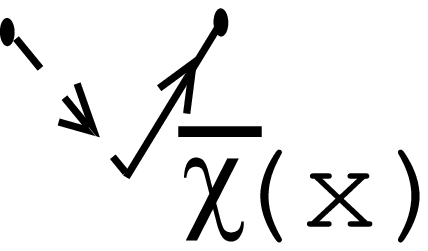}+\graph{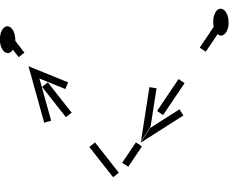}m(x)
+\graph{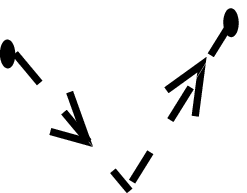}n(x)\nn
+\graph{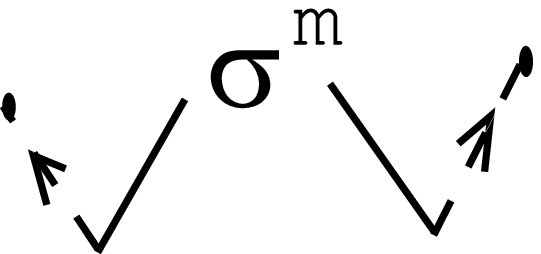}v_m(x)+\graph{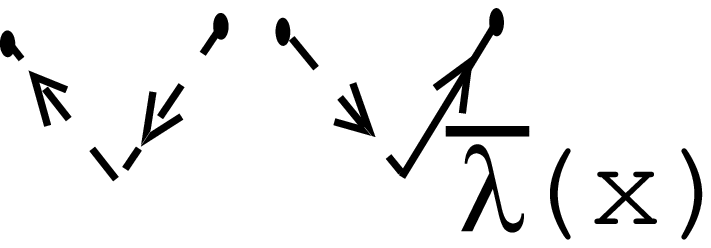}
+\graph{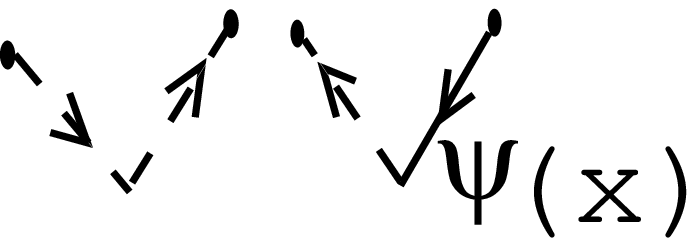}\nn
+\graph{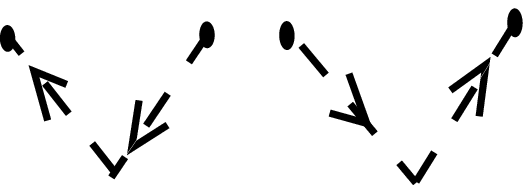}d(x)
\pr
\label{ss1}
\end{eqnarray}
The SUSY transformation generator $Q_\al$ and $\Qbar^\aldot$
are expressed as
\begin{eqnarray}
Q_\al=\frac{\pl}{\pl\sh^\al}-i\graph{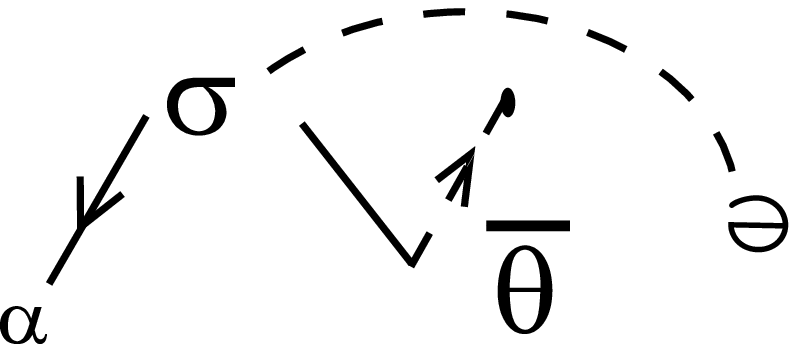}\com\nn
\Qbar^\aldot=\frac{\pl}{\pl\thbar_\aldot}+i\graph{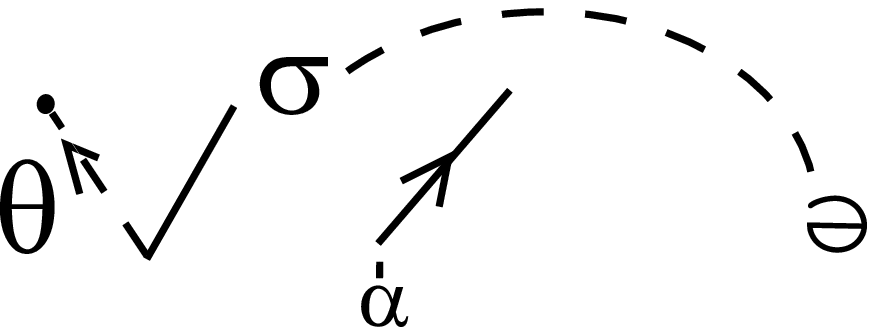}
\com
\label{ss2}
\end{eqnarray}
The SUSY derivative operators $D$ and $\Dbar$, which are
the conjugate partners of $Q$ and $\Qbar$, are expressed as
\begin{eqnarray}
D_\al=\frac{\pl}{\pl\sh^\al}+i\graph{F1ss2}\com\nn
\Dbar_\aldot=-\frac{\pl}{\pl\thbar^\aldot}-i\graph{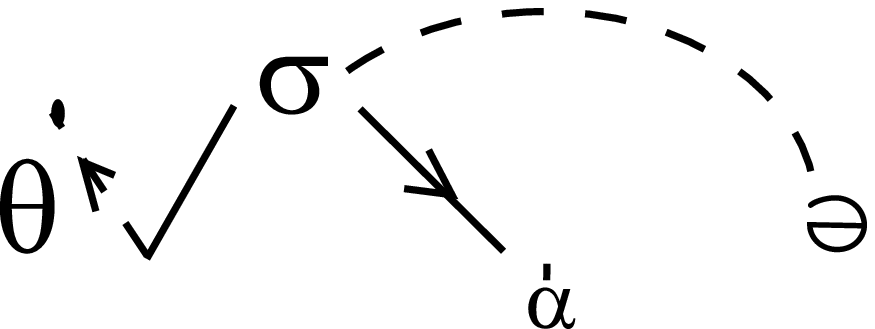}
\com
\label{ss3}
\end{eqnarray}
We can {\it graphically} confirm the SUSY algebra
by using the commutativity and anti-commutativity between
$\sh, \thbar, \frac{\pl}{\pl\sh}, \frac{\pl}{\pl\thbar}$ and $\pl_m$.
\begin{eqnarray}
\{Q_\al ,\Qbar_\aldot\}=2i\graph{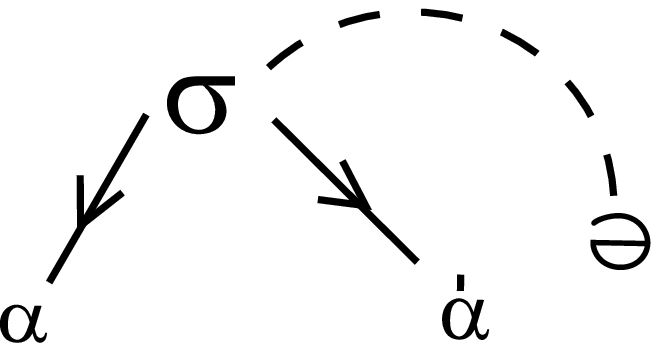}\com\nn
\{D_\al ,\Dbar_\aldot\}=-2i\graph{F1ss4}\pr
\label{ss4}
\end{eqnarray}

In the treatment of the chiral superfield, it is important to choose
appropriate coordinates:\ ($y^m=x^m+i\sh\si^m\thbar, \sh, \thbar$)
for the chiral field, and ($y^{\dag m}=x^m-i\sh\si^m\thbar, \sh, \thbar$)
for the anti-chiral one.
\begin{eqnarray}
y^m=x^m+i\graph{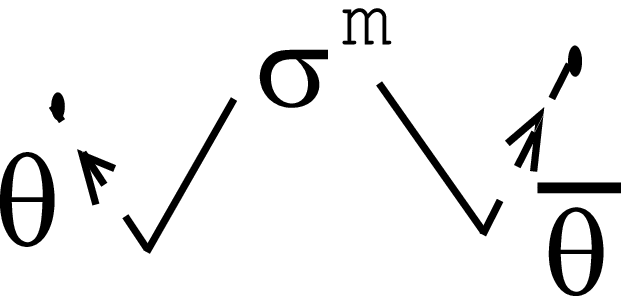}\com\nn
y^{\dag m}=x^m-i\graph{F1ss5}\pr
\label{ss5}
\end{eqnarray}
Because of the properties ($\Dbar_\aldot y^m=0, \Dbar_\aldot\sh=0$) and
($D_\al y^{\dag m}=0, D_\al\thbar=0$), the chiral superfield $\Phi$
($\Dbar_\aldot \Phi=0$) and the anti-chiral one $\Phi^\dag$
($D_\al\Phi^\dag=0$) are always written as
\begin{eqnarray}
\Phi(y,\sh,\thbar)=A(y)+\sqtwo\graph{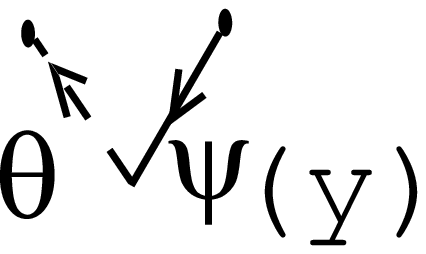}+\graph{F3ss1}F(y)\com\nn
\mbox{and}\q\q\q\q\q\q\q\q\q\q\q\q\nn
\Phi^\dag(y^\dag,\sh,\thbar)=A^*(y^\dag)+\sqtwo\graph{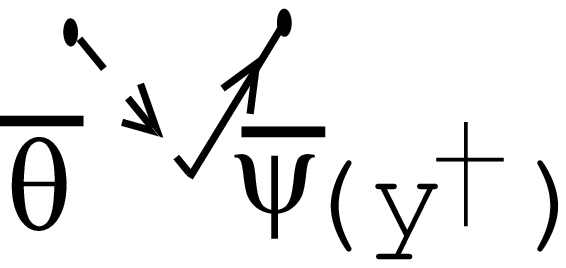}
+\graph{F4ss1}F^*(y^\dag)
\com
\label{ss6}
\end{eqnarray}
respectively. 
The SUSY differential operators are expressed as, in terms of
($y,\sh,\thbar$),
\begin{eqnarray}
D_\al=\frac{\pl}{\pl\sh^\al}+2i\graph{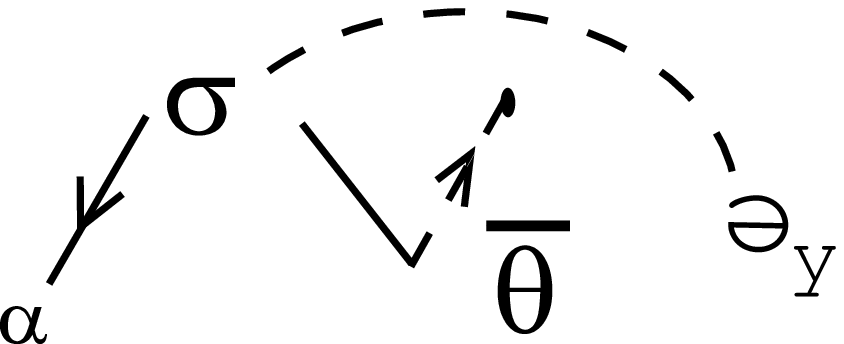}\com\q 
\Dbar_\aldot=-\frac{\pl}{\pl\thbar^\aldot}\com\nn
Q_\al=\frac{\pl}{\pl\sh^\al}\com\q
\Qbar_\aldot=-\frac{\pl}{\pl\thbar^\aldot}+2i\graph{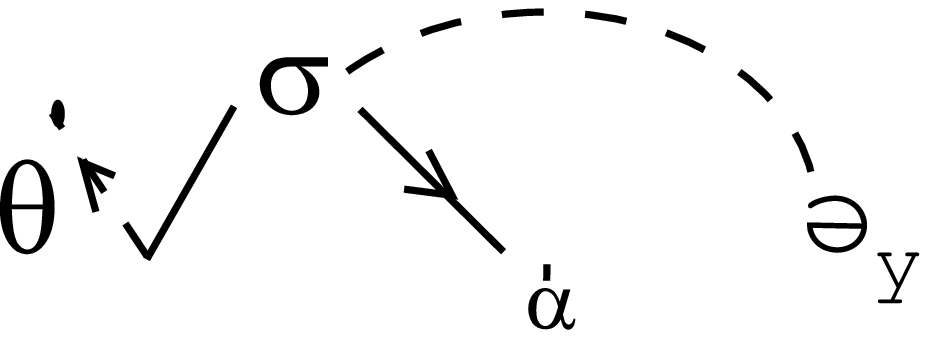}
\pr
\label{ss7}
\end{eqnarray}
and as, in terms of ($y^\dag,\sh,\thbar$),
\begin{eqnarray}
D_\al=\frac{\pl}{\pl\sh^\al}\com\q 
\Dbar_\aldot=-\frac{\pl}{\pl\thbar^\aldot}-2i\graph{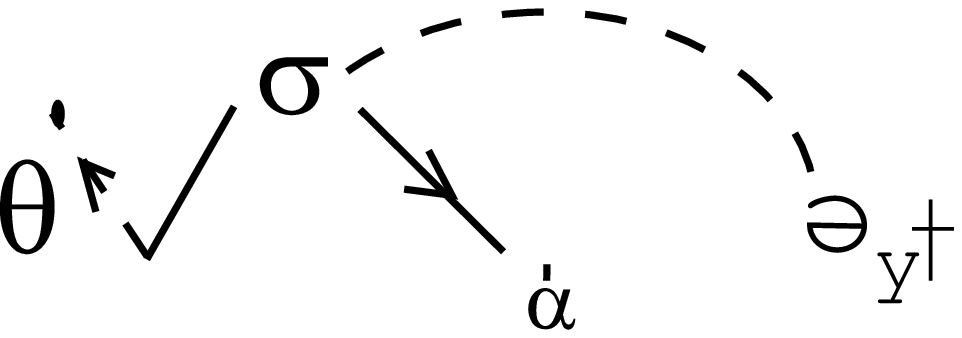}\com\nn
Q_\al=\frac{\pl}{\pl\sh^\al}-2i\graph{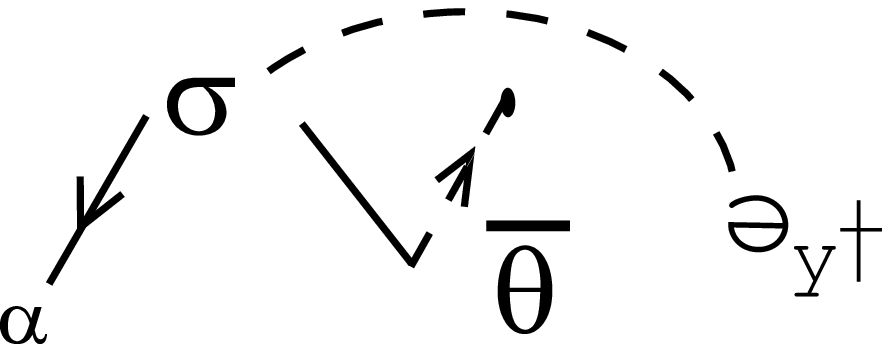}\com\q
\Qbar_\aldot=-\frac{\pl}{\pl\thbar^\aldot}
\pr
\label{ss8}
\end{eqnarray}

The superspace calculation can be performed {\it graphically}.
For example, the $\Phi^\dag\Phi$ calculation, in order to
find the 4D SUSY Lagragian, can be done using the graph
relations of Fig.17. An advantage, compared to the usual
one, is that the graphically expressed quantity is not
"obscured" by the dummy (contracted) suffixes.

\section{Application to 5D Supersymmetry}
In this section we apply the graphical method
to a recent subject \cite{MP97,Hebec0112}: 5D supersymmetry. 
Here {\it both} (4-components and 2-components spinors) {\it types
of representation appear} in relation to SUSY "decomposition". 
Stimulated by
the brane world physics, higher dimensional SUSY becomes
an important subject. In particular, 5 dimensional one is used
as a concrete extended model of the standard model. The simplest
one is the hypermultiplet $(A^i,\chi,F_i)$, where both $A^i$($i=1,2$) 
and $F^i$ 
are the SU(2)$_R$ doublet of complex scalars. $F^i$ are the auxiliary
fields. $\chi$ is a Dirac field. The SU(2)$_R$ suffix, $i$, is lowered
or raised by $\ep_{ij}$ and $\ep^{ij}$:\ 
$A_i=\ep_{ij}A^j, F^i=\ep^{ij}F_j$ 
where $\ep_{ij}$ and $\ep^{ij}$ are the same as (\ref{def1}). 

The doublet fields are graphically
represented as in Fig.18. 
\begin{figure}
\begin{center}
\includegraphics*[height=4cm]{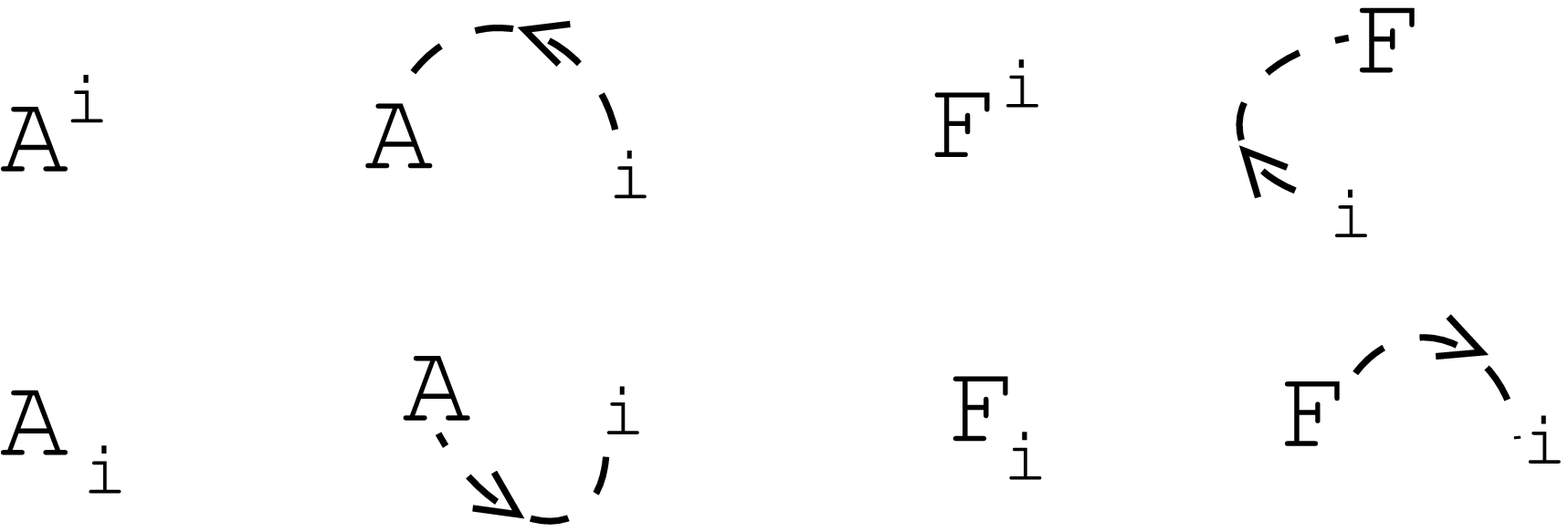}
\end{center}
   \caption{
The graphical representation for the SU(2)$_R$ doublet
fields, $A^i, A_i=\ep_{ij}A^j, F_i$ and $ F^i=\ep^{ij}F_j$.
The arrowed dotted line is depicted arbitrarily as far as
the one end is attached to the symbol and the correct
arrow direction is taken.
   }
\end{figure}
The SU(2)$_R$ suffix up-down is expressed by the arrow
direction:\ the flow-in direction for the up-suffix and the flow-out
direction for the down-suffix. (This representation of the suffix up-down
is different from the treatment taken in the spinor-suffix case of Sec.2.) 

As the 5D SUSY parameter, we take the symplectic Majorana spinors.
They are SU(2)$_R$ doublet of {\it Dirac} spinors $\xi^i$($i=1,2$) which
satisfy the symplectic Majorana condition ("reality" condition).
\begin{eqnarray}
\xi^i=\ep^{ij}C\xibar_j^T\com\q
C=
\left(\begin{array}{cc}
i\si^2 & 0 \\
0 & i\si^2 \end{array}
\right)
\pr\label{fds1}
\end{eqnarray}
From the number of the independent SUSY parameters,
we know the present system has 8 (counted in real) supercharges. 
We introduce the graphical
representation for $\xi^i$ and $\xibar_i$ as in Fig.19.
The Dirac spinor structure is graphically the same as the Majorana one
of Sec.5.

\begin{figure}
\begin{center}
\includegraphics*[height=4cm]{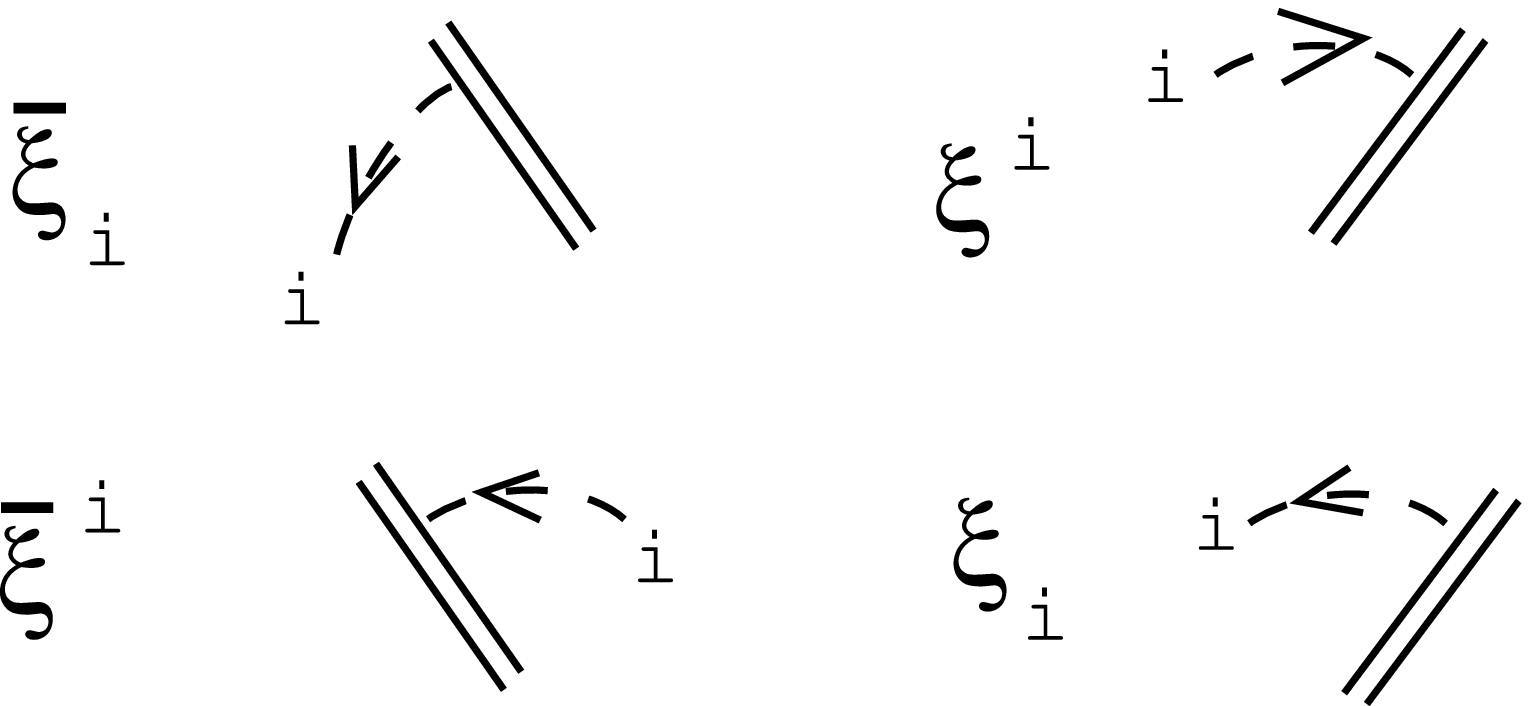}
\end{center}
   \caption{
The graphical representation for the 5D SUSY parameters, 
$\xibar_i, \xi^i, \xibar^i$ and $\xi_i$.
   }
\end{figure}

Then the 5D SUSY transformation is expressed as
\begin{eqnarray}
\del_\xi A^i=-\sqtwo\ep^{ij}\xibar_j\chi=-\sqtwo \graph{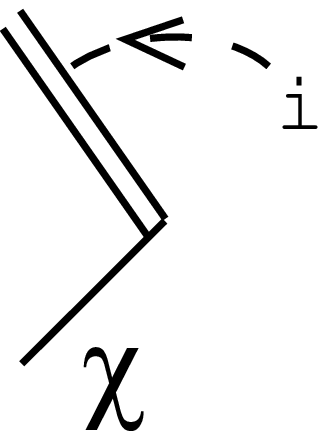}\com\nn
\del_\xi\chi=\sqtwo i\ga^M\pl_M A^i\ep_{ij}\xi^j+\sqtwo F_i\xi^i\nn
=\sqtwo i\graph{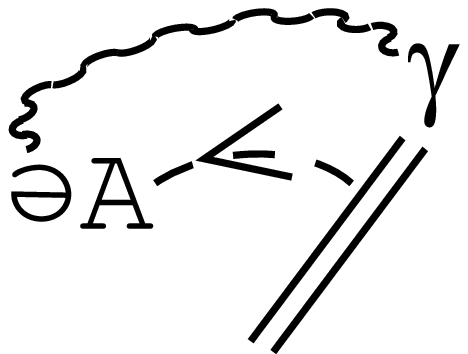}+\sqtwo\graph{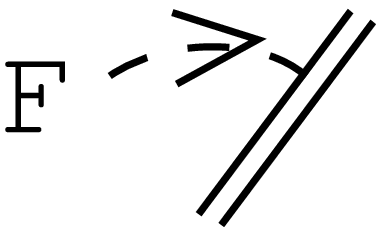}\com\nn
\del_\xi F_i=\sqtwo i\xibar_i\ga^M\pl_M\chi=\sqtwo i\graph{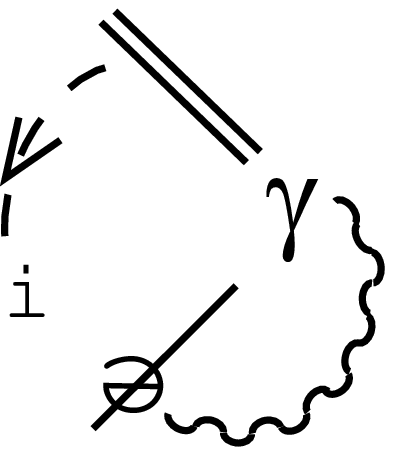}
\com\label{fds2}
\end{eqnarray}
where a wavy line is used to express the contraction of 
the 5D space-time suffixes ($M=0,1,2,3,5$).
\footnote{
5D Dirac gamma matrix is taken to be $(\ga^M)=(\ga^m,\ga^5)$ where
$\ga^m$ and $\ga^5$ are defined in (\ref{majo4}).
}
The complex conjugate one is given by
\begin{eqnarray}
\del_\xi A^*_i=\sqtwo\ep_{ij}\chibar\xi^j=\sqtwo \graph{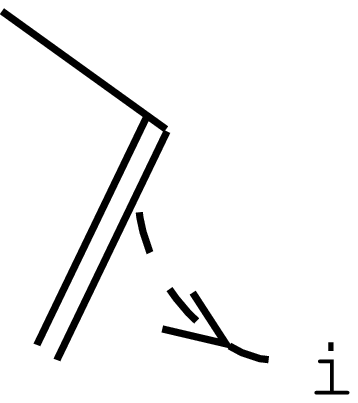}\com\nn
\del_\xi\chibar=-\sqtwo i\xibar_i\ga^M\pl_M A^*_j\ep^{ij}+\sqtwo \xibar_i F^{*i}\nn
=-\sqtwo i\graph{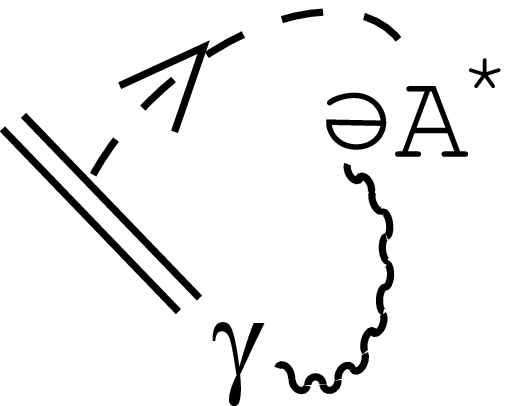}+\sqtwo\graph{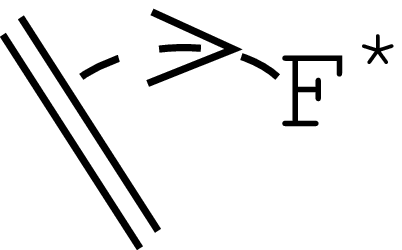}\com\nn
\del_\xi F^{*i}=-\sqtwo i\pl_M\chibar\ga^M\xi^i=-\sqtwo i\graph{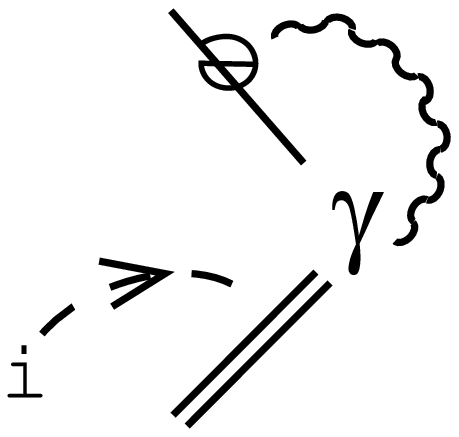}
\com\label{fds3}
\end{eqnarray}

The free Lagrangian is given by
\begin{eqnarray}
\Lcal=-i\chibar\ga^M\pl_M\chi-\pl^M A^*_i\pl_M A^i+F^{*i}F_i\nn
=-i\graph{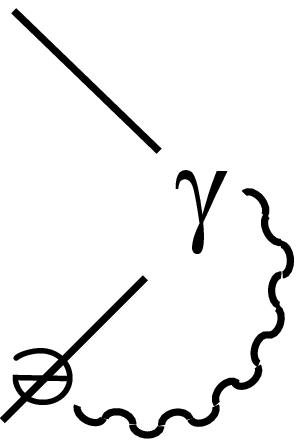}-\graph{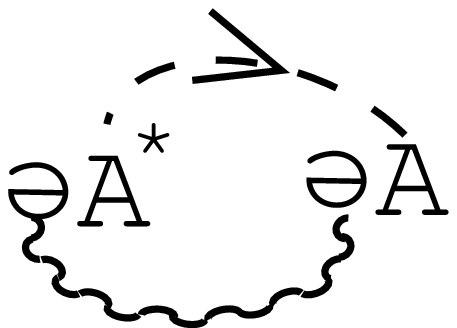}+\graph{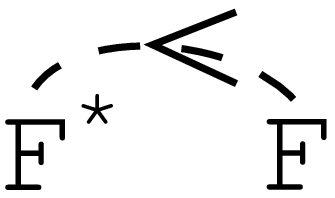}
\com\label{fds4}
\end{eqnarray}
Using the graphical rule of Fig.20 ($A^i(\ep_{ij}\xi^j)=-(\ep_{ji}A^i)\xi^j$),
and the basic spinor relation $\{\ga^M,\ga^N\}=-2\eta^{MN}$,
we can graphically confirm the SUSY invariance.
\footnote{
$\eta^{MN}=\mbox{diag}(-1,1,1,1,1)$
}
\begin{eqnarray}
\del_\xi\Lcal
=\pl_M\left\{-\sqtwo i\graph{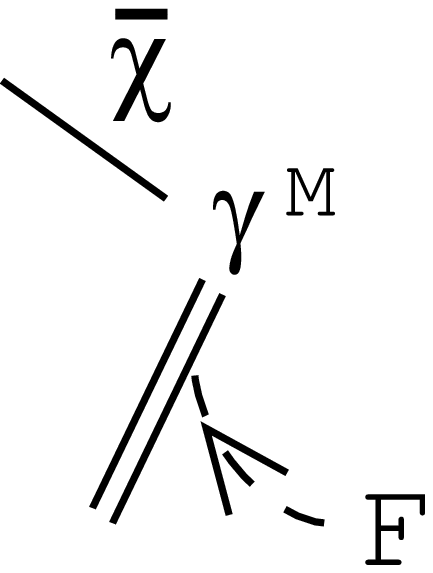}-\sqtwo\graph{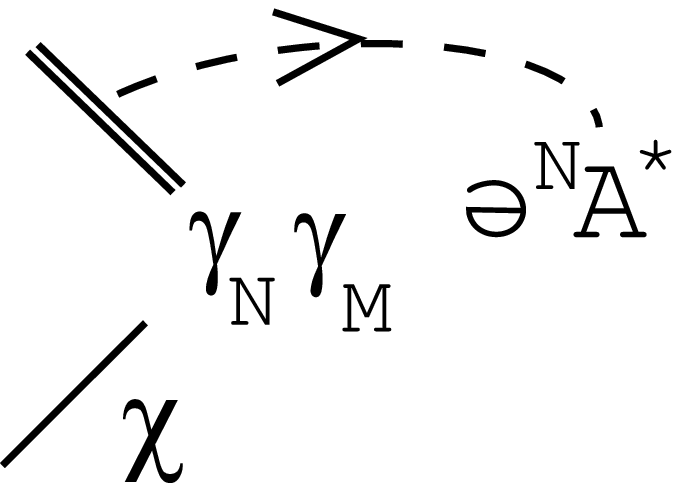}\right.\nn
\left.+\sqtwo\graph{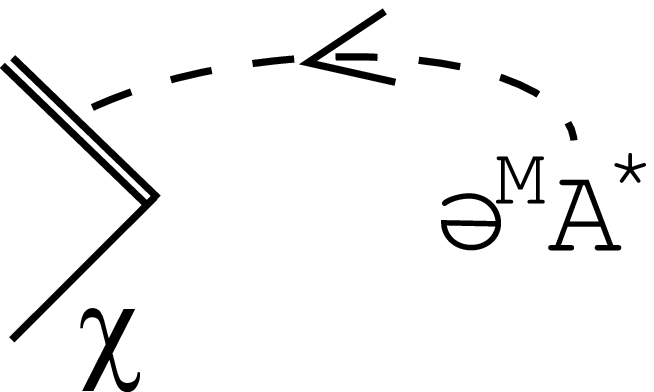}
-\sqtwo\graph{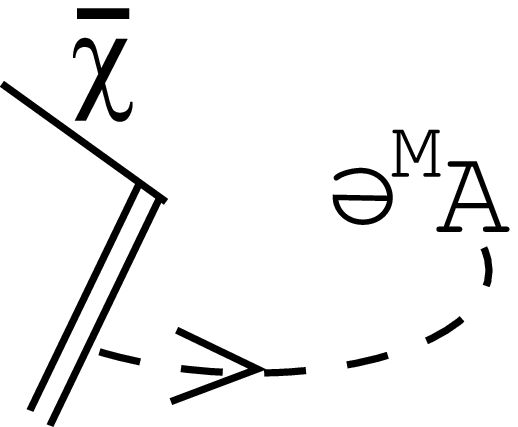}\right\}
\pr\label{fds5}
\end{eqnarray}

\begin{figure}
\begin{center}
\includegraphics*[height=2cm]{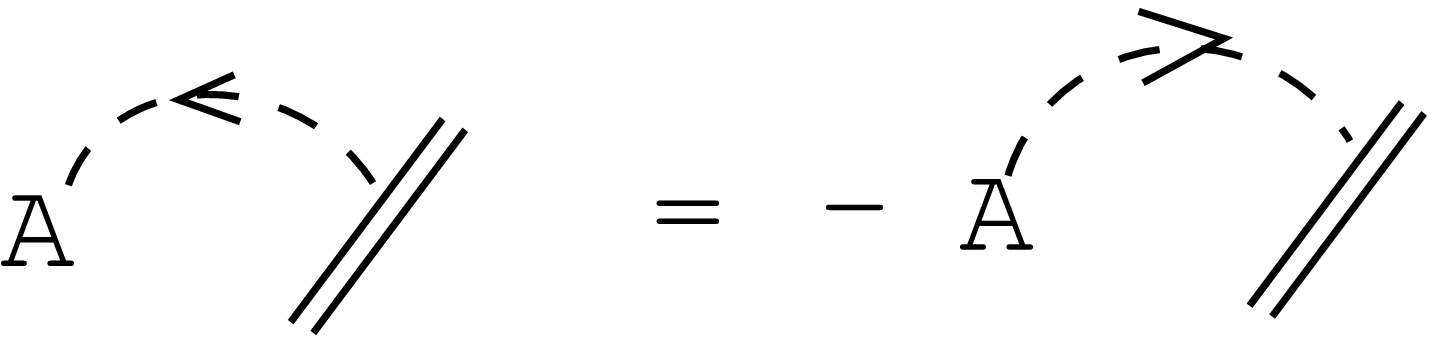}
\end{center}
   \caption{
The graphical rules for the relation:\ 
$A^i(\ep_{ij}\xi^j)=-(\ep_{ji}A^i)\xi^j$.
   }
\end{figure}

In relation to the SUSY decomposition, we rewrite the Dirac fields
($\chi,\chibar,\xi^i,\xibar^i$) in terms of Weyl spinors.
\begin{eqnarray}
\chi=\left(\begin{array}{c} (\chi_L)_\al \\ (\chibar_R)^\aldot
           \end{array}
     \right) \com\q
\chibar=\left( (\chi_R)^\al,(\chibar_L)_\aldot \right)\com\nn
\xi^1=\left(\begin{array}{c} (\xi_{1L})_\al \\ (\xibar_{1R})^\aldot
           \end{array}
     \right) 
=\left(\begin{array}{c} (\xi_{1L})_\al \\ (\xibar_{2L})^\aldot
           \end{array}
     \right) 
=\left(\begin{array}{c} \graph{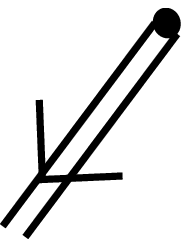} \\ \graph{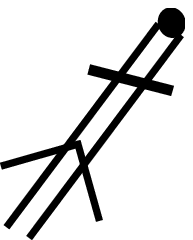}
           \end{array}
     \right) 
\com\nn
\xibar_1=\left( (\xi_{1R})^\al,(\xibar_{1L})_\aldot \right)
=\left( (\xi_{2L})^\al,(\xibar_{1L})_\aldot \right)
=\left( \graph{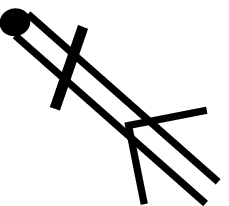},\graph{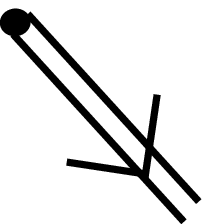} \right)\com\nn
\xi^2=\left(\begin{array}{c} (\xi_{2L})_\al \\ (\xibar_{2R})^\aldot
           \end{array}
     \right) 
=\left(\begin{array}{c} (\xi_{2L})_\al \\ -(\xibar_{1L})^\aldot
           \end{array}
     \right) 
=\left(\begin{array}{c} \graph{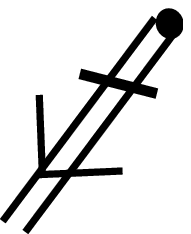} \\ -\graph{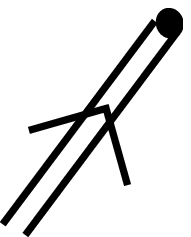}
           \end{array}
     \right) 
\com\nn
\xibar_2=\left( (\xi_{2R})^\al,(\xibar_{2L})_\aldot \right)
=\left( -(\xi_{1L})^\al,(\xibar_{2L})_\aldot \right)
=\left( -\graph{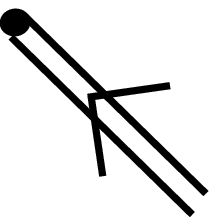},\graph{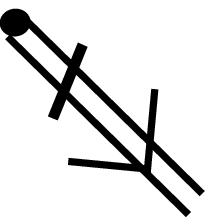} \right)\com
\label{fds6}
\end{eqnarray}
where the reality condition is used to express the SUSY parameters
by 8 (counted in real) independent quantities: $\xi_{1L}$, $\xi_{2L}$ and their
conjugates.
Then the 5D SUSY symmetry (\ref{fds2}) is decomposed as follows.
For the bosonic part, they are given by
\begin{eqnarray}
\begin{array}{cc}
(1)\ 
\frac{1}{\sqtwo}\del_\xi A^1  = 
-\graph{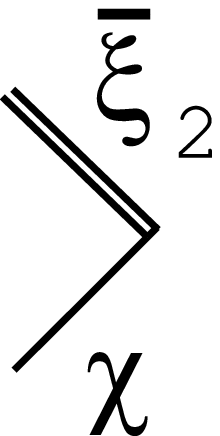} =& \\
\graph{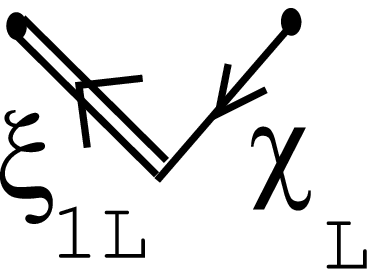}   &   -\graph{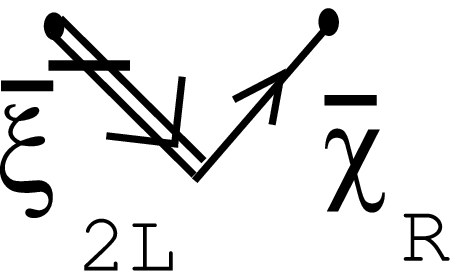}\ .\\
(2)\ \frac{1}{\sqtwo}\del_\xi A^2=\graph{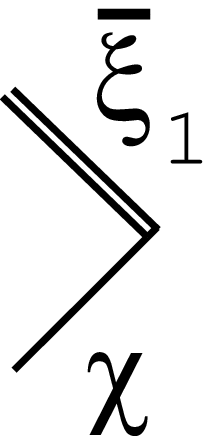}  = & \\
\graph{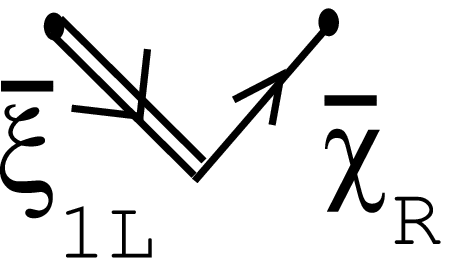}  & +\graph{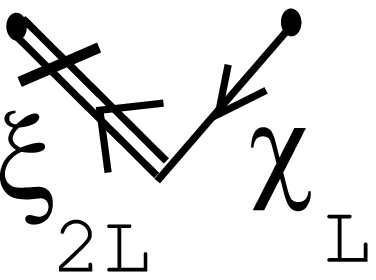}\ .\\
(3)\ \frac{1}{\sqtwo i}\del_\xi F_1=\graph{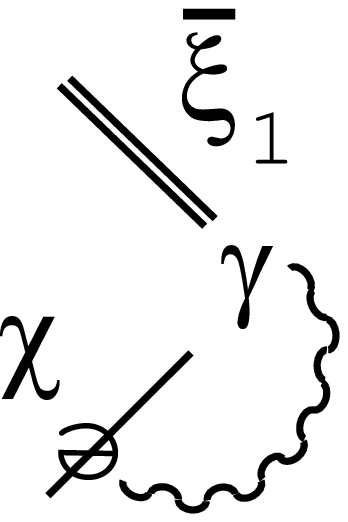}  =&  \\
\graph{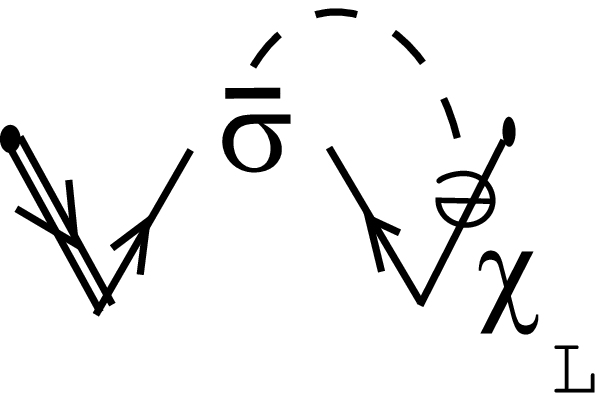}+i\graph{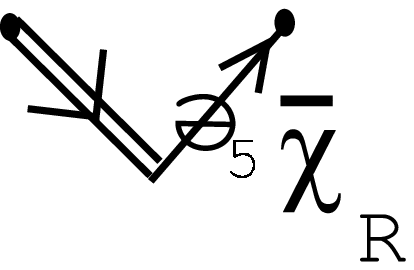} & 
+\graph{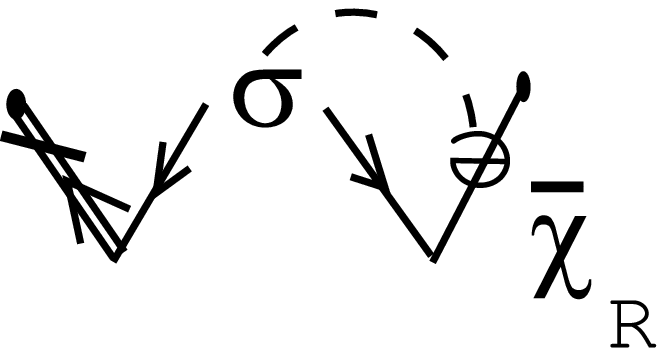}-i\graph{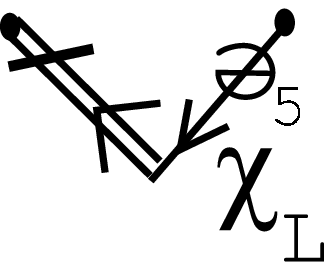}\ .\\
(4)\ \frac{1}{\sqtwo i}\del_\xi F_2=\graph{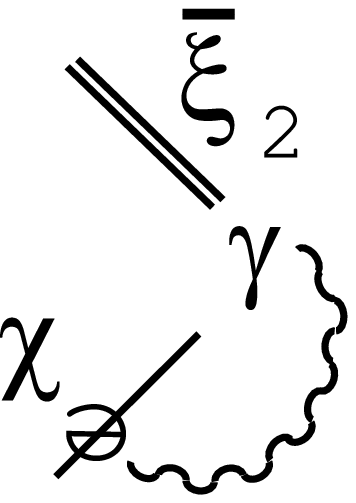} =& \\
-\graph{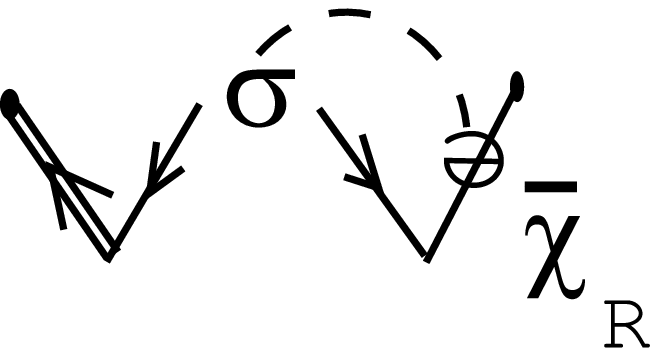}+i\graph{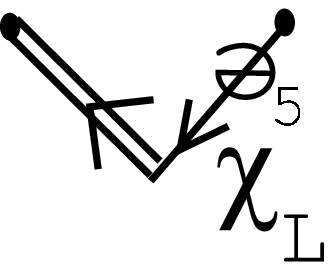}  &
+\graph{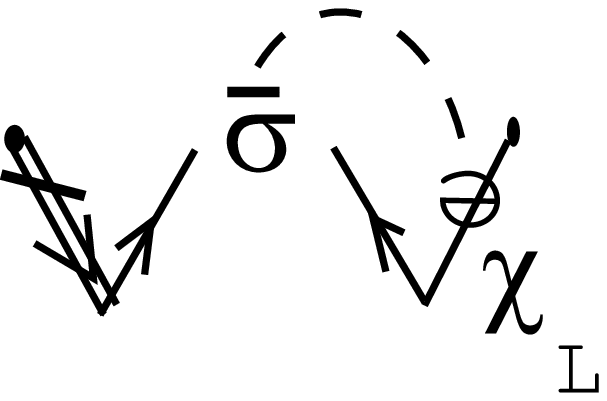}+i\graph{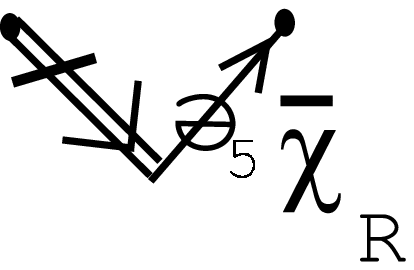}\ .
\end{array}
\label{fds7}
\end{eqnarray}

For the fermionic part they are given by
\begin{eqnarray}
(6)\ \frac{1}{\sqtwo}\del_\xi\chi_L = 
i\graph{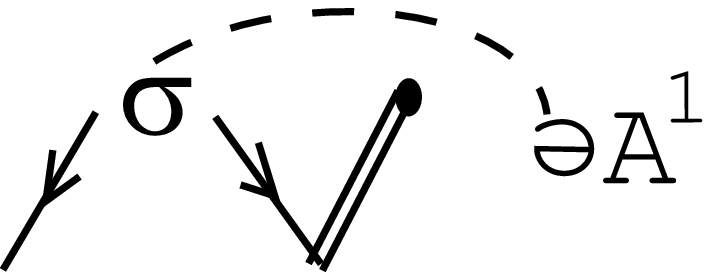}+(F_2+\pl_5 A^2)\graph{F20fds7} \nn
\q\q\q +i\graph{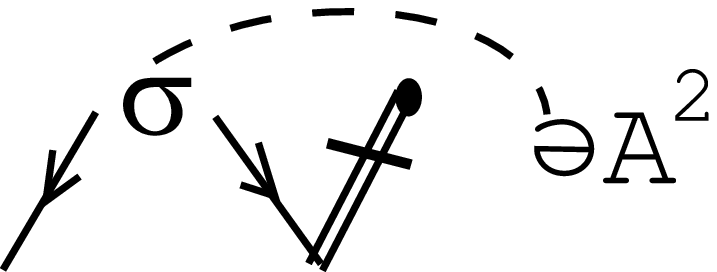}+(F_2-\pl_5 A^1)\graph{F22fds7}\ .\nn
(7)\ \frac{1}{\sqtwo}\del_\xi\chibar_R =
i\graph{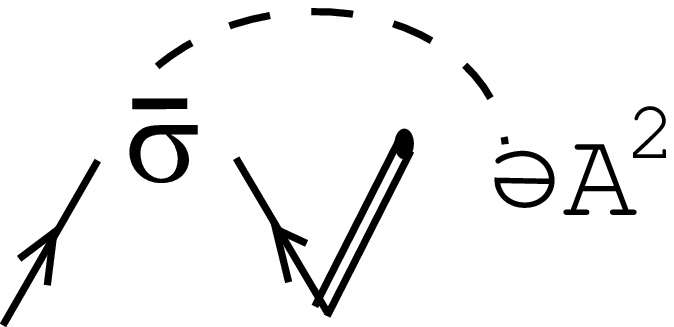}-(F_2+\pl_5 A^1)\graph{F26fds7} \nn 
\q\q\q -i\graph{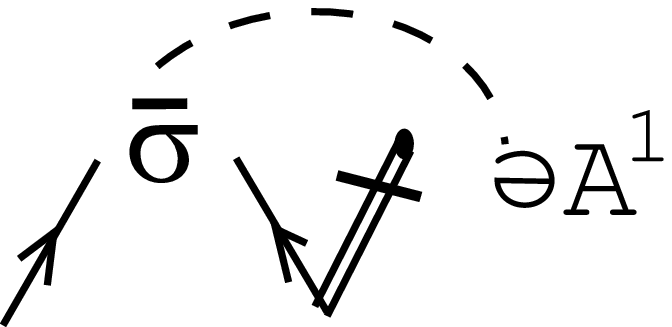}+(F_1-\pl_5 A^2)\graph{F24fds7}\ .\nn
\mbox{where}\q
(5)\ \frac{1}{\sqtwo}\del_\xi\chi=i\graph{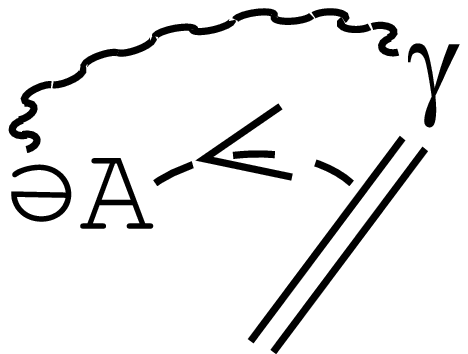}+\graph{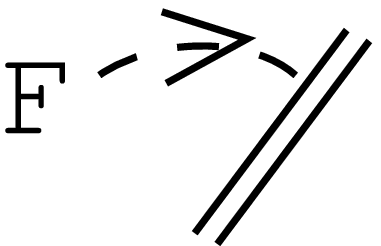}
=\left(
\begin{array}{c}\del_\xi\chi_L \\ \del_\xi\chibar_R\end{array}
  \right)\ .
\label{fds7b}
\end{eqnarray}
\footnote{The graph equations in
(\ref{fds7}) are, in the ordinary expression, as follows.\nl
$
(1)\ \del_\xi A^1/\sqtwo=\del_\xi A_2/\sqtwo=-\xibar_2\chi=
(\xi_{1L})^\al(\chi_L)_\al-(\xibar_{2L})_\aldot (\chibar_R)^\aldot;\nl
(2)\ \del_\xi A^2/\sqtwo=-\del_\xi A_1/\sqtwo=\xibar_1\chi
=\xibar_{1L}\chibar_R+\xi_{2L}\chi_L;\nl 
(3)\ \del_\xi F_1/(\sqtwo i)=-\del_\xi F^2/(\sqtwo i)=
\xibar_{1L}\sibar^m\pl_m\chi_L +
i(\xibar_{1L})_\aldot (\pl_5\chibar_R)^\aldot +
\xi_{2L}\si^m\pl_m\chibar_R
-i(\xi_{2L})^\al(\pl_5\chi_L)_\al;\nl 
(4)\ \del_\xi F_2/(\sqtwo i)=\del_\xi F^1/(\sqtwo i)=
-\xi_{1L}\si^m\pl_m\chibar_R
+i(\xi_{1L})^\al(\pl_5\chi_L)_\al
+\xibar_{2L}\sibar^m\pl_m\chi_L+
i(\xibar_{2L})_\aldot (\pl_5\chibar_R)^\aldot;\nl 
(5)\ \del_\xi \chi/\sqtwo=i\ga^M\pl_MA^i\ep_{ij}\xi^j+F_i\xi^i
=(\del_\xi\chi_L , \del_\xi\chibar_R)^T;\nl 
(6)\ \del_\xi\chi_L/\sqtwo=i\si^m\xibar_{1L}\pl_mA^1+(F_1+\pl_5A^2)\xi_{1L}
+i\si^m\xibar_{2L}\pl_mA^2+(F_2-\pl_5A^1)\xi_{2L};\nl 
(7)\ \del_\xi\chibar_R/\sqtwo=
i\sibar^m\xi_{1L}\pl_mA^2-(F_2+\pl_5A^1)\xibar_{1L}
-i\sibar^m\xi_{2L}\pl_mA^1+(F_1-\pl_5A^2)\xibar_{2L}.
$
}
Let us compare the above result with the decomposition structure 
in the case of 4D: Majorana (4-comonents) to Weyl (2-components). 
In the 4D case, the decomposition 
is done with respect to chirality (left versus right), and
$\gago$ controls it. 
In the present case of 5D, the decomposition is done
with respect to $(\xi_{1L},\xibar_{1L})$ and $(\xi_{2L},\xibar_{2L})$, 
 the labels 1 and 2 control it. 
Here we note that there is no "chiral matrix" in 5D
in the sense that $\prod_M \ga^M \propto 1$. 
Next we explain what symmetry plays the role
of separating 1 and 2. 


In relation to the decomposition (to $\Ncal=1$\,SUSY) procedure, 
we introduce Z$_2$-symmetry,
that is the {\it reflection} in the origin in the (fifth) extra coordinate.
\begin{eqnarray}
x^5\q\change\q -x^5
\pr\label{fds8}
\end{eqnarray}
We {\it assign} the Z$_2$-parity to all fields in a consistent way
with the decomposition relations (\ref{fds7}). A choice is 
given in Table 4.

$$
\begin{array}{|c|c|c|}
\hline
             &   P=+1\ ,\ \xi_{1L}         & P=-1\ ,\ \xi_{2L}   \\
\hline
A^i  & A^1 & A^2 \\
\hline
\chi  & \chi_L & \chi_R \\
\hline
F_i   & F_1 & F_2 \\
\hline
\multicolumn{3}{c}{\q}                                                 \\
\multicolumn{3}{c}{
\mbox{Table\ 4}\ \ Z_2-\mbox{parity~ assignment.}
                  }\\
\end{array}
$$

Then the the 5D SUSY symmetry is decomposed to two $\Ncal=1$ chiral
multiplets;\ one is for $P=+1$ states, the other is for $P=-1$ states.
Up to now, the SUSY decomposition is not directly related with 
the space-time dimensional reduction because all fields depend
both on the 4D coordinates $x^m$ and on the extra one $x^5$. 
Let us consider the case that the present 5D SUSY system 
has the {\it localized} configuration around the origin in
the extra coordinate.  
Then we can naturally suppose that 
the $Z_2$-symmetry, which is required from the configuration,
restricts the boundary condition of the fields and
the whole system 
decomposes into even-parity and odd-parity fields.
Then the dimensional reduction occurs.

\section{Discussion and conclusion}
The use of graphs is popular in the history of mathematics and theoretical
physics. 
Penrose \cite{Pen71}, with the similar motivation described in the
introduction, proposed a diagrammatical (graphical) notation
in the tensorial and spinorial calculation.
Feynman diagram is the most familiar graph method to represent a scattering
amplitude.
The diagram tells us, without the explicit calculation, 
important features such as
the coupling dependence, mass dependence and the
divergence degree. 
Nakanishi\cite{Nakanishi} analysed the Feynman amplitude
using the graph theory in mathematics. 
In this sense
quite a large part of mathematical physics relies on
the use of the graph. 

As an application of the present approach, supergravity is interesting.
Relegating the full treatment to a separate work\cite{SI03gr}, we indicate a graphical
advantage here. There appears the following quantity in the supergravity.:\ 
$\psi_{\del\deldot\ga\gadot\al}=(\si^d)_{\del\deldot}(\si^c)_{\ga\gadot}
e_d^{~n}e_c^{~m}(\psi_{nm})_\al,\ (\psi_{nm})^\al=\pl_n\psi_m^\al+\cdots  
-n\change m.$ Graphically it is expressed as
\begin{eqnarray}
\begin{array}{c}\mbox{
\includegraphics*[height=2cm]{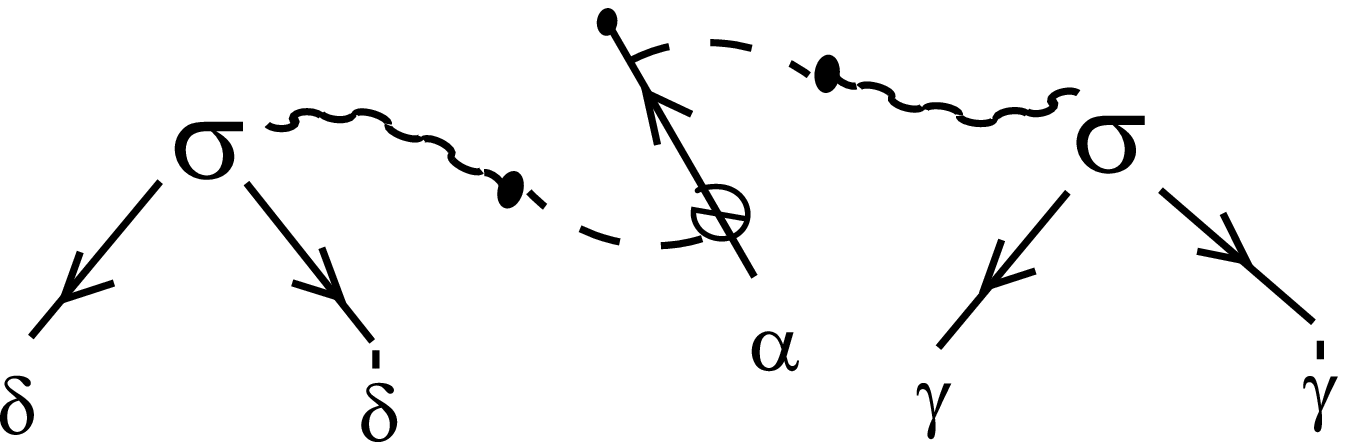}
}\end{array}
\q+\cdots\com
\label{conc1}
\end{eqnarray}
where we introduce the graphical representation for the vier-bein $e^n_{~a}$
and the Rarita-Schwinger field $\psi_m^{~\al}$ as
\begin{eqnarray}
e^n_{~a}\ :\q
\begin{array}{c}\mbox{
\includegraphics*[height=1cm]{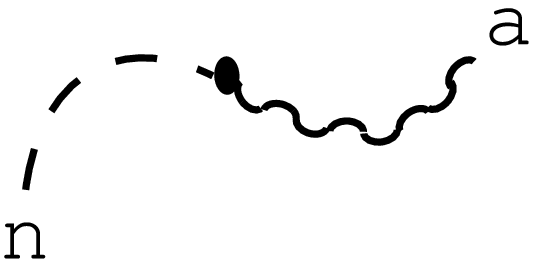}
}\end{array}\q\q
\psi_m^{~\al}\ :\q
\begin{array}{c}\mbox{
\includegraphics*[height=1cm]{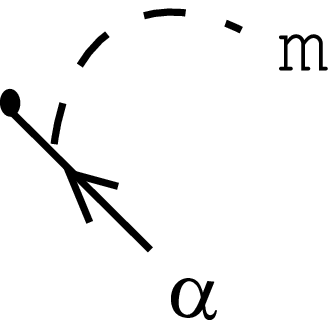}
}\end{array}
\label{conc2}
\end{eqnarray}
The set of indices, which specifies the above graph (\ref{conc1}),
is given as follows:\ 
$(LCN,RCN)=(3/2,1), (LUDN,RUDN)=(-1/2,-1), (LWN,RWN)=(0,0), 
DIM=5/2, DIF=1, SIG=2
$.

%

We have presented a graphical representation
of the supersymmetric theory. It has some
advantages over the conventional description.
The applications are diverse. Especially
the higher dimensional suspergravities
are the interesting physical models
to apply the present approach. 
In the ordinary approach, it has a technical problem which
hinders analysis. 
The theory is so "big" that it is rather hard 
in the conventional approach.
The present graphical description is
expected to resolve or reduce the technical
but an important problem. We point out that the
present representation is suitable for coding
SUSY calculations. 
The application has recently been done in \cite{SI06UW08} where
the transformation from the superfield expression
to the component one is demonstrated. 
\footnote{ 
See also \cite{SI98IJMPC,II97JMP} for the C-language program
and graphical calculation for the product of Riemann tensors.
}


\vs {5}
\begin{flushleft}
{\bf Acknowledgment}
\end{flushleft}
The basic idea of this work was born 
during the author's stay at Albert Einstein Institute (fall of 1999) . 
The author thanks the hospitality at the institute. 
He also thanks M. Abe, N. Ikeda, T. Kugo and N. Nakanishi
for comments and criticism in the RIMS(Kyoto Univ.) workshop(2002.9.30-10.2).
This work is completed in the present form in the author's stay
at DAMTP, Univ. of Cambridge(Winter of 2003) and ITP, Universit{\"a}t Wien(Spring of 2006). 
He thanks the hospitality there.
The author thanks G.W. Gibbons for comments and reference information. 
He also thanks A. Bartl for carefully reading the manuscript, and 
S. Rankin for the help in computer work. 

Finally the author thanks the governor of the Shizuoka prefecture for
the financial support.


\vs {5}


\end{document}